\documentclass[aps,10pt,prd,superscriptaddress,preprintnumbers,%
showkeys,nofootinbib,showpacs,twocolumn]{revtex4-2}
\usepackage{amsmath}
\usepackage{graphicx}
\usepackage{bm}
\usepackage{hyperref}
\begin{document}  
\title{Chiral-odd generalized parton distributions in the large-$N_{c}$ limit of QCD: \\
Spin-flavor structure, polynomiality, sum rules} 
\author{June-Young Kim}
\email[E-mail: ]{jykim@jlab.org}
\affiliation{Theory Center, Jefferson Lab, Newport News, VA 23606, USA}
\author{Christian Weiss}
\email[E-mail: ]{weiss@jlab.org}
\affiliation{Theory Center, Jefferson Lab, Newport News, VA 23606, USA}
\date{\today}
\begin{abstract}
We study the nonperturbative properties of the nucleon's chiral-odd generalized parton distributions
(transversity GPDs) in the large-$N_c$ limit of QCD.
This includes the parametric ordering of the spin-flavor components,
the polynomiality property of the moments, and the sum rules connecting the GPDs with the tensor form factors.
A multipole expansion in the transverse momentum transfer is used to enumerate and interpret the structures
in the nucleon matrix element of the chiral-odd partonic operator, including monopole, dipole and
quadrupole terms. The $1/N_c$ expansion of the GPDs is performed using the abstract mean-field picture of baryons
in the large-$N_c$ limit and its symmetries.
We derive a large-$N_c$ relation between the flavor-nonsinglet GPDs $E_T^{u-d}$ and $\tilde H_T^{u-d}$
and test it with recent lattice QCD results. We show that the polynomiality property and sum rules of the GPDs
are fulfilled with the restricted realization of translational and rotational invariance in the mean-field picture.
The results provide a basis for the phenomenological analysis of chiral-odd GPDs and hard exclusive
processes in the large-$N_c$ limit, and for calculations in specific dynamical models.
\end{abstract}
\maketitle
\tableofcontents
\section{Introduction}
\setlength{\parskip}{2ex}
Generalized parton distributions (GPDs) are an essential instrument for expressing nucleon structure
in QCD. They describe the nonforward nucleon matrix elements of partonic QCD operators representing
correlation functions of the quark and gluon fields at light-like distances.
The GPDs unify the concepts of the nucleon parton densities and elastic form factors (FFs)
and open new possibilities of characterizing the dynamical system; see
Refs.~\cite{Goeke:2001tz,Diehl:2003ny,Belitsky:2005qn,Boffi:2007yc} for a review.
The transverse coordinate representation of the GPDs describes the spatial distribution
of quarks/antiquarks/gluons in the nucleon and allows one to represent it as an extended object
in space \cite{Burkardt:2000za,Burkardt:2002hr,Diehl:2002he}.
The moments of the GPDs ($x$-weighted integrals) describe nucleon FFs of local QCD operators
of twist 2 and spin $n \geq 1$, which arise from the expansion of the partonic operator in powers of
the light-like distance. These local operators contain the QCD energy-momentum tensor ($n = 2$),
whose FFs describe the distribution of mass, angular momentum, and forces in the nucleon
and allow one to quantify its internal mechanical
properties \cite{Leader:2013jra, Polyakov:2018zvc, Lorce:2018egm, Burkert:2023wzr}.

The quark GPDs come in two types, determined by the chiral properties of the partonic operator
$\bar\psi (-z/2) \Gamma \psi (z/2)$, where $\psi$ and $\bar\psi$ are the QCD quark fields,
$z$ is the light-like 4-vector distance, $z^2 = 0$, and $\Gamma$ is a bispinor matrix.
The chiral-even operators with $\Gamma = \gamma^+$ or $\gamma^+\gamma^5$ define the so-called chiral-even GPDs.
They reduce to the unpolarized and helicity-polarized quark densities in the forward limit, and their first
moments are the FFs of the vector and axial vector current.
The chiral-odd operators with $\Gamma = \sigma^{+\perp}$ define the chiral-odd GPDs.
They correspond to the transversity-polarized quark densities in forward limit, and their
first moments are related to the FFs of the local tensor operator $\bar\psi (0) \sigma^{\mu\nu} \psi (0)$,
which include the tensor charge FF defining the tensor charge at zero momentum transfer.
(Here $+$ and $\perp$ are the light-cone 4-vector components along the direction of $z$;
definitions are given below.)

The chiral-even GPDs have been extensively studied regarding their general properties,
applications to nucleon structure, and model
predictions \cite{Goeke:2001tz,Diehl:2003ny,Belitsky:2005qn,Boffi:2007yc}.
The chiral-odd GPDs remain relatively unexplored, with many of their basic properties still unknown.
The chiral-odd GPDs possess unique features and present new probes of nucleon structure.
Chiral-odd quark operators do not mix with gluon operators (they are non-singlets) and depend only
weakly on normalization scale, which makes them effective probes of nonperturbative nucleon structure;
see Ref.~\cite{Barone:2001sp} for a review.
The hadronic matrix elements of chiral-odd operators can be regarded as manifestations of the
dynamical breaking of chiral symmetry in QCD, a fundamental effect governing the long-range behavior of the
effective dynamics and the structure of light hadrons. Comparison of chiral-even and chiral-odd structures reveals
relativistic effects in the context of the quark-model picture of the nucleon. The chiral-odd GPDs
describe the effect of transverse quark polarization on the spatial distributions \cite{Burkardt:2005hp,Diehl:2005jf}.
Their moments give access not only to the nucleon tensor charge but also to higher multipoles of the
tensor operator, such as the anomalous tensor magnetic moment and the quadrupole moment.
Studies of the properties of the chiral-odd GPDs are needed to enable these applications to nucleon structure.

GPDs parametrize the nucleon structure sampled in hard exclusive processes in lepton-nucleon scattering
in the context of QCD factorization \cite{Goeke:2001tz,Diehl:2003ny,Belitsky:2005qn,Boffi:2007yc}.
In the asymptotic regime (energy and momentum transfer $\gg$ hadronic scale) the scattering process
takes place on a single quark, whose emission and absorption by the nucleon is described by the GPDs.
This makes it possible to extract information on the GPDs from observables in exclusive processes.
Chiral-even GPDs appear in the factorization of deeply-virtual Compton scattering, $l + N \rightarrow
l + \gamma + N$, and in the asymptotic twist-2 mechanism of exclusive meson production, 
$l + N \rightarrow l' + M + N$.
Chiral-odd GPDs appear in the twist-3 mechanism of pseudoscalar
meson production, where they are combined with the chiral-odd meson distribution amplitudes
\cite{Ahmad:2008hp, Goloskokov:2009ia, Goloskokov:2011rd, Goldstein:2012az}.
While theoretically subleading, this mechanism produces large amplitudes at energy/momentum
transfers $\sim$ few GeV and predicts observables consistent with measurements at
JLab 6 GeV \cite{CLAS:2012cna,CLAS:2014jpc,CLAS:2017jjr}.
This opens the prospect of probing chiral-odd GPDs in pseudoscalar meson production experiments
at JLab 12 GeV (see Ref.~\cite{CLAS:2023wda} for first results), and at higher energies in the
CERN COMPASS experiment \cite{COMPASS:2019fea} and the future Electron-Ion Collider (EIC)
\cite{AbdulKhalek:2021gbh}.
The same chiral-odd mechanism can be applied to the time-reversed process of dilepton production
in pion-nucleon scattering, $\pi N \rightarrow l^+l^- + N$, which can be measured at J-PARC \cite{Aoki:2021cqa}.
Chiral-odd GPDs are also probed in exclusive high-mass pair production processes
\cite{Enberg:2006he,ElBeiyad:2010pji}.
Quantitative predictions for chiral-odd GPDs are needed to analyze the data and
simulate future experiments.

The $1/N_c$ expansion is a powerful method for analyzing nucleon structure in QCD.
The limit of a large number of colors corresponds to a semiclassical limit of QCD, in which the
dynamics simplifies in characteristic ways yet retains essential nonperturbative features
such as the dynamical breaking of conformal and chiral symmetry
\cite{tHooft:1973alw,Witten:1979kh,Coleman:1980mx}.
The $N_c$ scaling of hadron masses, couplings, and FFs can be established on general grounds.
Baryons appear as mean-field solutions with mass $\mathcal{O}(N_c)$ and size $\mathcal{O}(N_c^0)$,
analogous to solitons in classical field theories \cite{Witten:1979kh}.
The mean field couples spatial and flavor degrees of freedom and imposes a characteristic
spin-flavor symmetry on baryon properties \cite{Gervais:1983wq,Dashen:1993jt}.
Baryon states with spin-isospin quantum numbers
arise from quantization of the rotational zero modes, with the
$N$ and $\Delta$ appearing as rotational states with $S = T =$ 1/2 and 3/2 \cite{Witten:1979kh}.
Matrix elements of QCD operators such as GPDs can be analyzed according to their $N_c$ scaling.
The different spin-isospin components of the matrix elements exhibit different $N_c$ scaling,
dictated by the symmetries of the mean field and the quantization of the zero modes.

Large-$N_c$ methods can be applied to the analysis of nucleon matrix elements of QCD operators
at two levels. (i) Deriving the $N_c$ scaling of the spin-flavor components of the matrix elements
and relations between them. This can be done using only the symmetries of the mean field
and zero mode quantization. It does not require dynamical input, and the results are model-independent.
(ii) Making quantitative predictions for matrix elements. This is possible with dynamical models
that generate a specific mean-field solution and predict the expectation value of QCD operator
in the mean field. The zero-mode quantization then generates quantitative predictions for the
spin-flavor components of matrix elements. An example is the chiral quark-soliton model based on
the effective dynamics of massive quarks coupled
to a chiral meson field \cite{Diakonov:1987ty,Wakamatsu:1990ud,Christov:1995vm}.

In this work we study the chiral-odd GPDs of the nucleon in the large-$N_c$ limit of QCD.
The investigation is divided in two parts.
In the first part (reported in the present article) we derive the general properties of the chiral-odd GPDs
in large-$N_c$ limit, including the multipole expansion, the $N_c$ scaling of spin-flavor components,
the realization of polynomiality, and the sum rules and connection with the tensor FFs.
Here we use only model-independent features of baryon structure in the large-$N_c$ limit.
In the second part (reported in a subsequent article)
we obtain numerical estimates in the chiral quark-soliton model and discuss
the dynamical properties of the chiral-odd GPDs, such as the magnitude of the spin-flavor components,
valence and sea quark distributions, and the role of long-distance chiral dynamics.

A general technique for performing the $1/N_c$ expansion of GPDs was described in Ref.~\cite{Goeke:2001tz}
and applied to the analysis of chiral-even GPDs. Applications to chiral-odd GPDs were
considered in Ref.~\cite{Schweitzer:2016jmd}.
The present work extends the treatment of the chiral-odd GPDs in several aspects
and revises some of the earlier results. New elements are: 

\textit{Multipole expansion:} We systematically employ the multipole expansion to characterize the
momentum transfer and spin dependence of the nonforward matrix elements.
The method allows us to classify the structures and explain the observed dependence.
The use of multipoles is natural in the context of the $1/N_c$ expansion of baryon matrix elements,
as the $N_c$ scaling of structures
is determined by $t$-channel spin-isospin quantum numbers (so-called $I = J$ rule)
\cite{Mattis:1988hf, Mattis:1988hg, Lebed:2006us}.
In GPDs the multipole expansion is performed as an expansion in the transverse momentum transfer
for fixed longitudinal momentum variables. We compare this 2-dimensional (2D) multipole expansion with the
3-dimensional (3D) multipole expansion in the Breit frame to explain the observed structures.

\textit{Partonic and local operators:} We consider in parallel the matrix elements of the chiral-odd
partonic operator and the local tensor operator ($n = 1$ moment). The multipole expansion and $1/N_c$
expansion are performed for both operators. This approach helps us to motivate the structures appearing
in the multipole expansion of the partonic operator and explain their $N_c$ scaling. We find
that the decomposition of the partonic operator includes a quadrupole structure that was omitted
in the analysis of Ref.~\cite{Schweitzer:2016jmd}. With the complete decomposition, we are able to
derive the $N_c$ scaling of all the spin-flavor components and obtain a new non-trivial large-$N_c$
relation between the leading chiral-odd GPDs $E_T^{u-d}$ and $\tilde H_T^{u-d}$, which can be tested
with model predictions or experimental data.

\textit{Polynomiality in the large-$N_c$ limit:} GPD moments are matrix elements of local spin-$n$
tensor operators, whose dependence on the momentum transfer is constrained by relativistic covariance.
When expressed in partonic variables, this leads to the so-called polynomiality property,
that the GPD moments of spin $n$ are polynomials in the fractional light-cone momentum transfer
$\xi$ (so-called skewness) of degree not exceeding $n$
\cite{Goeke:2001tz,Diehl:2003ny,Belitsky:2005qn,Boffi:2007yc}. It represents a general constraint on
the GPDs as functions of the partonic variables and plays an important role in their analysis.
In the large-$N_c$ limit the realization of polynomiality is non-trivial, as relativistic
covariance is realized only order-by-order in the $1/N_c$ expansion. This is connected with the
fact that the baryon mean field at large $N_c$ breaks translational and rotational invariance,
and that they are restored by quantization of the zero modes. 
The problem of ``symmetry breaking by the mean field'' is well known from nonrelativistic nuclear physics 
and has been studied in that context; see Ref.~\cite{Sheikh:2019qdz} for a review.
Polynomiality of the GPDs in the large-$N_c$ limit was
demonstrated in the chiral-even sector in Ref.~\cite{Goeke:2007fp}, using the chiral quark-soliton model
as a specific realization. We show here that polynomiality in the large-$N_c$ limit
is also realized in the chiral-odd sector.
We present a more general formulation using an abstract mean-field picture and identify what symmetries
and other properties of the mean field bring about polynomiality. This approach can help to generalize
the large-$N_c$ analysis of GPDs and enable a formulation using group-theoretical methods.

Lattice QCD calculations of chiral-odd GPDs have been performed using local operators (moments)
\cite{Gockeler:2005cj,QCDSF:2006tkx,Park:2021ypf} and recently developed methods based on high-momentum
Euclidean correlation functions \cite{Alexandrou:2021bbo,Alexandrou:2022dtc}. The results can be used
to test the large-$N_c$ relations for the spin-flavor components of the FFs and GPDs.
The $1/N_c$ expansion can explain the hierarchy of structures observed in lattice calculations
and use lattice results to predict unknown spin-flavor components.

Section~\ref{sec:2} presents the definition of the chiral-odd GPDs and their general properties.
Section~\ref{sec:3} discusses the multipole expansion of the chiral-odd GPDs and the local tensor operator.
Section~\ref{sec:4} analyzes the spin-flavor structure of the chiral-odd GPDs at large $N_c$.
The general method for the $1/N_c$ expansion of GPDs is presented,
using the abstract mean-field picture of baryons in the large-$N_c$ limit.
The method is then applied to the $1/N_c$ expansion of the chiral-odd GPDs.
The $N_c$ scaling of the spin-flavor components and the large-$N_c$ relation between
the leading chiral-odd GPDs are derived.
The large-$N_c$ relation is then tested with recent lattice QCD results. 
Section~\ref{sec:5} covers the realization of polynomiality of chiral-odd
GPDs at large $N_c$. The GPDs are represented in a first-quantized form
within the abstract mean-field picture, and polynomiality is demonstrated
using general properties of the quark single-particle operators. 
Section~\ref{sec:6} establishes the connection between the chiral-odd GPDs
and the FFs of the local tensor operator in the large-$N_c$ limit.
The $1/N_c$ expansion of the FFs is performed, and the realization of the
sum rules in the mean-field picture is discussed.
Section~\ref{sec:7} summarizes the conclusions and suggestions for further studies.
Appendix~\ref{app:a} presents the proof of polynomiality of
the higher multipoles of the chiral-odd GPDs.
Appendix~\ref{app:sumrules} presents the proof of the sum rules for the higher multipoles
of the chiral-odd GPDs and tensor FFs.
\section{Chiral-odd GPDs \label{sec:2}}
The matrix element of the chiral-odd partonic QCD operator between nucleon states is defined as
\begin{align}
& \mathcal{M}_{\mathrm{GPDs}}[i\sigma^{+j}] =
P^{+}\int \frac{d z^{-}}{2\pi} e^{i x P^{+} z^{-} } 
\nonumber \\
& \times \langle p', s' | \bar{\psi} \left(-\frac{z}{2}\right) 
\left[-\frac{z}{2}, \frac{z}{2} \right] i\sigma^{+j} \psi \left(\frac{z}{2}\right) |  p, s \rangle 
\bigg{|}_{z^{+}, \bm{z}_\perp = 0} .
\label{eq:General_ME}
\end{align}
$\psi$ and $\bar\psi$ are the quark fields, $\sigma^{\mu\nu} \equiv (i/2) [\gamma^\mu,\gamma^\nu]$, 
and the light-cone 4-vector components are $v^\pm \equiv (v^0 \pm v^3)/\sqrt{2}$ and $v_\perp = (v^1, v^2)$. 
The space-time separation $z$ of the fields is light-like, and $[-z/2, z/2]$ denotes the gauge link
(Wilson line) along the straight lightlike line connecting the points.
The initial and final nucleon 4-momenta are $p$ and $p'$, and $P \equiv (p' + p)/2$ is the average 4-momentum.
The matrix element Eq.~(\ref{eq:General_ME}) is parametrized as
\begin{align}
& \mathcal{M}_{\mathrm{GPDs}}[i\sigma^{+j}] = 
\bar{u}' \left[ i \sigma^{+j} \, H_{T}
+ \frac{P^{+} \Delta^{j} - \Delta^{+} P^{j}}{M^{2}_{N}} \, \tilde{H}_{T} \right.
\nonumber \\
& \left. + \frac{\gamma^{+}\Delta^{j} -\Delta^{+} \gamma^{j} }{2M_{N}} \, E_{T}
+ \frac{\gamma^{+} P^{j} - P^{+} \gamma^{j}}{M_{N}} \, \tilde{E}_{T} 
\right] u.
\label{eq:General_ME_GPDs}
\end{align}
$u \equiv u(p, s)$ and $\bar u' \equiv \bar u(p', s')$ are the bispinor wave functions of initial and final nucleon,
normalized as $\bar{u} u = \bar u' u' = 2M_{N}$, where $M_{N}$ is the nucleon mass. $s$ and $s'$ are the
spin quantum numbers; the choice of spin states 
will be specified in following. $\Delta \equiv p'-p$ is the 4-momentum transfer. The functions $H_{T}$, 
$\tilde{H}_{T}$, $E_{T}$, and $\tilde{E}_{T}$ are the chiral-odd GPDs. They depend on the partonic variable $x$,
the fractional light-cone momentum transfer between the nucleon states,
$\xi \equiv -\Delta^+/2P^+$ (skewness); and the invariant momentum transfer
$t \equiv \Delta^2 $. The GPDs also depend on the normalization scale of the QCD operator;
this dependence will not be indicated explicitly.
The four terms in Eq.~(\ref{eq:General_ME_GPDs}) represent independent covariant structures;
their interpretation will be discussed in following.

The QCD operator in Eq.~(\ref{eq:General_ME}) and the GPDs in Eq.~(\ref{eq:General_ME_GPDs}) depend
on the quark flavor $f = (u, d)$. In the following we consider flavor distributions as well as
isoscalar/isovector combinations $u \pm d$.  The dependence on the quark flavor and nucleon isospin
will be indicated as needed.

Because of time reversal invariance the GPDs have definite parity under the transformation $\xi \to -\xi$,
\begin{subequations}
\label{xi_parity}
\begin{align}
&F(x, \xi, t) = +F(x, -\xi, t) \quad \mathrm{for} \quad F=H_{T},\tilde{H}_{T},E_{T},
\\
&F(x, \xi, t) = -F(x, -\xi, t) \quad \mathrm{for} \quad F=\tilde{E}_{T}.
\end{align}
\end{subequations}
The partonic operator in Eq.~\eqref{eq:General_ME} can be expanded in powers of the light-like separation,
generating a series of local operators of increasing spin. The local operator of spin $m$ can be expressed 
as the $x^{m-1}$-weighted integral of the non-local operator over $x$ (so-called $m$-th moment),
\begin{align}
& (P^{+})^{m}\int dx \, x^{m-1} \int \frac{dz^{-}}{2\pi} \, e^{i x P^{+} z^{-}} 
\nonumber \\
& \times \bar{\psi}   \left(-\frac{z}{2}\right) \left[-\frac{z}{2}, \frac{z}{2} \right] 
i\sigma^{+j} \psi \left(\frac{z}{2}\right)_{z^+, \bm{z}_\perp=0}
\nonumber \\[.5ex]
&=  \bar{\psi}(0) (i \overleftrightarrow{D}^{+})^{m-1} i\sigma^{+j} \psi \left(0\right),
\label{eq:op_polynomiality}
\end{align}
where $\overleftrightarrow{D}^{\mu} = (\overrightarrow{D}^{\mu} - \overleftarrow{D}^{\mu})/2$ is the QCD 
covariant derivative. Using this operator relation it can be shown that the moments of the GPDs are
polynomials in $\xi$ \cite{Hagler:2004yt},
\begin{subequations}
\label{eq:polynomiality}
\begin{align}
&\int^{1}_{-1} dx \, x^{m-1} H_{T}(x,\xi,t) 
= \sum^{m-1}_{\substack{i=0 \\ \mathrm{even}}} (-2\xi)^{i} A_{T m,i}(t), 
\\
&\int^{1}_{-1} dx \, x^{m-1} \tilde{H}_{T}(x,\xi,t) 
= \sum^{m-1}_{\substack{i=0 \\ \mathrm{even}}} (-2\xi)^{i} \tilde{A}_{T m,i}(t),
\\
&\int^{1}_{-1} dx \, x^{m-1} E_{T}(x,\xi,t) 
= \sum^{m-1}_{\substack{i=0 \\ \mathrm{even}}} (-2\xi)^{i} B_{T m,i}(t), 
\label{polynomiality_E_T}
\\ 
&\int^{1}_{-1} dx \, x^{m-1} \tilde{E}_{T}(x,\xi,t) 
= \sum^{m-1}_{\substack{i=0 \\ \mathrm{odd}}} (-2\xi)^{i} \tilde{B}_{T m,i}(t).
\end{align}
\end{subequations}
The polynomials are even or odd in $\xi$ according to Eq.~(\ref{xi_parity}). The degree of the polynomials 
representing the $m$-th moment is $\leq m - 1$, with the exact degree depending on the even/oddness 
of $m - 1$ and on the GPD. The coefficients of the polynomials depend only on $t$ and are called the
generalized tensor FFs. The polynomiality property Eq.~(\ref{eq:polynomiality}) plays a fundamental role 
in the structure of the GPDs.

The first moments of the chiral-odd GPDs are connected with the nucleon tensor FFs. The nucleon matrix element 
of the local tensor operator is parametrized as
\begin{align}
& \mathcal{M}_{\mathrm{FFs}}[i\sigma^{\mu \nu}]
= \langle p', s' | \bar{\psi}(0) i\sigma^{\mu \nu} \psi (0) | p, s \rangle
\nonumber \\[1ex]
&= \bar{u}' \left[ i \sigma^{\mu \nu} \, H_{T}(t)
+ \frac{P^{\mu} \Delta^{\nu} - \Delta^{\mu} P^{\nu}}{M^{2}_{N}} \, \tilde{H}_{T}(t)
\right.
\nonumber \\
& \left. + \; \frac{\gamma^{\mu}\Delta^{\nu} -\Delta^{\mu} \gamma^{\nu} }{2M_{N}} \, E_{T}(t) \right] u.
\label{eq:General_ME_tensor}
\end{align}
Here the tensor decomposition is covariant and does not reference the light-cone direction.
We denote the tensor FFs in Eq.~(\ref{eq:General_ME_tensor}) by the same symbols as the GPDs,
to emphasize the correspondence of the structures; the functions can be distinguished by their arguments.
Comparing Eq.~\eqref{eq:General_ME} and Eq.~\eqref{eq:General_ME_tensor}, one has
\begin{subequations}
\label{eq:first_Mel}
\begin{align}
&\int^{1}_{-1} dx H_{T}(x, \xi, t) = A_{T10}(t) 
\equiv H_{T}(t), 
\\
&\int^{1}_{-1} dx \tilde{H}_{T}(x, \xi, t) = \tilde{A}_{T10}(t) \equiv \tilde{H}_{T}(t),
\\
&\int^{1}_{-1} dx E_{T}(x, \xi, t) = B_{T10}(t) \equiv 
E_{T}(t), 
\\
&\int^{1}_{-1} dx \tilde{E}_{T}(x, \xi, t)  = 0.
\label{E_T_tilde_first_moment}
\end{align}
\end{subequations}
The vanishing first moment of $\tilde{E}_{T}$ is due to its antisymmetry in $\xi$. 

In the forward limit, $\xi \rightarrow 0$ and $|t| \rightarrow 0$,
the chiral-odd GPD $H_{T}$ reduces to the transversity 
parton distribution function
\begin{align}
H_{T} (x, \xi=0, t=0) = h_{1} (x).
\end{align}
Its first moment is known as the nucleon's tensor charge
\begin{align}
& \int^{1}_{-1} dx \, H_{T} (x, \xi=0, t=0) 
\nonumber \\
&= \int^{1}_{-1} dx \, h_{1} (x) =  H_{T}(0) = g_{T}.
\end{align}
In applications to exclusive pseudoscalar meson production processes one introduces the linear
combination of GPDs
\begin{align}
\bar{E}_{T} \equiv E_{T}+2\tilde{H}_{T}.
\end{align}
Its first moment in the forward limit is known as the nucleon's anomalous tensor magnetic moment
\begin{align}
& \int^{1}_{-1} dx \, \bar{E}_{T} (x, \xi=0, t=0) 
\nonumber \\
& = E_{T}(0)+2\tilde{H}_{T}(0) = \kappa_{T}.
\end{align}
Thus the chiral-odd GPDs provide information on fundamental characteristics of the nucleon
derived from the local tensor operator.

\section{Multipole expansion \label{sec:3}}
\subsection{Multipole expansion of chiral-odd GPDs}
\label{subsec:multipole_gpds}
To analyze the structure of the chiral-odd GPDs and FFs, it is useful to perform a multipole expansion
of the matrix elements of the chiral-odd operators. 
The multipoles allow one to enumerate the independent structures in the matrix element and exhibit
their spin and orbital angular momentum content.
The multipole expansion prepares the matrix element for the $1/N_c$ expansion, where the multipoles
have definite $N_c$ scaling, determined by their spin and isopspin quantum numbers.

We consider here the multipole expansion of both GPDs (non-local partonic operator) and FFs (local operator).
The multipole expansion of the GPDs is performed in light-front variables, as an expansion in the 2D transverse 
momentum transfer $\bm{\Delta}_\perp$, constrained by 2D rotational invariance (conservation of angular momentum
along the 3-direction, called longitudinal angular momentum).
The multipole expansion of the FFs can be performed either as an expansion in the 2D light-front 
momentum transfer, in same way as for the GPDs, or as an expansion in the 3D momentum transfer $\bm{\Delta}$
in the Breit frame, constrained by 3D rotational invariance (conservation of all components of the angular momentum).
Comparison of the two expansions provides insight into the origin of the 2D light-front structures and implements
the constraints on the light-front multipoles arising from 3D rotational invariance (so-called angular conditions)
and discrete symmetries.

The multipole expansion of the matrix element of the chiral-odd partonic operator is performed
using light-front momentum variables for the nucleon states.
We consider the parametrization Eq.~(\ref{eq:General_ME_GPDs}) in a class of reference frames
where the average nucleon 4-momentum has zero transverse component, $\bm{P}_\perp = 0$,
and the momentum transfer has nonzero component $\bm{\Delta}_\perp \neq 0$.
In these frames the light-front 4-vector components of $P$ and $\Delta$ are given by
(in the notation $v=[v^{+},v^{-},\bm{v}_{\perp}]$)
\begin{subequations}
\label{eq:GPD_frame}
\begin{align}
P &= \left[ P^{+}, \frac{ M^{2}_{N} +|\bm{\Delta}_{\perp}|^{2}/4}{2P^{+} (1-\xi^{2})}, \bm{0}_{\perp} \right], 
\\
\Delta &= \left[-2\xi P^{+}, \frac{ \xi (M^{2}_{N} + |\bm{\Delta}_{\perp}|^{2}/4)}{P^{+}(1-\xi^{2}) }, 
\bm{\Delta}_{\perp} \right] .
\end{align}
\end{subequations}
The mass shell conditions of the initial and final nucleon 4-momenta imply that
\begin{align}
\Delta \cdot P = 0, \quad  P^{2}+\frac{\Delta^{2}}{4}=M^{2}_{N}.
\label{eq:on_shell}
\end{align}
The condition $\bm{P}_\perp = 0$ does not determine a unique frame but an equivalence class of frames related
by longitudinal boosts; the value of $P^+$ remains arbitrary and specifies a particular frame in the class.
The nucleon spin states are chosen as light-front helicity states, prepared from rest-frame spin states
by a sequence of light-front boosts \cite{Brodsky:1997de}. Explicit expressions for the light-front bispinors
in terms of rest-frame 2-component spinors are given e.g.\ in Ref.~\cite{Cosyn:2020kwu}.

In the frames of Eq.(\ref{eq:GPD_frame}) the only transverse vector characterizing the matrix element is
the momentum transfer $\bm{\Delta}_\perp$. The matrix element can therefore be expanded in 2D multipole
structures in $\bm{\Delta}_\perp$. The 2D rank-$n$ irreducible tensors are defined as \cite{Kim:2022wkc}
\begin{subequations}
\label{2D_orbital}
\begin{alignat}{2}
&X_{0} = 1 && (L_3 = 0),
\\[1ex]
&X^{i}_{1} = \bm{n}_\perp^i && (L_3 = \pm 1),
\\
&X^{ij}_{2} = \bm{n}_\perp^i \bm{n}_\perp^j - \frac{1}{2}\delta^{ij} \hspace{2em}
&& (L_3 = \pm 2),
\label{orbital_quadrupole}
\end{alignat}
\end{subequations}
where $\bm{n}_\perp \equiv \bm{\Delta}_\perp / |\bm{\Delta}_\perp|$ and $i,j=1,2$.
The tensors correspond to structures with longitudinal orbital angular momentum $L_3 = 0, \pm 1, \pm 2$,
respectively, as indicated in Eq.~(\ref{2D_orbital}).
The orbital structures are accompanied by spin structures formed from the spin wave functions of the
initial and final nucleon.
The spin structures appear as bilinear forms in the 2-component spinors describing the spin wave function
of each nucleon in its rest frame,
\begin{align}
\chi^\dagger (s') \, \hat{O} \, \chi (s).
\label{spin_operators}
\end{align}
The quantization axis of the spinors can be chosen along any direction in the rest frame;
the spin quantum numbers $s$ and $s'$ are then defined as the spin projections along this axis.
In the following the quantization axis is chosen as the 3-axis, and the spinors are eigenspinors
of the 3-component of the spin operator,
\begin{align}
\frac{\sigma^3}{2} \chi (S_3) = S_3 \, \chi(S_3), \hspace{2em} S_3 = \pm 1/2,
\end{align}
and the spin quantum numbers are the spin projections along the 3-axis in the rest frame,\footnote{
The quantization axis can also be chosen as the transverse 1-axis, by defining the spinors
as eigenspinors of $\sigma^1/2$. The transversely polarized spinors are linear combinations
of the longitudinally polarized spinors. The subsequent arguments concerning conservation of
the longitudinal angular momentum refer to the angular momentum of the bilinear form
of Eq.~(\ref{spin_operators}) and do not depend on the choice of spin states.}
\begin{align}
s \equiv S_3, \hspace{2em} s' \equiv S_3'.
\label{spin_S_3}
\end{align}
The operator $\hat O$ in Eq.~(\ref{spin_operators}) can be the unit operator $\bm{1}$ or a
component of the spin operator $\bm{S} = \bm{\sigma}/2$. In the present context the independent structures 
characterizing the transition are (here $i = 1,2$)
\begin{subequations}
\label{2D_spin}
\begin{alignat}{2}
\hat O = \; & \bm{1}  && (L_3 = 0),
\\
&\sigma^3 \hspace{4em} && (L_3 = 0),
\\
&\sigma^i && (L_3 = \pm 1).
\label{spin_dipole}
\end{alignat}
\end{subequations}
These scalar and vector structures have longitudinal angular momentum $L_3 = 0, 0, \pm 1$, respectively,
as indicated in Eq.~(\ref{2D_spin}).
Expanding the matrix element Eq.~\eqref{eq:General_ME_GPDs} in the structures of
Eqs.~(\ref{2D_orbital}) and (\ref{2D_spin}), we obtain
\begin{align}
& \mathcal{M}_{\mathrm{GPDs}}[i\sigma^{+j}]
\nonumber \\[1ex]
& 
 = \frac{2P^{+}}{\sqrt{1-\xi^{2}}} \left\{ 
 i \epsilon^{3jm} \sigma^{m} \, X_0 \phantom{\frac{0}{0}}
\right. 
\nonumber \\
& \hspace{2em} \times 
\left[ (1 - \xi^2) H_{T} + \frac{|\bm{\Delta}_{\perp}|^{2}}{4M^{2}_{N}} \tilde{H}_{T} 
- \xi^{2} E_{T} + \xi \tilde{E}_{T} \right]
\nonumber \\[1ex]
& + \bm{1} \, X^{j}_{1} \, \frac{|\bm{\Delta}_{\perp}|}{ 2M_{N}}
\left( 2 \tilde{H}_{T} + E_T - \xi \tilde{E}_{T} \right)
\nonumber \\[1ex]
&+ i \epsilon^{3jm} \sigma^{3} \, X^{m}_{1} \, \frac{|\bm{\Delta}_{\perp}|}{2M_{N}}
\left( - \xi E_{T} + \tilde{E}_{T}  \right)
\nonumber \\[1ex]
&+ \left. i \epsilon^{3ml} \sigma^{l} \, X^{mj}_{2} \, \frac{|\bm{\Delta}_{\perp}|^{2}}{4M^{2}_{N}}
\left(  2\tilde{H}_{T} \right) \right\} .
\label{eq:multipole_GPDs}
\end{align}
Here it is implied that the spin operators are contracted with the nucleon rest-frame
spinors as in Eq.~(\ref{spin_operators}), and that the matrix element is a function of $S_3$ and $S_3'$.
One sees that the four structures in Eq.~(\ref{eq:multipole_GPDs}) contain one orbital monopole,
two dipoles, and one quadrupole.

The number of independent structures in Eq.~(\ref{eq:multipole_GPDs}) can be explained by the
addition of longitudinal angular momentum in the light-front representation. 
The matrix element of the partonic operator
with $\sigma^{+j}$ is a transverse vector and has components with $L_3 = \pm 1$.
The four structures in Eq.~(\ref{eq:multipole_GPDs}) are those that can be formed by combining the orbital
structures in Eq.~(\ref{2D_orbital}) and the spin structures in Eq.~(\ref{2D_spin}) such that $L_3$
adds up to $\pm 1$.
The parity of the structures involving the spin operators $\sigma^3$ and $\sigma^i$ is adjusted
by the 2D pseudotensor $\epsilon^{3ij}$; the existence of this tensor is specific to the light-front
representation with the preferred 3-direction.

The multipole expansion of the chiral-odd matrix element Eq.~(\ref{eq:multipole_GPDs}) 
includes a term with the orbital quadrupole $X_{2}$. It appears from the coupling of the spin dipole
Eq.~(\ref{spin_dipole}) with the orbital quadrupole Eq.~(\ref{orbital_quadrupole}).
In this way the orbital quadrupole can be present even though nucleon-to-nucleon matrix elements
cannot support a spin quadrupole structure. The presence of the orbital quadrupole structure
is confirmed by the multipole expansion of the local tensor operator in the Breit
frame (see Sec.~\ref{sec:aa}). It plays an important role in the $1/N_c$ expansion of the
chiral-odd GPDs (see Sec.~\ref{sec:4}). The quadrupole structure was not included in the analysis of 
Ref.~\cite{Schweitzer:2016jmd}.

The quadrupole structure appears only in the chiral-odd, not in the chiral-even GPDs. In the chiral-even GPDs
the operators $\gamma^+$ and $\gamma^+\gamma_5$ have $L_3 = 0$. In this case the addition of
longitudinal angular momentum produces four independent structures formed from the $L_3 = 0$ and $\pm 1$
structures in Eq.~(\ref{2D_orbital}) and (\ref{2D_spin}), but cannot involve the $L_3 = 2$ orbital structure.
This causes an essential difference between
the non-forward matrix elements of the chiral-odd and chiral-even operators,
which is not apparent from the forward limit.

The combinations of GPDs appearing in Eq.~(\ref{eq:multipole_GPDs}) can be referred to as the ``multipole GPDs''.
They represent an alternative definition of the chiral-odd GPDs with an obvious physical interpretation and
appear naturally in the $1/N_c$ expansion (see Sec.~\ref{sec:4}). Their physical properties and connection
with observables could be explored also independently of the $1/N_c$ expansion.

\subsection{Multipole expansion of tensor FFs}
The multipole expansion of the matrix element of the local tensor operator can be performed in light-front
momentum variables, in the same way as for the nonlocal partonic operator. We consider the matrix element
Eq.~(\ref{eq:General_ME_tensor}) in the class of frames defined by Eq.~(\ref{eq:GPD_frame})
with $\xi = 0$, where the 4-vector components are
\begin{subequations}
\label{eq:2D_DYF}
\begin{align}
P &= 
\left[P^{+}, \frac{ M^{2}_{N} + |\bm{\Delta}_{\perp}|^{2}/4}{2P^{+} }, \bm{0}_{\perp} \right],
\\
\Delta &= \left[ 0, 0, \bm{\Delta}_{\perp} \right],
\end{align}
\end{subequations}
and where
\begin{align}
t = - |\bm{\Delta}_\perp|^2 .
\end{align}
These are the so-called Drell-Yan-West frames used in the analysis of FFs in light-front
quantization \cite{Brodsky:1997de}. In these frames the matrix element can be expanded in the 2D multipoles
in $\bm{\Delta}_\perp$. The result for the local tensor operator can be obtained by setting $\xi = 0$
in Eq.~(\ref{eq:multipole_GPDs}) and using the relations Eq.~(\ref{eq:first_Mel}) between the moments
of the GPDs and the tensor FFs. We obtain
\begin{align}
& \mathcal{M}_{\mathrm{FFs}}[i\sigma^{+j}] 
\nonumber \\[1ex]
&=  2P^+ \left[  i \epsilon^{3jm} \sigma^{m} \, X_0 
\left( H_{T} - \frac{t}{4M^{2}_{N}} \tilde{H}_{T}  \right) \right.
\nonumber \\
&+ \bm{1} \, X^{j}_{1} \, \frac{\sqrt{-t}}{2M_{N}} \left( 2 \tilde{H}_{T} + E_{T} \right)
\nonumber \\[1ex]
&\left. +i \epsilon^{3jl} \sigma^{m} \, X^{lm}_{2} \,
\frac{t}{4M^{2}_{N}} \left( 2\tilde{H}_{T} \right) \right] ,
\label{eq:2DLF_multipole_FF}
\end{align}
where we have set $|\bm{\Delta}_\perp| = \sqrt{-t}$. One sees that the
local tensor operator generates only three structures.
The absence of the orbital dipole structure $i \epsilon^{3jm} X^{m}_{1} \sigma^{3}$ compared to the nonlocal
partonic operator Eq.~(\ref{eq:multipole_GPDs}) is due to time reversal invariance, which causes the vanishing
of the first moment of the GPD $\tilde E_T$; see Eq.~(\ref{eq:first_Mel}). The 2D multipole expansion of the
local operator is thus constrained by considerations beyond longitudinal angular momentum conservation.

The multipole expansion of the local tensor operator can also be performed using the 3D vector components,
preserving 3D rotational invariance. We consider the matrix element in the Breit frame, where the average
nucleon 3-momentum is zero, $\bm{P} = 0$, and where the 4-momentum components are given by
[in the notation $v=(v^0, \bm{v})$]
\begin{align}
&P= (\bar M, \bm{0}), \hspace{2em} \Delta= (0, \bm{\Delta}),
\label{eq:rot_sym}
\end{align}
with
\begin{align}
|\bm{\Delta}|^2 = -t, \hspace{2em} \bar M \equiv {\textstyle\sqrt{M_N^2 - t/4}}.
\end{align}
The nucleon spin states are now chosen as canonical spin states, obtained by canonical boosts from
the rest-frame spin states. In this setup the matrix elements of tensor operators are constrained
by 3D angular momentum conservation. The 3D rank-$n$ irreducible tensors are defined as
\begin{subequations}
\label{3D_orbital}
\begin{alignat}{2}
&Y_{0} = 1 && (L = 0),
\\[1ex]
&Y^{i}_{1} = \bm{n}^i && (L = 1),
\\
&Y^{ij}_{2} = \bm{n}^i \bm{n}^j - \frac{1}{3}\delta^{ij} \hspace{2em}
&& (L = 2),
\end{alignat}
\end{subequations}
where $\bm{n} \equiv \bm{\Delta} / |\bm{\Delta}|$; $i,j=1,2,3$; and
where we have indicated the orbital angular momentum of the structures.
The spin operators appearing in the 3D expansion are
\begin{subequations}
\begin{alignat}{2}
&\bm{1}, \hspace{4em}  && L = 0,
\\
&\sigma^i, && L = 1.
\label{3D_spin}
\end{alignat}
\end{subequations}
The expansion of the matrix element Eq.~\eqref{eq:General_ME_tensor}
is now performed separately for the $0j$- and $ij$-components ($i,j = 1,2,3$).
We obtain
\begin{subequations}
\label{eq:general_tensor_FF_largeNc}
\begin{align}
&\mathcal{M}_{\mathrm{FFs}}[i\sigma^{0j}] 
\nonumber \\
&= 2M_N \left\{  \bm{1} \, Y^{j}_{1} \, \frac{\sqrt{-t}}{2M_N} \left[ H_{T} 
+ \frac{2 \bar{M}^2}{M_{N}^2} \tilde{H}_{T} + E_{T}  \right] \right\} , 
\label{eq:general_tensor_FF_largeNc_a}
\\
&\mathcal{M}_{\mathrm{FFs}}[i\sigma^{ij}] 
\nonumber \\
&= 2M_N \left\{ i \epsilon^{ijk} \sigma^{k} \, Y_0 \,
\left[ \frac{1}{3}\left( \frac{\bar{M}}{M_N} + 2\right) H_{T} + \frac{t}{6M_{N}^2}  E_{T} \right] \right.
\nonumber \\
&\left. +i \epsilon^{ijk} \sigma^{m} \, Y^{km}_{2} \,
\left[  \left( \frac{\bar{M}}{M_N} - 1 \right) H_{T} - \frac{t}{4M_{N}^2}  E_{T} \right] \right\} .
\end{align}
\end{subequations}
The multipole structures are now constrained by angular momentum conservation following from 3D rotational
invariance. They are also constrained by the discrete symmetries of parity and time reversal, which act on
the 3D vectors in a simple manner. Note that the orbital quadrupole structure $Y_2$ is present in the 3D expansion.

The 2D and 3D multipole expansions can be compared directly, in a way that the structures appearing in
both expansions can be matched with each other.
This is done by going to the special light-front frame that is identical to the Breit frame \cite{Kim:2023xvw},
namely the frame of Eq.~(\ref{eq:2D_DYF}) with $P^+ = \bar M/\sqrt{2}$, where
\begin{align}
P^+ = P^- = \frac{\bar M}{\sqrt{2}},
\hspace{2em}
P^0 = \frac{1}{\sqrt{2}} (P^+ - P^-) = \bar M.
\end{align}
In this frame both expansions are valid, and the expressions can be equated.
The 2D light-front components are obtained as the linear combination of the 3D components,
\begin{align}
\sigma^{+j}  = \frac{1}{\sqrt{2}}(\sigma^{0j} + \sigma^{3j}).
\label{eq:admix}
\end{align}
Comparing the expansions Eq.~(\ref{eq:2DLF_multipole_FF}) and
Eq.~(\ref{eq:general_tensor_FF_largeNc}) in this way, we observe the following:

(i) The 2D multipole structure Eq.~(\ref{eq:2DLF_multipole_FF}) is induced by the 3D multipole 
structure Eq.~(\ref{eq:general_tensor_FF_largeNc}) as follows:
\begin{subequations}
\begin{align}
i \epsilon^{ijk} \sigma^{k} \, Y_0
&\to
i \epsilon^{3jk} \sigma^{k} \, X_0 ,
\\
\bm{1} \, Y^{j}_{1}
&\to
\bm{1} \, X^{j}_{1},
\\
i \epsilon^{ijk} \sigma^{m} \, Y^{km}_{2}
&\to  
i \epsilon^{3jk} \sigma^{m} \, X^{km}_{2}, \; i \epsilon^{3jk} \sigma^{k} \, X_0 .
\end{align}
\end{subequations}
The 3D monopoles and dipoles reduce directly to their 2D counterparts. 
The 3D quadrupole reduces to a 2D quadrupole and a 2D monopole.
Such behavior is expected, as the 2D projection of a 3D traceless tensor can be a 
2D traceless or traceful tensor.
This phenomenon has been investigated in the context of Abel tomography for spin-1 particles \cite{Kim:2022bia}. 
It shows that the 3D quadrupole structure in the tensor matrix element affects not only the results for the
2D quadrupole but also the monopole.

(ii) The $t$-dependent functions accompanying the multipoles in the 2D and 3D expansions
show differences at the level of terms $\propto \sqrt{-t}/M_N$. 
They are due to the fact that the 2D expansion was performed using light-front helicity states, while the 3D 
expansion was performed using canonical spin states.
This effect has been discussed in connection with charge and current densities in hadrons
and is well understood \cite{Lorce:2020onh}. It could be corrected by performing the spin rotation
transforming the canonical into light-front bispinors (Melosh rotation) in the 3D expression
before matching with the 2D expression. In the large-$N_c$ limit the effect of the spin rotation
is suppressed by $1/N_c$ and can be neglected when computing the multipole FFs in leading order.

\section{Spin-flavor structure at large $N_{c}$  \label{sec:4}}
\subsection{$1/N_c$ expansion in mean-field picture \label{sec:aa}}
The $1/N_c$ expansion of nucleon matrix elements of QCD composite operators (local or partonic)
can be performed using a method based on the mean-field picture of baryons in the large-$N_c$ limit.
Baryons are characterized by a mean field with contracted spin-flavor symmetry, from which spin-isospin
states emerge through quantization of the zero modes. The QCD operators are first evaluated in the mean field,
which imposes the spin-flavor symmetry on the expectation value; the transition matrix elements between
spin-isospin are then obtained from the quantization of the zero modes.
The method uses only abstract features of the mean-field picture (symmetries, parametric scaling)
and does not refer to any specific dynamics.
The results for the $N_c$ scaling of nucleon matrix elements
are model-independent and equivalent to those obtained with
group-theoretical approaches \cite{Dashen:1993jt}.
The mean-field method of the $1/N_c$ expansion is particularly convenient for partonic operators
has been used extensively in the analysis of GPDs \cite{Goeke:2001tz}.

In the large-$N_c$ limit baryon states are classified by the emergent spin-flavor symmetry.
The $N$ and $\Delta$ appear in the representation with spin-isospin $S = T = 1/2$ and $3/2$.
The baryon states are characterized by their spin-isospin quantum numbers
$B \equiv \{S = T, S_3, T_3\}$. The baryon masses are $M_{N, \Delta} = \mathcal{O}(N_c)$, and the 
splitting is $M_N - M_\Delta = \mathcal{O}(N_c^{-1})$. The $1/N_c$ expansion of baryon matrix 
elements is performed in a class of frames where the initial and final baryons have
3-momenta and energies of the order
\begin{subequations}
\label{nc_momenta}
\begin{align}
|\bm{p}|, |\bm{p}'| &= \mathcal{O}(N_{c}^0),
\\
p^0, p^{\prime 0} &= M_N + \mathcal{O}(1/N_c) = \mathcal{O}(N_c).
\end{align}
\end{subequations}
The baryon states are normalized as
\begin{align}
\label{Eq-R:norm}
& \langle \bm{p}^\prime, B^\prime | \bm{p}, B \rangle
= 2p^0(2\pi)^3\delta^{(3)}(\bm{p}^\prime - \bm{p})
\,\delta_{B'B},
\nonumber \\[1ex]
& \delta_{B'B} \equiv 
\delta_{S'S}\,\delta_{S_3'S_3} \, \delta_{T'T}\,\delta_{T_3'T_3} .
\end{align}
In this study we consider $N \rightarrow N$ matrix elements; the following discussion
can easily be extended to $N \rightarrow \Delta$ matrix elements.

One considers the matrix element of a partonic QCD operator of the form
\begin{align}
\langle \bm{p}^\prime, B^\prime | 
\, \bar\psi_{\alpha'f'} (-z/2) [-z/2, z/2] \psi_{\alpha f}(z/2) \, | \bm{p}, B \rangle ,
\label{largen_operator}
\end{align}
where $z = (z^0, \bm{z})$ is the light-like separation of the fields, $z^2 = (z^0)^2 - |\bm{z}|^2 = 0$.
Summation over color indices is implied; $[-z/2,z/2]$ is the gauge link along the straight line
connecting the points $-z/2$ and $z/2$; it will be omitted in the following expressions for brevity
but is always assumed to be present. $\alpha$ and $\alpha'$ are the bispinor indices of the fields;
it is assumed that the matrix element Eq.~(\ref{largen_operator}) will be contracted with an external
bispinor matrix,
\begin{align}
\Gamma_{\alpha'\alpha} \, \langle ..| \, \bar\psi_{\alpha'f'} \, \psi_{\alpha f} \, | .. \rangle
= \langle ..| \, \bar\psi_{f'} \Gamma \psi_{f} \, | .. \rangle .
\label{largen_operator_contracted}
\end{align}
$f$ and $f'$ are the flavor indices; we assume two light flavors $(u, d)$ and exact isospin symmetry.
The $1/N_c$ expansion of the matrix element Eq.~(\ref{largen_operator}) is performed as follows.

In the first step one takes the expectation value of the operator in the mean-field state
of the large-$N_c$ baryon, with the mean field centered at the origin,
\begin{align}
&\langle \textrm{mf}\, | \, \bar\psi_{\alpha' f'} (y - z/2) \, \psi_{\alpha f} (y + z/2) \,
| \textrm{mf} \, \rangle
\nonumber \\
& = \; \mathcal{F}_{\alpha'\! f' \! , \, \alpha f}(z^0, \bm{z} \, | \, \bm{y}).
\label{operator_mf}
\end{align}
Here $y = (y^0, \bm{y})$ is the center coordinate of the partonic operator; the displacement from
the center of the mean field is necessary to account for the momentum transfer to the baryon (see below).
The expectation value Eq.~(\ref{operator_mf}) defines a function of the space-time
coordinates and spinor/flavor indices of the operator.
This function is regarded as an abstract object (or parametrization): its specific form is governed
by dynamics and can only be determined in models, but its symmetries can be established on general grounds.
The mean field is localized in space and breaks translational invariance in space; the expectation value
of the operator therefore depends on both coordinates $\bm{z}$ and $\bm{y}$.
The mean field is time-independent (static) and preserves translational invariance in time;
the expectation value therefore depends only on the time difference $z^0$ of the fields,
not on the average time $y^0$.

%
%
\begin{figure}[t]
\includegraphics[scale=0.43]{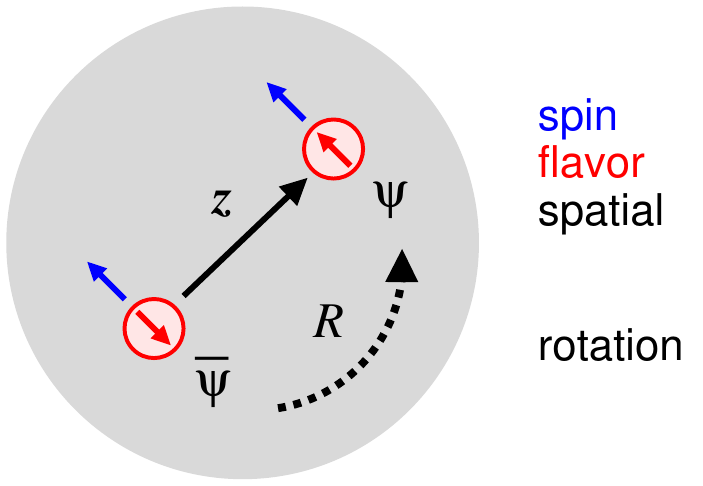}
\caption{Visualization of the spin-flavor symmetry of the mean-field expectation value of the
partonic operator Eq.~(\ref{operator_mf}). The mean-field expectation value is invariant under
combined spatial, spin, and flavor rotations of the operator, Eq.~(\ref{Eq-R:hedgehog-symmetry}).}
\label{fig:spinflavor}
\end{figure}
Most importantly, the mean field possesses the spin-flavor symmetry characteristic of baryons
in the large-$N_c$ limit (``hedgehog symmetry'') \cite{Witten:1983tx}. It implies that the expectation value
Eq.~(\ref{operator_mf}) is invariant under combined spatial, spin, and flavor rotations of the operator
(see Fig.~\ref{fig:spinflavor}),
\begin{align}
& S^{-1}_{\beta' \! \alpha'} \, S_{\alpha\beta} \,
R^{-1}_{g'\! f'} \, R_{fg} \, \mathcal{F}_{\beta' \! g' \! , \, \beta g}(z^0, O \bm{z} \, | \, O \bm{y})
\nonumber \\
& = \mathcal{F}_{\alpha'\! f' \!, \, \alpha f} (z^0, \bm{z} \, | \, \bm{y}).
\label{Eq-R:hedgehog-symmetry}
\end{align}
Here $R$ is an $SU(2)$ rotation matrix (applied to the flavor rotations),
$O \equiv O(R)$ is the associated $O(3)$ rotation matrix (applied to the spatial rotations),
\begin{align}
O^{ab} \equiv \tfrac{1}{2} \, {\rm tr} [R^{-1} \tau^a R \tau^b]
\hspace{2em} (a, b = 1, 2, 3),
\label{O_def}
\end{align}
and $S \equiv S(R)$ is the associated bispinor rotation matrix \cite{Berestetskii:1982qgu}
(applied to the spin rotations).
When the bispinor indices are contracted with an external matrix $\Gamma$ of 3D vector/tensor
character as in Eq.~(\ref{largen_operator_contracted}),
the vector/tensor indices on the matrix rotate with the $O(3)$ rotation matrix
Eq.~(\ref{O_def}). Equation~(\ref{Eq-R:hedgehog-symmetry}) imposes the spin-flavor symmetry of large-$N_c$
baryons on the operator matrix elements and plays a central role in the $1/N_c$ expansion.

In the second step one quantizes the translational zero mode of the mean field and
projects on baryon states with definite momenta. In leading order of $1/N_c$ this is done
by shifting the center of the mean field to position $\bm{X}$ and integrating over the collective
coordinate with the wave functions $e^{i\bm{p}\bm{X}}$ and $e^{-i\bm{p}'\bm{X}}$.
Because of translational invariance the mean field expectation value depends only on the
difference between the center coordinate of the operator and the mean field, $\bm{y} - \bm{X}$.
The projection can therefore be done equivalently by leaving the mean field centered at $\bm{X} = 0$ and
integrating over $\bm{y}$, with the corresponding change of variables in the wave functions.
We define
\begin{align}
&\langle \bm{p}^\prime, \textrm{mf} \, |\, \bar\psi_{\alpha'f'}(-z/2) \psi_{\alpha f}(z/2) \,
| \bm{p}, \textrm{mf} \, \rangle 
\nonumber \\[1ex]
&\equiv 2M_{N} \, \int d^3y \, e^{i(\bm{p}' - \bm{p})\cdot\bm{y}} \;
\mathcal{F}_{\alpha'\! f'\! , \, \alpha f}(z^0, \bm{z} \, | \, \bm{y})
\nonumber \\[1ex]
&= 2M_{N} \, \int d^3y \, e^{i(\bm{p}' - \bm{p})\cdot\bm{y}}
\nonumber\\
&\times 
\langle \textrm{mf} \, | \, \bar\psi_{\alpha'f'}(-z^0/2, \bm{y} - \bm{z}/2) \,
\psi_{\alpha f}(z^0/2, \bm{y} - \bm{z}/2) \, | \, \textrm{mf} \rangle .
\nonumber \\
\label{quantization_momentum}
\end{align}
Equation~(\ref{quantization_momentum})
represents the matrix element of the operator between large-$N_c$ baryon states with
definite momenta but as yet indefinite spin-isospin and is referred to as the mean-field matrix element
(or soliton matrix element). This ``intermediate'' object has interesting properties, and several
aspects of large-$N_c$ baryon structure can already be discussed at this at this level
(see the application to GPDs in Sec.~\ref{subsec:gpds_largenc}).

In the third step one quantizes the rotational zero mode of the mean field in position and flavor space
and projects on baryon states with the desired spin-isospin quantum numbers. In leading order of $1/N_c$
this is done by integrating over the flavor rotations $R$, Eq.~(\ref{Eq-R:hedgehog-symmetry}), with
rotational wave functions $\phi_{B^{ }}(R)$ and $\phi^\ast_{B^\prime}(R)$ describing baryon states
with spin-isospin quantum numbers $B$ and $B'$ \cite{Witten:1979kh,Pobylitsa:2000tt},
\begin{align}
&\langle \bm{p}^\prime, B^\prime \, | \, \bar\psi_{\alpha'f'}(-z/2) \psi_{\alpha f}(z/2) \,
| \, \bm{p}, B \rangle 
\nonumber \\[1ex]
&= \int dR\;\phi^\ast_{B^\prime}(R)\,\phi_{B^{ }}(R) \;
\nonumber\\
&\times \; R^{-1}_{g'\! f'} R_{fg} \,
\langle \bm{p}^\prime, \textrm{mf} \, | \, \bar\psi_{\alpha g'}(-z/2) \psi_{\alpha g}(z/2)
\, | \, \bm{p}, \textrm{mf} \, \rangle .
\label{quantization_spinisospin}
\end{align}
The rotational wave functions are given by the Wigner finite-rotation matrices as \cite{Goeke:2001tz}
\begin{align}
\phi_B (R) & \equiv \phi_{S_3T_3}^{S=T}(R)
\nonumber \\[1ex]
& = \sqrt{2S+1}\;(-1)^{T_{ }+T_3} D_{-T_3,S_3}^{S=T}(R) .
\label{rotational_wave_function}
\end{align}
Equations~(\ref{quantization_momentum}) and (\ref{quantization_spinisospin}) represent
the transition matrix element of the
QCD operator in leading non-vanishing order of the $1/N_c$ expansion.
The spin-flavor symmetry of the mean field, Eq.(\ref{Eq-R:hedgehog-symmetry}), restricts
the spin-isospin structures emerging from the rotational integral and determines the 
$N_c$ scaling of the spin-flavor components of the matrix element.

In the analysis here it is assumed that the partonic operator is renormalized at a scale
$\mu^2 \gg \Lambda_{\rm QCD}^2$, sufficiently large to allow for perturbative treatment of
the scale dependence (evolution). The renormalization scale is subsumed in the definition
of the mean-field expectation value Eq.~(\ref{operator_mf}) and does not affect the $N_c$
scaling of the spin-flavor components of the matrix element derived from it, which is
the object of study here. The value of the renormalization scale becomes relevant only
when one attempts to calculate the mean-field expectation value of the operator in
models of nonperturbative dynamics.
\subsection{$1/N_c$ expansion of chiral-odd GPDs}
\label{subsec:gpds_largenc}
In order to set up the $1/N_c$ expansion of the chiral-odd GPDs, one has to specify the
parametric order of the kinematic variables in the non-forward matrix element Eq.~(\ref{eq:General_ME}). 
The $1/N_c$ expansion is performed
with nucleon 3-momenta $\mathcal{O}(N_c^0)$, Eq.~(\ref{nc_momenta}). This assignment
defines a class of frames related by boosts, which includes the frames used in the
2D and 3D multipole expansions (see Sec.~\ref{sec:3}). The average and difference of
the initial and final momenta are of the order ($i = 1,2,3$)
\begin{subequations}
\label{P_Delta_parametric}
\begin{alignat}{2}
&P^{i} = \mathcal{O}(N^{0}_{c}), \hspace{2em} &&P^{0} = M_{N} + \mathcal{O}(N^{-1}_{c}),
\\
&\Delta^{i} = \mathcal{O}(N^{0}_{c}), &&\Delta^{0} = \mathcal{O}(N^{-1}_{c}),
\end{alignat}
\end{subequations}
which implies that the momentum transfer variables in the GPDs are of the order
\begin{align}
\xi = \mathcal{O}(N^{-1}_{c}), \hspace{2em} |t| = \mathcal{O}(N^{0}_{c}).
\label{xi_t_parametric}
\end{align}
The partonic variable $x$ is taken to be of the order
\begin{align}
x = \mathcal{O}(N^{-1}_{c}),
\label{x_parametric}
\end{align}
which is the standard regime considered in the $1/N_c$ expansion
of nucleon parton distributions \cite{Diakonov:1996sr},
corresponding to nonexceptional configurations with quark momenta $\mathcal{O}(N_c^0)$
in the nucleon rest frame.

We now perform the $1/N_c$ expansion of the matrix element of the chiral-odd partonic operator,
Eq.~(\ref{eq:General_ME}), using the method of Sec.~\ref{sec:aa}.
The mean-field matrix element of the partonic operator between large-$N_c$ baryon states with momenta $\bm{p}$
and $\bm{p}'$ is defined according to Eq.~(\ref{quantization_momentum}).
This matrix element can be parametrized in the form
\begin{align}
& \frac{M_N}{\sqrt{2}} \int \frac{d z^{-}}{2\pi} e^{i x P^{+} z^{-} }
\nonumber \\
& \times \langle \bm{p}', \textrm{mf}\, | \, \bar{\psi}_{f'}
\left(-z/2\right) i\sigma^{+j} \psi_{f} \left(z/2\right)
\, | \, \bm{p}, \textrm{mf} \, \rangle \bigg{|}_{z^{+}, \bm{z}_\perp=0}
\nonumber \\
&= \frac{2M_{N}}{\sqrt{2}} \bigg{[} - 3 i \epsilon^{3jm} \, (\tau^{m})_{f'\! f} \, X_0 \, G_{\textrm{mf}, 0}
\nonumber \\
&+ \bm{1}_{f'\!f} \, X^{j}_{1} \,
\frac{|\bm{\Delta}_{\perp}|}{2 M_{N}} \, G_{\textrm{mf}, 1}
\nonumber \\
&- 3 i \epsilon^{3jm}  (\tau^{3})_{f'\! f} \, X^{m}_{1} \,
\frac{|\bm{\Delta}_{\perp}|}{2 M_{N}} \,
\tilde{G}_{\textrm{mf}, 1}
\nonumber \\
&- 3 i \epsilon^{3jl} \, (\tau^{m})_{f'\! f} \, X^{lm}_2 \,
\frac{|\bm{\Delta}_{\perp}|^{2}}{4M^{2}_{N}} \,
G_{\textrm{mf}, 2} \bigg{]} .
\label{eq:soliton_M}
\end{align}
The decomposition mirrors the multipole expansion of Eq.~(\ref{eq:multipole_GPDs}) but takes into
account the spin-flavor symmetry of the mean field. The flavor dependence is carried by the
isoscalar and isovector structures $\bm{1}$ and $\tau^a$. The terms in Eq.~(\ref{eq:soliton_M})
are invariant under combined flavor and spatial rotations, Eq.~(\ref{Eq-R:hedgehog-symmetry}),
which includes rotation of the spatial component of the light-like separation $z$. The functions
\begin{align}
G_{\textrm{mf}, 0}, \, G_{\textrm{mf}, 1}, \, \tilde{G}_{\textrm{mf}, 1}, \, G_{\textrm{mf}, 2}
\; = \; \textrm{functions}(x,\xi, t)
\label{mean_field_gpds_variables}
\end{align}
are the mean-field GPDs and depend on the variables $x$, $\xi$, and $t$
in the parametric domain of Eqs.~(\ref{x_parametric}) and (\ref{xi_t_parametric}).
The subscript denotes the multipole order in the momentum transfer $\bm{\Delta}_{\perp}$.
Note that Eq.~(\ref{eq:soliton_M}) contains a term with the quadrupole structure $X_2$
and a quadrupole mean-field GPD $G_{\textrm{mf}, 2}$; this structure was not included
in the analysis of Ref.~\cite{Schweitzer:2016jmd}.

The projection on spin-isospin states is done by integrating over the flavor rotations of
the mean-field matrix element Eq.~(\ref{eq:soliton_M}) according to Eq.~(\ref{quantization_spinisospin}).
The rotations of the structures in Eq.~(\ref{eq:soliton_M}) are
performed as
\begin{align}
\bm{1} \rightarrow \bm{1},
\hspace{2em} \tau^a \rightarrow O^{ab} \tau^b,
\label{eq:two_com}
\end{align}
where $O \equiv O(R)$ is the $O(3)$ rotation matrix Eq.~(\ref{O_def}). We obtain
\begin{align}
& \frac{M_N}{\sqrt{2}} \int \frac{d z^{-}}{2\pi} e^{i x P^{+} z^{-} }
\nonumber \\
&\times \langle \bm{p}', B'\, | \, \bar{\psi}_{f'}
\left(-z/2\right) i\sigma^{+j} \psi_{f} \left(z/2\right)
\, | \, \bm{p}, B \, \rangle \bigg{|}_{z^{+}, \bm{z}_\perp=0}
\nonumber \\
&= \frac{2M_{N}}{\sqrt{2}} \left\{
\bm{1}_{f'\! f} \langle 1\rangle_{B'B} \bigg{[}X^{j}_{1}
\frac{|\bm{\Delta}_{\perp}|}{2M_N} G_{\textrm{mf}, 1} \bigg{]}
\right.
\nonumber \\
& - 3i \epsilon^{3jl} (\tau^k)_{f'\! f} \langle O^{km} \rangle_{B'B} \bigg{[}  \delta^{ml} X_0 \,
G_{\textrm{mf}, 0} 
\nonumber \\
&  +  \delta^{m3} X^{l}_{1} \,
\frac{|\bm{\Delta}_{\perp}|}{2M_N} \,
\tilde{G}_{\textrm{mf}, 1} \left. +  X^{lm}_2  \, 
\frac{|\bm{\Delta}_{\perp}|^{2}}{4 M_{N}^2} \,
G_{\textrm{mf}, 2} \bigg{]} \right\} .
\label{largenc_general}
\end{align}
The integral over rotations with the wave function Eq.~(\ref{rotational_wave_function})
is denoted as
\begin{align}
&\langle ... \rangle_{B'B} \equiv \int dR \, \phi_{S'_3T'_3}^{*S'=T'}(R) \, ... \, \phi_{S_3T_3}^{S=T}(R)
\label{eq:collect_integral}
\end{align}
and evaluates to
\begin{subequations}
\label{collective_spin_isospin}
\begin{align}
\langle 1 \rangle_{B'B} &= \delta_{B'B},
\\
\langle O^{km} \rangle_{B'B} &=
- \sqrt{\frac{2S + 1}{2S' + 1}} \; \langle S S_{3}, 1 b | S' S^{\prime}_{3}\rangle \;
\nonumber \\
& \times \langle T T_3, 1 a | T^{\prime} T^{\prime}_{3} \rangle U^{\rm SC}_{ak} U^{\rm SC}_{bm},
\label{O_rotational}
\end{align}
\end{subequations}
where $a, b = 0, \pm 1$ and $k, m = 1, 2, 3$ are the spherical and Cartesian 3-vector components,
respectively, and $U^{\rm SC}$ is the transformation matrix from the spherical to the Cartesian
coordinates \cite{Varshalovich:1988ifq}. Equation~(\ref{largenc_general}) is the general
result for the chiral-odd GPD in baryon states in the $S = T = 1/2$ or $3/2$ representation in leading order
of the $1/N_c$ expansion. It covers diagonal and non-diagonal flavor operators ($f' = f$ and $f' \neq f$)
and $N \rightarrow N, N \rightarrow \Delta$, and $\Delta \rightarrow \Delta$ transitions.

The general large-$N_c$ matrix element Eq.~(\ref{largenc_general}) is now evaluated for the spin-flavor
quantum number of the proton, $S = T = S' = T = 1/2$ and $T_{3}=T'_{3}=1/2$. In this case only
the flavor-diagonal matrices $\bm{1}$ and $\tau^3$ contribute to the baryon matrix element, and the
rotational matrix element Eq.~(\ref{O_rotational}) is given by
\begin{align}
\langle O^{3i} \rangle_{B'B} &=- \frac{1}{3} (\tau^{3})_{T'_{3} T_{3}} (\sigma^{i})_{S'_{3}S_{3}}
=- \frac{1}{3} (\sigma^{i})_{S'_{3}S_{3}} .
\label{eq:proton_wf}
\end{align}
We obtain the proton matrix element in the notation of Eq.~(\ref{eq:General_ME}) as 
\begin{subequations}
\label{eq:model_spin_flavor_2}
\begin{align}
\mathcal{M}^{u+d}_{\mathrm{GPDs}}[i\sigma^{+j}] = & \frac{2 M_N}{\sqrt{2}} \left[ \bm{1} \, X^{j}_{1} \, 
\frac{|\bm{\Delta}_{\perp}|}{2  M_N} G_{\textrm{mf}, 1} \right],
\\
\mathcal{M}^{u-d}_{\mathrm{GPDs}}[i\sigma^{+j}] = & \frac{2 M_N}{\sqrt{2}}
\biggl{[} i \epsilon^{3jm} \sigma^{m} \, X_0  \,
G_{\textrm{mf}, 0}
\nonumber \\
&+i \epsilon^{3jm} \sigma^{3} \, X^{m}_{1} \,
\frac{|\bm{\Delta}_{\perp}|}{2  M_N} \,
\tilde{G}_{\textrm{mf}, 1}
\nonumber \\
&+ i \epsilon^{3jl} \sigma^{m} \, X^{lm}_2 \,
\frac{|\bm{\Delta}_{\perp}|^{2}}{4M_N^2} \,
G_{\textrm{mf}, 2} \biggr{]},
\end{align}
\end{subequations}
where it is understood that $\bm{1}$ and $\sigma^i$ are matrices in the spin quantum numbers $S_3$ and $S_3'$
[see Eqs.~(\ref{spin_operators})--(\ref{spin_S_3}) and Sec.~\ref{subsec:multipole_gpds}].
Equation~(\ref{eq:model_spin_flavor_2}) expresses the proton matrix
in terms of the mean-field GPDs.
Note that $G_{\textrm{mf}, 0}, \tilde G_{\textrm{mf}, 1}$ and $G_{\textrm{mf}, 2}$ multiply the
isovector structures emerging from the mean field, and $G_{\textrm{mf}, 1}$ the isoscalar structure.
In order to connect the mean-field GPDs with the conventional proton GPDs,
we compare Eq.~(\ref{eq:model_spin_flavor_2}) with the multipole expansion of the proton matrix element
Eq.~(\ref{eq:General_ME_GPDs}). In the large-$N_c$ limit the latter becomes
\begin{align}
&\mathcal{M}^{u\pm d}_{\mathrm{GPDs}}[i\sigma^{+j}]
\nonumber \\
& = \frac{2M_{N}}{\sqrt{2}} \bigg{\{}i \epsilon^{3jm} \sigma^{m} \, X_0 
\left[ H^{u\pm d}_{T} - \left( \frac{|\bm{\Delta}_{\perp}|^{2}}{8M^{2}_{N}} + \xi^{2} \right) E^{u\pm d}_{T}\right.
\nonumber \\
& \left. \hspace{1em} \phantom{\frac{0}{0}} 
+ \xi \tilde{E}^{u\pm d}_{T} \right]
\nonumber \\
&+ \bm{1} \, X^{j}_{1} \,
\frac{|\bm{\Delta}_{\perp}|}{2 M_{N}} \left( H^{u\pm d}_{T} + 2 \tilde{H}^{u\pm d}_{T} + E^{u\pm d}_{T}\right)
\nonumber \\
&+ i \epsilon^{3jm} \sigma^{3} \, X^{m}_{1} \,
\frac{|\bm{\Delta}_{\perp}|}{2 M_{N}}
\left( - \frac{\xi}{2} H^{u\pm d}_{T} + \tilde{E}^{u\pm d}_{T} - \xi E^{u\pm d}_{T} \right)
\nonumber \\
&+ i \epsilon^{3jl} \sigma^{m} \, X^{lm}_2 \,
\frac{|\bm{\Delta}_{\perp}|^{2}}{4 M_N^2} \left( \frac{1}{2} H^{u\pm d}_{T} + E^{u\pm d}_{T} \right) \bigg{\}},
\label{eq:Multipole}
\end{align}
where we have simplified the expressions using the $N_c$ scaling of the kinematic variables
Eq.~(\ref{xi_t_parametric}). [Equation~(\ref{eq:Multipole}) is the multipole expansion with
canonical nucleon spin states, which differs from the one with light-front helicity spin
states, Eq.~(\ref{eq:multipole_GPDs}), by terms of the order $|\bm{\Delta}_\perp|/M_{N}$;
the difference is irrelevant in
leading order of $1/N_c$ but becomes relevant when considering subleading corrections.]
Equation~(\ref{eq:Multipole}) applies to both the flavor-nonsinglet and singlet matrix elements.
Comparing Eq.~(\ref{eq:Multipole}) with Eq.~(\ref{eq:model_spin_flavor_2}) we obtain
relations between the proton GPDs and the mean-field GPDs in the large-$N_c$ limit.
In the flavor-nonsinglet sector,
\begin{subequations}
\label{eq:largeNc_relation_nonsinglet}
\begin{align}
H^{u-d}_{T} + \left( \frac{t}{8M^{2}_{N}} - \frac{\xi^2}{2}  \right) E^{u-d}_{T}
+ \xi \tilde{E}^{u-d}_{T}
&= G_{\textrm{mf}, 0},
\label{relation_0}
\\
H^{u-d}_{T} + 2 \tilde{H}^{u-d}_{T} + E^{u-d}_{T}
&= Z_1,
\label{relation_1}
\\[.5ex]
- \frac{\xi}{2} H^{u-d}_{T} - \xi E^{u-d}_{T} + \tilde{E}^{u-d}_{T} 
&= \tilde{G}_{\textrm{mf}, 1},
\label{relation_1tilde}
\\
\frac{1}{2} H^{u-d}_{T} + E^{u-d}_{T}
&= G_{\textrm{mf}, 2}.
\label{relation_2}
\end{align}
\end{subequations}
In the flavor-singlet sector
\begin{subequations}
\label{eq:largeNc_relation_singlet}
\begin{align}
H^{u+d}_{T} + \left( \frac{t}{8M^{2}_{N}} - \frac{\xi^2}{2}  \right) E^{u+d}_{T} + \xi \tilde{E}^{u+d}_{T}
&= Z_0,
\label{relation_singlet_0}
\\
H^{u+d}_{T} + 2 \tilde{H}^{u+d}_{T} + E^{u+d}_{T}
&= G_{\textrm{mf}, 1},
\label{relation_singlet_1}
\\[.5ex]
- \frac{\xi}{2} H^{u+d}_{T} - \xi E^{u+d}_{T} + \tilde{E}^{u+d}_{T} 
&= \tilde{Z}_1,
\label{relation_singlet_1tilde}
\\
\frac{1}{2} H^{u+d}_{T} + E^{u+d}_{T}
&= Z_2.
\label{relation_singlet_2}
\end{align}
\end{subequations}
Here we have used that in the large-$N_c$ limit [see Eq.~(\ref{xi_t_parametric})]
\begin{align}
\frac{|\bm{\Delta}_{\perp}|^{2}}{8M^{2}_{N}} + \xi^{2}
= -\frac{t}{8M^{2}_{N}} + \frac{\xi^2}{2} + \mathcal{O}(N_c^{-3}).
\end{align}
The functions $Z_1$ in Eq.~(\ref{eq:largeNc_relation_nonsinglet}) and $Z_0, \tilde{Z}_1$ and $Z_2$
Eq.~(\ref{eq:largeNc_relation_singlet}) represent isovector and isoscalar multipoles that are zero
in leading order of the $1/N_c$ expansion (i.e., in the static mean field) and become nonzero only
at next-to-leading order (due to rotations of the mean field). The order in $1/N_c$ of these
``zero functions'' will be established in the following and will result in a set of equations
that can be solved consistently within the $1/N_c$ expansion.
\subsection{Scaling behavior and relations \label{sec:IVC}}
Using the relations Eq.~(\ref{eq:largeNc_relation_nonsinglet}) and (\ref{eq:largeNc_relation_singlet})
we can now establish the $N_c$ scaling of the proton GPDs and derive relations between them.
The primary input to this is the $N_c$ scaling of the mean-field GPDs.

In the parametrization of the mean-field matrix element Eq.~(\ref{eq:soliton_M}), the mass dimension of
the powers of $|\bm{\Delta}_\perp|$ is compensated by inverse powers of $M_N$, following standard practice as
e.g.\ in Eq.~(\ref{eq:General_ME_GPDs}). In the large-$N_c$ limit $M_N = \mathcal{O}(N_c)$, and the
powers of $M_N$ influence the $N_c$ scaling of the mean-field GPDs multiplying the structures.
It can easily be seen that the functions exhibiting ``natural'' $N_c$ scaling are
\begin{align}
G_{\textrm{mf}, 0}, \; \frac{G_{\textrm{mf}, 1}}{M_N}, \; \frac{\tilde{G}_{\textrm{mf}, 1}}{M_N},
\; \frac{G_{\textrm{mf}, 2}}{M_N^2},
\label{ncscaling_natural}
\end{align}
which have dimension $\textrm{(mass)}^{-n}$ for the multipoles of order $n =$ 1 and 2.
The physical scale governing these functions is the baryon radius, which is $\mathcal{O}(N_c^0)$
and stable in the large-$N_c$ limit. This circumstance should be kept
in mind in the following. To facilitate comparison with the conventional GPDs we present the $N_c$ scaling
in terms of the dimensionless functions
$G_{\textrm{mf}, 0}, G_{\textrm{mf}, 1}, \tilde{G}_{\textrm{mf}, 1}$ and $G_{\textrm{mf}, 2}$,
even though their scaling behavior is influenced by the powers of $M_N$.

The $N_c$ scaling of the mean-field GPDs is posited as
\begin{align}
&\left\{ G_{\textrm{mf}, 0}, \, \frac{G_{\textrm{mf}, 1}}{M_N}, \, \frac{\tilde{G}_{\textrm{mf}, 1}}{M_N},
\, \frac{G_{\textrm{mf}, 2}}{M_N^2} \right\}(x,\xi,t)
\nonumber \\
& \sim N_c^2 \times \mathrm{function}(N_{c}x,N_{c}\xi,t).
\label{largenc_dimensionful}
\end{align}
The scaling function does not depend on $N_c$ and is stable in the large-$N_c$ limit; the
form of the arguments follows from the scaling of the $x$ and $\xi$ variables,
Eqs.~(\ref{xi_t_parametric}) and (\ref{x_parametric}) \cite{Diakonov:1996sr}.
[It is understood that each GPD has its own scaling function; Eq.~(\ref{largenc_dimensionful})
and the following formulas indicate the scaling behavior in a compact notation.]
The power of $N_c$ multiplying the function is based on several arguments and observations:
(i)~The power of $N_c$ of the GPDs can be inferred from the $N_c$ scaling of the tensor FFs
representing the first moments of the GPDs (see Secs.~\ref{sec:2} and \ref{sec:5}).
(ii)~The analysis of GPDs in the general mean-field picture of large-$N_c$ baryons in Sec.~\ref{sec:5}
shows that the functions Eq.~(\ref{largenc_dimensionful}) arise as the sum of quark single-particle matrix
elements and have the indicated $N_c$ scaling. (iii) Calculation of the GPDs
in the chiral quark-soliton model confirms the $N_c$ scaling Eq.~(\ref{largenc_dimensionful}).

Based on Eq.~(\ref{largenc_dimensionful}), the $N_c$ scaling of the dimensionless mean-field
GPDs is obtained as
\begin{align}
&\{G_{\textrm{mf}, 0}, \, G_{\textrm{mf}, 1}, \, \tilde{G}_{\textrm{mf}, 1}, \, G_{\textrm{mf}, 2}\}(x,\xi,t)
\nonumber \\
& \sim \ \{N^{2}_{c}, \, N^{3}_{c}, \, N^{3}_{c}, \, N^{4}_{c} \} \times \mathrm{function}(N_{c}x,N_{c}\xi,t).
\label{eq:Nc_GPDs_1}
\end{align}
The functions $Z_1$ in Eq.~(\ref{eq:largeNc_relation_nonsinglet}) and $Z_0, \tilde{Z}_1$ and $Z_2$
in Eq.~(\ref{eq:largeNc_relation_singlet}) parametrize terms in the matrix elements with flavor structure
opposite to that of the leading structures parametrized by the mean-field GPDs $G_1$ and
$G_0, \tilde{G}_1$ and $G_2$, respectively. The scaling behavior in $1/N_c$ of these functions
is therefore suppressed by at least one power of $1/N_c$ relative to that of the mean-field GPDs
in Eq.~(\ref{eq:Nc_GPDs_1}),
\begin{align}
& \{Z_0, \, Z_1, \, \tilde Z_1, \, Z_2 \}(x,\xi,t)
\nonumber \\
& \sim  \{N^{1}_{c}, \, N^{2}_{c}, \, N^{2}_{c}, \, N^{3}_{c} \}
\times \mathrm{function}(N_{c}x,N_{c}\xi,t).
\label{eq:Nc_zero_functions}
\end{align}

Using the scaling assignments of Eqs.~(\ref{eq:Nc_GPDs_1}) and (\ref{eq:Nc_zero_functions})
and the systems of equations Eqs.~(\ref{eq:largeNc_relation_nonsinglet}) and (\ref{eq:largeNc_relation_singlet}),
we can now derive the $N_c$ scaling of the proton GPDs. In order to isolate the individual proton GPDs a careful
analysis is needed, combining the equations in a manner consistent with the parametric order of the terms.
In the flavor-nonsinglet sector, combining Eqs.~(\ref{relation_1tilde}) and (\ref{relation_2})
we first obtain that $\tilde{E}_T \sim N_c^3$.
Using this result, and assuming that $E_T$ is at most $\sim N_c^4$ as allowed by
Eq.~(\ref{relation_2}) (excluding unnatural cancellations), we then determine
from Eq.~(\ref{relation_0}) that $H_T \sim N_c^2$.
Carrying this into Eq.~(\ref{relation_2}), we in turn obtain
$E_T \sim N_c^4$. Using this in Eq.~(\ref{relation_1}), we finally obtain
$2 \tilde{H}_T + E_T \sim N_c^2$. Because $\tilde{H}_T$ and $E_T$ are individually
$\sim N_c^4$, the latter implies the nontrivial relation
$2 \tilde{H}_T = - E_T + \mathcal{O}(N_c^2)$. Altogether, we obtain the scaling behavior
of the leading chiral-odd proton GPDs in the large-$N_c$ limit as
\begin{align}
& \{H^{u-d}_{T}, \, \tilde{H}^{u-d}_{T}, \, E^{u-d}_{T}, \, \tilde{E}^{u-d}_{T}\}(x,\xi,t)
\nonumber \\
& \sim \ \{ N^{2}_{c}, \, N^{4}_{c}, \, N^{4}_{c}, \, N^{3}_{c} \} \times \mathrm{function}(N_{c}x,N_{c}\xi,t),
\label{eq:Ncscaling}
\end{align}
with the non-trivial relation
\begin{align}
2\tilde{H}^{u-d}_{T}(x, \xi, t) = - E^{u-d}_{T}(x, \xi, t),
\label{eq:largeNc_relation_1}
\end{align}
which is valid up to terms $\sim N_c^2$, i.e., up to relative corrections $\sim 1/N_c^2$ to the
functions on each side.\footnote{The parametric order of the corrections to the large-$N_c$ relation
Eq.~(\ref{eq:largeNc_relation_1}) depends on the details of the implementation of the mean-field picture
beyond the leading order. In the present calculation with independent translational and rotational
zero modes, the corrections appear only at relative order $1/N_c^2$, i.e., suppressed by two powers
of $1/N_c$ relative to the leading order. Interplay of the translational and rotational zero modes
at subleading order may give rise to relative corrections $1/N_c$. The bag model calculation of
Ref.~\cite{Tezgin:2024tfh} finds corrections of order $1/N_c$. We therefore
only claim Eq.~(\ref{eq:largeNc_relation_1}) to be valid up to relative corrections of order $1/N_c$.
\label{footnote:corrections}
}
In the flavor-singlet sector, in a similar way we obtain
\begin{align}
& \{H^{u+d}_{T}, \, \tilde{H}^{u+d}_{T}, \, E^{u+d}_{T}, \, \tilde{E}^{u+d}_{T} \}(x,\xi,t)
\nonumber \\
& \sim \ \{N_{c}, \, N^{3}_{c}, \, N^{3}_{c}, \, N^{2}_{c} \} \times \mathrm{function}(N_{c}x,N_{c}\xi,t),
\label{eq:Ncscaling_singlet}
\end{align}
and there is no relation analogous to Eq.~(\ref{eq:largeNc_relation_1}).
Equations~(\ref{eq:Ncscaling})--(\ref{eq:Ncscaling_singlet}) establish the $N_c$ scaling of the
proton's conventional chiral-odd GPDs as defined by the parametrization Eq.~(\ref{eq:General_ME}).

Having determined the $N_c$ scaling of the GPDs, we can now ``reverse the logic'' and express the
mean-field GPDs in terms of the conventional GPDs. Simplifying the relations
Eqs.~(\ref{eq:largeNc_relation_nonsinglet}) and (\ref{eq:largeNc_relation_singlet}) by using the scaling
behavior of Eqs.~(\ref{eq:Ncscaling})--(\ref{eq:Ncscaling_singlet}), we obtain
\begin{subequations}
\label{eq:large_Nc_relations}
\begin{align}
G_{\textrm{mf}, 0} &=
H^{u-d}_{T}
+ \left( \frac{t}{8M^{2}_{N}} - \frac{\xi^2}{2}  \right) E^{u-d}_{T}
+ \xi \tilde{E}^{u-d}_{T},
\label{G0_conventional}
\\
G_{\textrm{mf}, 1} &=  2 \tilde{H}^{u+d}_{T} + E^{u+d}_{T},
\\[1ex]
\tilde{G}_{\textrm{mf}, 1} &= - \xi E^{u-d}_{T} + \tilde{E}^{u-d}_{T} ,
\\[1ex]
G_{\textrm{mf}, 2} &= E^{u-d}_{T}.
\end{align}
\end{subequations}
The combinations on the RHS can be regarded as alternative definitions of the chiral-odd GPDs
that have homogeneous $N_c$ scaling and coincide with the mean-field GPDs in the large-$N_c$ limit.
Supplemented with the corresponding expressions for the opposite flavor combinations, which are
suppressed by a power $1/N_c$, Eq.~(\ref{eq:large_Nc_relations}) provides an alternative definition
of the full set of chiral-odd GPDs. The new basis has a clear physical interpretation and can
be employed in the discussion of nucleon structure and the analysis of exclusive scattering processes.
Our subsequent analysis of the chiral-odd GPDs will be conducted in terms of these new GPDs.

The expression of the mean-field GPDs in terms of the conventional GPDs, Eq.~(\ref{eq:large_Nc_relations}),
combined with the parity in $\xi$ of the conventional GPDs, Eq.~(\ref{xi_parity}), implies that the
mean-field GPDs have definite parity in $\xi$. This property will be explored further in Sec.~\ref{sec:5}
and Appendix~\ref{app:a}.
\subsection{Comparison with lattice QCD results}
%
%
\begin{figure}[t]
\includegraphics[scale=0.6]{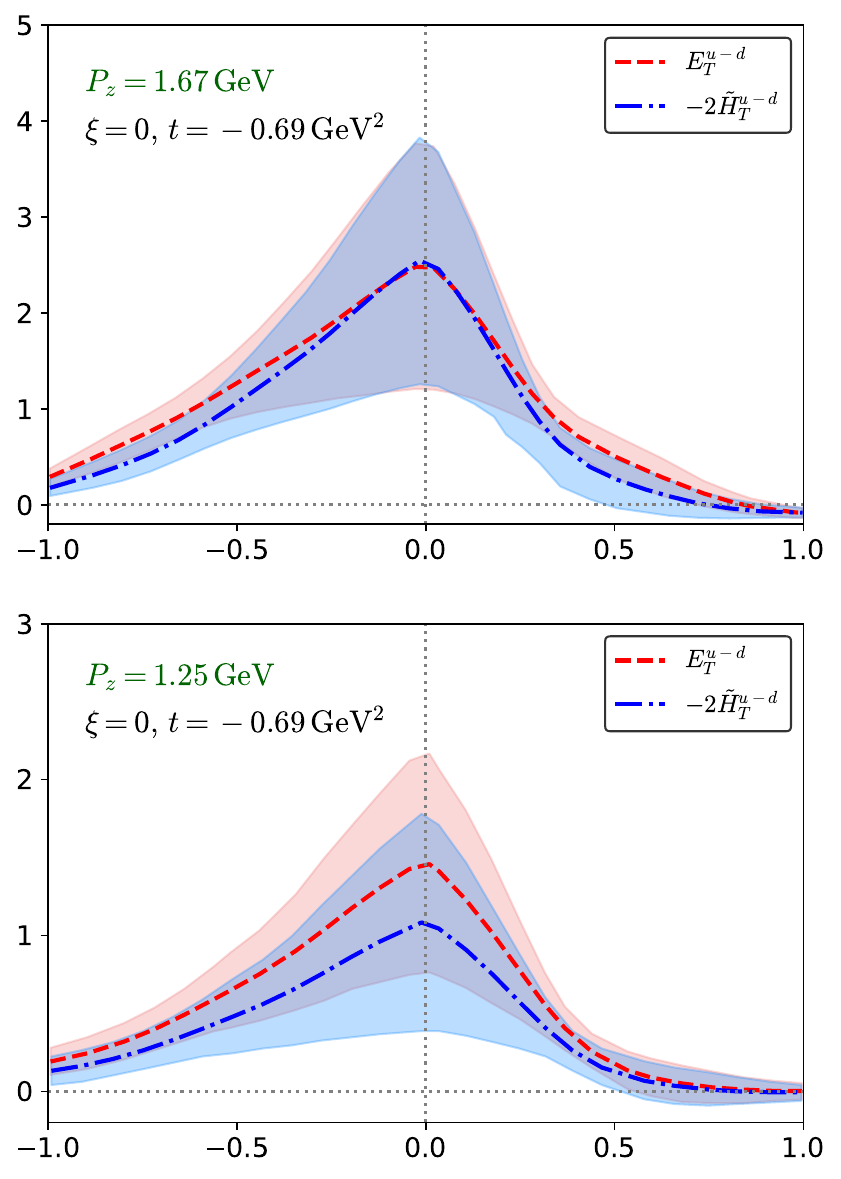}
\caption{Comparison between the GPDs $E^{u-d}_{T}$ (dashed lines)
and $-2\tilde{H}^{u-d}_{T}$ (dot-dashed lines) extracted from the lattice QCD
calculation of Ref.~\cite{Alexandrou:2021bbo}. The GPDs are shown as functions
of $x$, at $\xi=0$ and $t=-0.69 \, \mathrm{GeV}^{2}$, and were extracted from
correlation functions with nucleon momenta $P_{z}=1.67~\mathrm{GeV}$~(upper panel)
and $P_{z}=1.25~\mathrm{GeV}$~(lower panel). The $1/N_c$ expansion predicts that
the two GPDs are the same in leading order, see Eq.~(\ref{eq:largeNc_relation_1}).}
\label{fig:lattice}
\end{figure}
We now want to confront the results of the $1/N_c$ expansion with numerical estimates of chiral-odd GPDs.
The most conclusive test can be performed with the large-$N_c$ relation Eq.~(\ref{eq:largeNc_relation_1}).
It states that the functions $2\tilde H_T^{u-d}$ and $-E_T^{u-d}$ are individually large $\mathcal{O}(N_c^4)$
and equal at this order, while their difference is small $\mathcal{O}(N_c^3)$
(see Footnote~\ref{footnote:corrections}).
This prediction can be
compared with lattice QCD results for the tensor FFs and chiral-odd GPDs
\cite{Gockeler:2005cj,QCDSF:2006tkx,Park:2021ypf,Alexandrou:2021bbo,Alexandrou:2022dtc}.

Figure~\ref{fig:lattice} shows recent lattice QCD results for the flavor-nonsinglet
GPDs $E^{u-d}_{T}$ and $-2\tilde{H}^{u-d}_{T}$ \cite{Alexandrou:2021bbo}.
In this approach the GPDs are extracted by approximating the light-cone correlation function
of quark fields by an equal-time correlation function in a high-momentum nucleon state;
see references in Ref.~\cite{Alexandrou:2021bbo} for details. One sees that the functions $E^{u-d}_{T}$ and
$-2\tilde{H}^{u-d}_{T}$ are individually large and very close to each other, as
stated by the large-$N_c$ relation Eq.~(\ref{eq:largeNc_relation_1}). It shows that the
$1/N_c$ expansion can provide useful quantitative predictions for the GPDs at $N_c = 3$.
More stringent comparisons will become possible with more precise lattice QCD results,
and with $1/N_c$ corrections to the leading-order expansion.

The large-$N_c$ relation Eq.~(\ref{eq:largeNc_relation_1}) is also supported by the
numerical estimates of chiral-odd GPDs in the bag model~\cite{Tezgin:2024tfh} and a
light-front quark model \cite{Pasquini:2005dk}. It is also confirmed by the results of
the chiral quark-soliton model, which realizes the large-$N_c$ mean-field picture of baryons
with the effective dynamics emerging from chiral symmetry breaking \cite{inprep}.

Numerical tests of the large-$N_c$ hierarchy of different spin-flavor structures of the GPDs are less conclusive,
because the $1/N_c$ expansion predicts only the parametric order in $1/N_c$, and the numerical values
depend on coefficients of order unity. This has been observed in the large-$N_c$ analysis of matrix
elements of local operators such as the vector and axial vector currents.

\section{Polynomiality at large $N_c$ \label{sec:5}}
\subsection{GPDs in mean-field picture}
The chiral-odd GPDs satisfy the polynomiality relations Eq.~(\ref{eq:polynomiality}).
Their moments are polynomials of degree $\leq m - 1$ in $\xi$ and have definite parity in $\xi$.
These properties follow from the fact that the moments are matrix elements of local tensor operators,
which are constrained by relativistic covariance and reflection symmetries.
In the $1/N_c$ expansion these symmetries are not manifest, as the mean-field breaks translational
invariance and exhibits only a generalized form of rotational invariance. It is therefore
necessary to study how polynomiality is realized in the $1/N_c$ expansion.

Polynomiality in the mean-field picture is studied most easily by representing the
GPDs as matrix elements in quark single-particle states in the mean field.
The single-particle motion in the mean-field exhibits a generalized time reversal invariance,
which controls the behavior of the GPDs under reflection of $\xi$.
Polynomiality arises from combining this reflection symmetry with the spin-flavor rotational
symmetry of mean field. This can be demonstrated in abstract form, with general assumptions
about the single-particle motion in the mean field that do not depend on the specific dynamics.

Polynomiality of the GPDs at large $N_c$ was studied in Ref.~\cite{Goeke:2007fp} in the specific
dynamics of the chiral quark-soliton model, where the quarks possess a dynamical mass as a result of
chiral symmetry breaking, the mean field is a chiral meson field (``soliton"), and the spin-flavor
symmetry is realized by the ``hedgehog'' form of the chiral meson field. In the present study we
remain at an abstract level and use only general features that are independent of the specific dynamics
\cite{Witten:1979kh}.
Our aim is to isolate and identify the elements that are needed for ensuring polynomiality of the GPDs.
This is useful for generalizing to other realizations of the large-$N_c$ mean-field picture.

We make the following minimal assumptions:

1) The relativistic motion of the quarks in the mean field of the large-$N_c$ baryon is described by a
single-particle hamiltonian $\hat{H}$ in orbital, spin, and flavor degrees of freedom \cite{Witten:1979kh}.
The single-particle wave functions and energies are determined by the eigenvalue equation
(in bispinor representation)
\begin{align}
\hat{H} \Phi_n (\bm{x}) = E_n \Phi_n (\bm{x}).
\label{hamiltonian}
\end{align}
Here the mean field is centered at the origin and invariant under under spatial reflection (parity).

2) Combined flavor and spatial rotations are generated by the so-called grand spin operator
\begin{align}
\hat{\bm{K}} = \hat{\bm{L}} + \hat{\bm{\Sigma}} + \hat{\bm{T}},
\end{align}
where $\hat{\bm{L}} = \bm{\hat{x}} \times \bm{\hat{p}}$ is the orbital angular momentum operator,
$\hat{\bm{\Sigma}}$ the spin operator, and $\hat{\bm{T}}=\bm{\tau}/2$ the isospin operator.
The invariance of the mean field under such combined rotations implies that the hamiltonian commutes
with the grand spin operator,
\begin{align}
[\hat{\bm{K}}, \hat{H}] = 0.
\end{align}
The single-particle states are therefore classified by their grand spin eigenvalues $K$ and $K_3$,
in addition to the other quantum numbers characterizing the states (angular momentum, radial quantum number)
\begin{align}
| n  \rangle \equiv | K, K_{3}, \ldots \rangle .
\label{n_states}
\end{align}

3) In the ground state of the nucleon, a set of quark single-particle levels is occupied with $N_c$ 
quarks each, producing a state with total baryon number $B = 1$. Quantities such as the nucleon mass 
and other observables arise as sum over quark single-particle levels. The precise nature of single-particle 
spectrum (discrete or continuous, positive and negative energy-states) is not needed in the following.

4) The matrix elements of the leading-twist partonic operators are given by sums over the occupied quark
single-particle states, with the single-particle operators acting only on the single-particle degrees of
freedom (position/momentum, spin), without interactions with the rest of system.
The leading spin-flavor components of the matrix element in $1/N_c$ are given by single sums
over the quark levels. For the chiral-odd partonic operator,
\begin{align}
&
\left.
\begin{array}{l}
\mathcal{M}^{u+d}_{\mathrm{GPDs}}[i \sigma^{+j}]
\\[2ex]
\mathcal{M}^{u-d}_{\mathrm{GPDs}}[i \sigma^{+j}]
\end{array}
\right\}
= 2 M^{2}_{N} N_{c}
\left\{
\begin{array}{c}
\bm{1}
\\[2ex]
-\frac{1}{3} \sigma^{k}
\end{array}
\right\}
\nonumber \\[2ex]
& \times \sum_{n,\mathrm{occ}} \;
\int \frac{d\lambda}{2\pi}  e^{i \lambda ( xM_{N}-E_{n} ) } \int d^{3}y \, e^{i \bm{\Delta} \cdot \bm{y}}
\nonumber \\
&\times \Phi^{\dagger}_{n}(\bm{y} + \bm{e}_{3}\lambda/2)
\; \gamma^{0} i \sigma^{+j}
\left\{
\begin{array}{c}
1
\\[1ex]
\tau^{k}
\end{array}
\right\}
\Phi_{n}(\bm{y} - \bm{e}_{3}\lambda/2 ) .
\label{eq:general}
\end{align}
The form follows from the general expression of the mean-field matrix element
Eq.~(\ref{quantization_momentum}). The light-like separation $z$ in the operator parametrized as
\begin{align}
z^0 = \lambda, \hspace{2em} \bm{z} = \lambda \bm{e}_3 .
\end{align}
The dependence on $z^0$ is governed by the time evolution of the single-particle wave function,
which is determined by the single-particle energy $E_n$ in accordance with Eq.~(\ref{hamiltonian}).
``occ'' denotes the sum over occupied single-particle levels.
In Eq.~\eqref{eq:general} it is understood that $\bm{1}$ and $\sigma^k$
are matrices in the nucleon spin quantum numbers $S_3'$ and $S_3$, as in Eq.~(\ref{eq:model_spin_flavor_2}).
The matrix element is for the proton $T'_{3}=T_{3}=1/2$; for the neutron the isoscalar component remains 
the same, while the isovector component changes sign. 

From Eq.~\eqref{eq:general}, we obtain the mean-field GPDs in the
definition of Eqs.~(\ref{eq:soliton_M}) and (\ref{mean_field_gpds_variables}) as
\begin{subequations}
\label{eq:explicit}
\begin{align}
& G_{\textrm{mf}}(x, \xi, t)
\nonumber \\[.5ex]
& = 2 M_N N_c \sum_{n, \mathrm{occ}}
\int \frac{d\lambda}{2\pi} e^{i \lambda ( x M_{N} -E_{n} ) } 
\int d^{3}y \, e^{i \bm{\Delta} \cdot \bm{y}}
\nonumber \\
& \times \Phi^{\dagger}_{n}(\bm{y} + \bm{e}_{3}\lambda/2)
\, \hat{O} \,
\Phi_{n}(\bm{y} - \bm{e}_{3}\lambda/2), 
\end{align}
\begin{alignat}{2}
&
\hat{O} =
&&
\nonumber 
\\[2ex]
& -\frac{1}{12} (1 + \gamma^{0} \gamma^{3})   i (\bm{\gamma} \times \bm{\tau})^{3}
&&
(\textrm{for}\; G_{\textrm{mf}, 0}),
\nonumber 
\\[2ex]
& - \frac{1}{|\bm{\Delta}_{\perp}|^{2}} (1 + \gamma^{0}
 \gamma^{3}) (\bm{\gamma}_{\perp} \cdot \bm{\Delta_{\perp}})
&& \left( \frac{G_{\textrm{mf}, 1}}{M_N} \right),
\nonumber 
\\[2ex]
& - \frac{1}{3 |\bm{\Delta}_{\perp}|^{2}} (1 + \gamma^{0}
\gamma^{3}) i (\bm{\gamma} \times \bm{\Delta})^{3} \tau^{3}
&& \left( \frac{\tilde G_{\textrm{mf}, 1}}{M_N} \right),
\nonumber 
\\[2ex]
& - \frac{4}{3 |\bm{\Delta}_{\perp}|^{4}} (1 +
 \gamma^{0} \gamma^{3}) i \left[ (\bm{\gamma} \times
 \bm{\Delta})^{3}( \bm{\Delta_{\perp}}\cdot \bm{\tau_{\perp}}) \phantom{\frac{0}{0}}
\hspace{-2em}
\right.
&&
\nonumber \\
& \hspace{1em} \left. - \frac{1}{2} (\bm{\gamma} \times \bm{\tau})^{3}
 |\bm{\Delta_{\perp}}|^{2} \right].
&& \left(\frac{G_{\textrm{mf}, 2}}{M_N^2}\right).
\nonumber
\\[1ex]
\end{alignat}
\end{subequations}
The $3$-momentum transfer $\bm{\Delta}$ is related to the GPD variables as
\begin{align}
-2 M_N \xi = \Delta^3,
\hspace{2em}
t = - |\bm{\Delta}|^2,
\label{xi_t_Delta}
\end{align}
and the $N_c$ scaling of the variables is as specified in Eqs.~(\ref{P_Delta_parametric})--(\ref{x_parametric}).
Here the mean-field GPDs are represented in first-quantized form, as matrix elements of
quark single-particle operators acting only on single-particle variables.
The single-particle operators express the multipole character of the respective GPDs.
Note that the explicit expressions of the mean-field GPDs in Eq.~(\ref{eq:explicit})
exhibit the $N_c$ scaling summarized in Eq.~(\ref{largenc_dimensionful}).

Assumptions 1-4 and the first-quantized representation of the mean-field GPDs of Eq.~(\ref{eq:explicit})
are our basis for the analysis of polynomiality of the chiral-odd GPDs in the large-$N_c$ limit.
\subsection{Moments in mean-field picture}
From Eq.~(\ref{eq:explicit}) we obtain the moments of the mean-field GPDs in the
first-quantized representation as
\begin{align}
&\int dx \, x^{m-1} \, G_{\textrm{mf}, 0}(x,\xi, t) 
\nonumber \\
&= \frac{2N_{c} i^{m-1}}{M^{m-1}_N}  \sum_{n, \mathrm{occ}} \,
\int d^{3}y \, e^{i \bm{\Delta} \cdot \bm{y}} 
\left( \frac{d}{d\lambda} \right)^{m-1}
\nonumber \\
&\times   \left[e^{-i\lambda E_{n}}\Phi^{\dagger}_{n}(\bm{y} + \bm{e}_{3}\lambda/2)
\, \hat{O} \,
\Phi_{n}(\bm{y} - \bm{e}_{3}\lambda/2)\right] \bigg{|}_{\lambda=0}, \hspace{1em}
\label{G0_moment_original}
\end{align}
and similarly for other the multipoles (see Appendix~\ref{app:a}). The derivative with respect to the
light-cone distance $\lambda$ can be expressed as the action of single-particle momentum operators
on the wave functions, similar to the covariant derivatives in the QCD expression Eq.~(\ref{eq:polynomiality})
\cite{Schweitzer:2002nm,Schweitzer:2003ms,Goeke:2007fp}.
Representing the single-particle wave function in abstract form as
\begin{align}
\Phi_n(\bm{x}) \equiv \langle \bm{x}| n \rangle ,
\end{align}
and introducing the single-particle momentum operator $\hat{\bm{p}}$ conjugate to the
position operator $\hat{\bm{x}}$,
\begin{align}
[\hat{p}^i, \hat{x}^j] = -i \delta^{ij},
\end{align}
the shift in the argument of the single-particle wave function can be represented as
\begin{align}
\Phi_{n}(\bm{y}
- \bm{e}_{3}\lambda/2) = \langle \bm{y} | e^{-i \hat{p}^{3} \lambda/2
} | n \rangle .
\end{align}
Using the completeness relation
\begin{align}
\int d^{3}y \, e^{i \bm{\Delta} \cdot \bm{{y}}} | \bm{y}
\rangle \langle \bm{y} | = e^{i \bm{\Delta} \cdot \bm{\hat{x}}},
\label{completeness}
\end{align}
Eq.~(\ref{G0_moment_original}) can be converted to 
\begin{align}
&\int dx \, x^{m-1} G_{\textrm{mf}, 0}(x,\xi, t)
\nonumber \\
&=-\frac{1}{6} \frac{ M_{N} N_{c}}{M^{m}_{N}} \sum_{n,\mathrm{occ}}
\sum^{m-1}_{k=0} \left(\begin{array}{c} m-1 \\ k \end{array} \right)
\frac{E^{m-1-k}_{n}}{2^{k}} \sum^{k}_{j=0} \left(\begin{array}{c} k \\ j \end{array} \right) 
\nonumber \\
& \times \langle n | (1+\gamma^{0}\gamma^{3}) i (\bm{\gamma} \times \bm{\tau})^{3}
(\hat{p}^{3})^{j} e^{i \bm{\Delta} \cdot \bm{ \hat{x}}} (\hat{p}^{3})^{k-j} | n \rangle 
\label{G0_moment}
\end{align}
(for the other mean-field GPDs, see Appendix~\ref{app:a}).
Here the moments are expressed as single-particle matrix elements of local operators
formed from the single-particle momentum and position operators. A particular feature of
the mean-field description is that the position operator appears explicitly in the expression
of the matrix element, resulting from the breaking of translational invariance by the mean field.
The function $e^{i \bm{\Delta} \cdot \bm{ \hat{x}}}$ ``pins'' the momentum operators on
either side, giving rise to a unique structure of the matrix element.

We want to demonstrate that the large-$N_c$ expression of the moment Eq.~(\ref{G0_moment}) is
a polynomial in $\xi$ of required degree and parity. This can be done using the general features
of the mean-field picture listed in the previous subsection and the specific techniques described in the
following.

Note that in the moment Eq.~(\ref{G0_moment_original}) the integral over $x$ extends over $[-\infty, \infty]$.
In the large-$N_c$ limit, when considering partonic structure in the domain $x = \mathcal{O}(N_c^{-1})$,
the range of $x$ is not limited to $[-1, 1]$. Rather, the parton distributions are exponentially
small for values $|x| \sim 1 \gg 1/N_c$, so that the integration can be extended over the infinite
domain \cite{Diakonov:1996sr,Diakonov:1997vc}.
\subsection{Symmetries in mean-field picture}
The dynamical symmetries of the mean-field picture play an essential role in the realization of polynomiality.
Here we summarize the properties and techniques used in the following calculation.

{\textit{Time reversal ($G_{5}$ symmetry):}}
Time reversal and hermiticity not only impose constraints on the polynomial properties of the GPDs
in $\xi$ but also imply the vanishing of the first moments of the GPDs $\tilde{E}^{f}_{T}$,
Eq.~(\ref{E_T_tilde_first_moment}). In the context of the mean-field picture one considers
a combination of the standard time-reversal symmetry and an isospin rotation, the so-called
$G_{5}$ transformation \cite{Schweitzer:2003ms}. It is represented by the unitary matrix
\begin{align}
G_{5}= \gamma^{1}\gamma^{3}\tau^{2},
\end{align}
and the following identities hold:
\begin{subequations}
\label{eq:G5}
\begin{align}
& G_{5} \gamma^{\mu} G^{-1}_{5} = (\gamma^{\mu})^{T},
\hspace{2em}
G_{5} \hat{\Sigma}^{i} G^{-1}_{5} = -(\hat{\Sigma}^{i})^{T},
\\
&G_{5} \tau^{a} G^{-1}_{5} = -(\tau^{a})^{T},
\\
&G_{5} \hat{H} G^{-1}_{5} = (\hat{H})^{T}, \hspace{2em}
G_{5} \Phi_{n}(\bm{x}) = \Phi^{*}_{n}(\bm{x}).
\end{align}
\end{subequations}
Using these relations, we can examine the behavior of the matrix elements of the various single-particle
operators in Eq.~(\ref{G0_moment}). For a general matrix element,
\begin{align}
& \langle n |  \Gamma (\hat{p}^{3})^{l} F(\bm{\hat{x}}) (\hat{p}^{3})^{m} | n  \rangle
\nonumber \\
& = (-1)^{l+m} \langle n |  (G_{5} \Gamma  G^{-1}_{5})^{T} (\hat{p}^{3})^{m} F(\bm{\hat{x}}) 
(\hat{p}^{3})^{l} | n  \rangle,
\label{eq:Gparity}
\end{align}
where $\Gamma$ is a spin-flavor matrix and $F(\bm{\hat{x}})$ is a general function of the position operator.

{\textit{Parity ($\hat{\Pi}$ symmetry):}}
Parity constrains the form of the chiral-odd GPDs by restricting the allowed angular momentum values
in the partial-wave expansion of the single-particle operators. The parity transformation $\hat{\Pi}$
is defined as
\begin{align}
&\hat{\Pi} = \hat{\Pi}^{-1} \equiv \gamma^{0}\hat{\mathcal{P}}, \quad
\hat{\mathcal{P}} F(\hat{\bm{x}}) \hat{\mathcal{P}}^{-1} = F(-\hat{\bm{x}}) .
\end{align}
The mean-field centered at the origin is invariant under parity, so that the
hamiltonian commutes with the parity operator,
\begin{align}
[\hat{\Pi}, \hat{H}] = 0,
\end{align}
and the single-particle states are eigenstates of parity
\begin{align}
\hat{\Pi} | n \rangle = \pm | n \rangle.
\end{align}
The general matrix element Eq.~(\ref{eq:Gparity}) transforms as
\begin{align}
&\langle n |  \Gamma (\hat{p}^{3})^{l} F(\hat{\bm{x}}) (\hat{p}^{3})^{m} | n  \rangle
\nonumber \\
&= (-1)^{l+m} \langle n |  (\gamma^{0}
\Gamma   \gamma^{0}) (\hat{p}^{3})^{l} (\hat{\mathcal{P}} F(\hat{\bm{x}})
\hat{\mathcal{P}}^{-1}) (\hat{p}^{3})^{m} | n  \rangle.
\label{eq:parity}
\end{align}

{\textit{Partial-wave expansion:}}
The dependence of the moment Eq.~(\ref{G0_moment}) on the momentum transfer $\Delta$ is contained
in the operator function $e^{i \bm{\Delta} \cdot \hat{\bm{x}}}$ arising from the quantization of the
translational motion of the mean field. The dependence on $\xi$ emerges from the identification of
$\xi$ with $\Delta^3$ in the large-$N_c$ kinematics, Eq.~(\ref{xi_t_Delta}). In order to exhibit
this dependence, we perform a partial-wave expansion of the function $e^{i \bm{\Delta} \cdot \hat{\bm{x}}}$
with the 3-axis as quantization axis (see Refs.~\cite{Landau:1991wop,Schweitzer:2003ms} 
and Appendix F in Ref.~\cite{Goeke:2007fp}),
\begin{align}
e^{i \bm{\Delta} \cdot \hat{\bm{x}}} &= \sum^{\infty}_{l=0} i^{l} (2l+1) j_{l}(|\hat{\bm{x}}| |\bm{\Delta}|)
\nonumber \\
& \times 
P_{l}\left( \frac{\hat{x}^3}{|\hat{\bm{x}}|} \right) P_{l} \left( \frac{\Delta^3}{|\bm{\Delta}|} \right),
\end{align}
where $j_{l}$ are the spherical Bessel functions and $P_{l}$ the Legendre polynomials. In terms of the
GPD variables $\xi$ and $t$, Eq.~(\ref{xi_t_Delta}), this becomes
\begin{align}
e^{i \bm{\Delta} \cdot \hat{\bm{x}}} &= \sum^{\infty}_{l=0} i^{l} (2l+1) j_{l}(|\hat{\bm{x}}|\sqrt{-t})
\nonumber \\
& \times
P_{l}\left( \frac{\hat{x}^3}{|\hat{\bm{x}}|} \right) P_{l} \left( -\frac{2\xi M_{N}}{\sqrt{-t}} \right).
\label{partial_wave_xi_t}
\end{align}
In particular, when taking the limit $t\to 0$ while keeping $\xi\neq 0$ fixed, this reduces to
\begin{align}
\lim_{t\to 0, \ \xi \neq 0}
e^{i \bm{\Delta} \cdot \hat{\bm{x}}}
= \sum^{\infty}_{l=0} \frac{(-2i\xi M_{N} |\hat{\bm{x}} |)^{l}}{l!} 
P_{l}\left( \frac{\hat{x}^3}{|\hat{\bm{x}}|} \right).
\label{eq:t0limit}
\end{align}

{\textit{Grand spin selection rule:}}
The spin-flavor symmetry of the mean field provides that the quark single-particle states are eigenstates
of the grand spin operator, Eq.~(\ref{n_states}). The mean-field GPDs and the moments Eq.~(\ref{G0_moment})
are given by sums of the expectation values of certain single-particle operators $\hat{O}$
in the single-particle states,
\begin{align}
\sum_{\textrm{other}} \sum_{K,K_3}  \langle K,K_{3},\ldots | \hat{O} | K,K_{3},\ldots \rangle ,
\label{eq:single_sum}
\end{align}
where ``other'' denotes the summation over the other quantum numbers.
Here the left and right state have the same grand spin
quantum numbers $K$ and $K_3$, and the summation includes the grand spin projection $K_3$. This circumstance
implies certain selection rules for the matrix elements of the single-particle operator. For a spherical tensor
operator in grand spin quantum numbers, $\hat{O}_{K'K_3'}$, where $K'$ and $K'_3$ are the grand spin quantum
numbers characterizing the tensor components, the summation in Eq.~(\ref{eq:single_sum}) can be performed using
the Wigner-Eckart theorem,
\begin{align}
& \sum_{K,K_3}\langle K,K_{3},\ldots | \hat{O}_{K'K_3'} | K,K_{3},\ldots \rangle
\nonumber \\
&=  \sum_{K,K_3} (-1)^{2K'} \frac{ \langle K K_{3}, K'K_3' | K K_{3}\rangle}{\sqrt{2K+1}}
\langle K,\ldots || \hat{O}_{K'} || K,\ldots \rangle
\nonumber \\[1ex]
&=  \sum_{K}  \sqrt{2K+1}  \langle K,\ldots || \hat{O}_{K'} || K,\ldots \rangle \delta_{K'0} \delta_{K_3'0},
\label{eq:Wigner_Ekart}
\end{align}
where $\langle K,\ldots || \hat{O}_{K'} || K,\ldots \rangle$ denotes the reduced matrix element of
the operator. Only the spherical tensor component with total grand spin $K' = 0$ and projection $K'_3 = 0$
can contribute to the expectation value, a consequence of the generalized rotational invariance of the
mean field. Since the isospin of the quark single-particle operators is limited ($t$-channel isospin 0 or 1),
the grand spin selection rule restricts the angular momentum of tensor operators that can contribute
to the sum over single-particle levels in the mean field.
\subsection{Polynomiality in mean-field picture}
\label{subsec:polynomiality}
Using the symmetry relations from the previous subsection, we can now demonstrate the polynomiality
of the chiral-odd GPD moments in the mean-field picture. We present the proof for the monopole GPD 
$G_{\textrm{mf}, 0}$, Eq.~(\ref{G0_moment}); the extension to the higher multipole GPDs 
in Eq.~(\ref{eq:explicit}) is described in Appendix~\ref{app:a}. First, we apply the $G_{5}$ 
transformation of Eq.~\eqref{eq:Gparity} to the single-particle matrix element of Eq.~(\ref{G0_moment}) 
and obtain
\begin{align}
& \langle n | (1+\gamma^{0}\gamma^{3}) i (\bm{\gamma} \times \bm{\tau})^{3}
(\hat{p}^{3})^{j} e^{i \bm{\Delta} \cdot \bm{ \hat{x}}} (\hat{p}^{3})^{k-j} | n \rangle
\nonumber \\
&= \langle n | (\gamma^{0}\gamma^{3})^{k+1} i (\bm{\gamma} \times \bm{\tau})^{3}
(\hat{p}^{3})^{j} e^{i \bm{\Delta} \cdot \bm{ \hat{x}}} (\hat{p}^{3})^{k-j} | n \rangle.
\label{eq:mono_int_0}
\end{align}
Here and in the following we use the fact that
\begin{align}
(\gamma^{0}\gamma^{3})^{k+1} =
\left\{
\begin{array}{ll}
\gamma^{0}\gamma^{3} & \textrm{$k$ even},
\\
1 & \textrm{$k$ odd},
\end{array}
\right.
\label{gamma0_gamma_3_notation}
\end{align}
to write the formulas for even and odd $k$ in a compact form.
Next, we perform the partial-wave expansion of the operator function $e^{i \bm{\Delta} \cdot \hat{\bm{x}}}$
using Eq.~(\ref{partial_wave_xi_t}) and obtain
\begin{align}
&\sum^{\infty}_{l=0} i^{l} (2l+1) P_{l} \left(-\frac{2\xi M_{N}}{\sqrt{-t}} \right)
\nonumber \\
& \times \langle n | (\gamma^{0}\gamma^{3})^{k+1} i (\bm{\gamma} \times \bm{\tau})^{3} 
\nonumber \\[1ex]
&\hspace{1em} \times (\hat{p}^{3})^{j} j_{l} (|\hat{\bm{x}}| \sqrt{-t})
P_{l}\left( \frac{\hat{x}^3}{|\hat{\bm{x}}|} \right)
 (\hat{p}^{3})^{k-j} | n \rangle.
\label{eq:mono_int_1}
\end{align}
Next, applying the parity transformation Eq.~\eqref{eq:parity}, we conclude that only even
partial waves ($l=0,2,4,...$) are allowed and replace
\begin{align}
\sum^{\infty}_{l=0}  [...] \to \sum^{\infty}_{l=0,2,4...} [...].
\end{align}
Next, we determine the maximum value of $l$ from the grand spin selection rule
Eq.~(\ref{eq:Wigner_Ekart}).
The spin-flavor part of the single-particle operator in Eq.~(\ref{eq:mono_int_1})
can be rewritten as
\begin{subequations}
\begin{alignat}{2}
& (\gamma^{0}\gamma^{3})^{k+1} i (\bm{\gamma} \times \bm{\tau})^{3}
&&
\nonumber \\[1ex]
&= \gamma^{0} (\bm{\Sigma} \cdot \bm{\tau} - \Sigma^3 \tau^3)
= \gamma^{0} (\bm{\Sigma}_{\perp} \cdot \bm{\tau}_{\perp}) \hspace{1em} && (\text{$k$ even}),
\label{spin_flavor_keven}
\\[1ex]
&= -  i (\bm{\Sigma} \times \bm{\tau})^{3} \gamma^{0} \gamma^{5}
&& (\text{$k$ odd}),
\label{spin_flavor_kodd}
\end{alignat}
\end{subequations}
where we have used
\begin{align}
&\bm{\Sigma} = - \gamma^0 \bm{\gamma} \gamma_5,
\hspace{2em}
\gamma_5 \equiv i \gamma^0 \gamma^1 \gamma^2 \gamma^3 .
\end{align}
The expression for even $k$, Eq.~(\ref{spin_flavor_keven}), corresponds to a sum of structures with $t$-channel
grand spin 0 and 2; the expression for odd $k$, Eq.~(\ref{spin_flavor_kodd}), is a structure with grand spin 1.
The orbital part of the single-particle operator in Eq.~(\ref{eq:mono_int_1}), consisting of the functions
of the momentum and position operators, has to be coupled to the spin-flavor part to acheive total $t$-channel
grand spin 0, in order to satisfy the selection rule Eq.~(\ref{eq:Wigner_Ekart}). The products of powers
of $\hat{p}^3$ and $\hat{x}^3$ in Eq.~(\ref{eq:mono_int_1}) amount to a set of tensor operators with
maximum rank $k + l$. The condition that these tensors be reducible to total spin 2 (for even $k$) or
spin 1 (for odd $k$) fixes the maximum possible value of $l$ for a given $k$ as
\begin{subequations}
\label{l_max_k}
\begin{alignat}{2}
l_{\mathrm{max}}(k) &= k + 2 \hspace{2em} && (\text{$k$ even}),
\\
&= k +1 && (\text{$k$ odd}).
\end{alignat}
\end{subequations}
In the representation of the $m$-th moment in Eq.~(\ref{G0_moment}), the values of $k$ are summed over
the range $0\leq k\leq m-1$. For a given $m$, the maximal value of $l$ attained in the partial-wave 
expansion Eq.~(\ref{eq:mono_int_1}) is given by $l_{\rm max}(k = m - 1)$, which
according to Eq.~(\ref{l_max_k}) is
\begin{subequations}
\label{l_max_m}
\begin{alignat}{2}
l_{\mathrm{max}}(m) &= m + 1 \hspace{2em} && (\text{$m$ odd}),
\\
&= m && (\text{$m$ even}).
\end{alignat}
\end{subequations}
Next, knowing the limits of $l$ in the partial-wave expansion directly in terms of $m$,
we interchange the order of summation over $k$ and $l$ in Eqs.~(\ref{G0_moment}) and
(\ref{eq:mono_int_1}),
\begin{align}
\sum^{m-1}_{k=0} \sum^{l_{\mathrm{max}}(m)}_{l=0,2,4...} 
= \sum^{l_{\mathrm{max}}(m)}_{l=0,2,4...} \sum^{m-1}_{k=0}.
\end{align}
Finally, after these steps, we can present the moment of the mean-field GPD in Eq.~(\ref{G0_moment}) in the form
\begin{align}
&\int dx \, x^{m-1} \,
G_{\textrm{mf}, 0}(x,\xi, t)
\nonumber \\
&= \sum^{l_{\mathrm{max}}(m)}_{l=0,2,4...} (2l+1) P_{l} \left(-\frac{2\xi M_{N}}{\sqrt{-t}} \right)
C_{\textrm{mf}, 0}^{ml}(t), 
\label{eq:mono_poly}
\end{align}
where
\begin{align}
& C_{\textrm{mf}, 0}^{ml} (t)
\nonumber \\
& =-\frac{1}{6} \frac{ M_{N} N_{c}}{M^{m}_{N}} \sum_{n,\mathrm{occ}} \sum^{m-1}_{k=0} 
\left(\begin{array}{c} m-1 \\ k \end{array} \right) 
\frac{E^{m-1-k}_{n}}{2^{k}} \sum^{k}_{j=0} \left(\begin{array}{c} k \\ j \end{array} \right)
\nonumber \\
&\times \langle n | (\gamma^{0}\gamma^{3})^{k+1} i (\bm{\gamma} \times \bm{\tau})^{3} 
\nonumber \\[1ex]
& \hspace{1em} \times i^{l} (\hat{p}^{3})^{j} j_{l} (|\hat{\bm{x}}| \sqrt{-t})
P_{l}\left( \frac{\hat{x}^3}{|\hat{\bm{x}}|} \right) (\hat{p}^{3})^{k-j} | n \rangle.
\label{eq:mono_poly_2}
\end{align}
These functions are the mean-field generalized FFs in the partial-wave representation of the GPD moments,
where the $\xi$ dependence is contained in the Legendre polynomials of angular momentum $l$.

One observes that the $m$-th moment of the chiral-odd mean-field GPD $G_{0, {\rm mf}}$ is an even polynomial in
$\xi$ with degree $m + 1$ (for odd $m$) or $m$ (for even $m$). These findings agree with the
general polynomiality properties Eq.~\eqref{eq:polynomiality} if the mean-field GPD is identified with
the conventional chiral-odd GPDs according to Eq.~(\ref{G0_conventional}). In particular, the
maximum power of $\xi$ in $G_{0, {\rm mf}}$ is consistent with the presence of the
$\xi^{2} E^{u-d}_{T}$ term in Eq.~(\ref{G0_conventional}), which raises the degree of the polynomial
by two compared to Eq.~(\ref{polynomiality_E_T}).

The polynomiality properties of the dipole and quadrupole mean-field GPDs are demonstrated in a similar
manner in Appendix~\ref{app:a}. Altogether our analysis shows that the polynomiality properties of the
chiral-odd GPDs are correctly reproduced in the mean-field picture of the nucleon at large $N_c$.

\section{Sum rules at large $N_c$ \label{sec:6}}
\subsection{Spin-flavor structure of tensor FFs}
The chiral-odd GPDs are connected with the nucleon tensor FFs through the sum rules
Eq.~(\ref{eq:first_Mel}). They follow from the connection of the chiral-odd partonic operator 
with the the local tensor operator and the constraints imposed by relativistic covariance.
It is interesting to study how the sum rules are obtained in the large-$N_c$ limit, 
with the restricted realization of rotational invariance in the mean-field picture.
This will also provide explicit expressions for the mean-field tensor FFs and
their representation in quark single-particle operators.

The $1/N_c$ expansion of the matrix element of the tensor operator is performed in the 
Breit frame Eq.~(\ref{eq:rot_sym}) using canonical spin states. In analogy to the study of the
partonic operator in Sec.~\ref{subsec:gpds_largenc}, we start from 
the matrix element of the local tensor operator between large-$N_{c}$ baryon states 
with definite momenta but as yet indefinite spin-isospin, referred to as the 
mean-field matrix element, given by Eq.~\eqref{quantization_momentum} with $z = 0$.
The mean-field matrix elements of the $0i$ and $ij$ components of the local
tensor operator are parametrized as
\begin{subequations}
  \label{eq:correlator_2}
\begin{align}
& \langle \bm{p}', \textrm{mf}\, | \bar{\psi}_{f'}
\left(0\right) i\sigma^{0 i} \psi_{f} \left(0\right)
| \bm{p}, \textrm{mf} \, \rangle 
\nonumber \\
&=2 M_{N} \, \bm{1}_{f'\! f} \, Y^{i}_{1} \, \frac{\sqrt{-t}}{2M_{N}} F_{\textrm{mf}, 1},
\\[1ex]
& \langle \bm{p}', \textrm{mf}\, | \bar{\psi}_{f'}
\left(0\right) i\sigma^{ij} \psi_{f} \left(0\right)
| \bm{p}, \textrm{mf} \, \rangle 
\nonumber \\
&=2 M_{N}\left[ - 3 i \epsilon^{ijm} \, (\tau^{m})_{f'\! f} \, Y_0 \, F_{\textrm{mf}, 0}
\phantom{\frac{0}{0}} \nonumber \right.\\
&\left. + 3 i \epsilon^{ijl} \, (\tau^{m})_{f'\! f} \, Y^{lm}_{2} \, 
\frac{t}{4M^{2}_{N}} F_{\textrm{mf}, 2} \right],
\end{align}
\end{subequations}
where 
\begin{align}
F_{\textrm{mf}, 0}, \, F_{\textrm{mf}, 1}, \, F_{\textrm{mf}, 2} \; = \; \textrm{functions}(t)
\end{align}
are the mean-field FFs, and the subscript denotes the order of the multipole structure 
in the 3D momentum transfer $\bm{\Delta}$.
Compared to the mean-field GPDs Eq.~(\ref{eq:soliton_M}), there is one less multipole structure. 
This is because time-reversal invariance for the mean-field GPDs imposes only even/oddness in $\xi$, 
while for the mean-field FFs it eliminates one of the multipole structures.

In the next step we project the mean-field matrix elements on definite baryon spin-isospin states,
by performing the integral over flavor rotations as in Eqs.~(\ref{eq:collect_integral}) 
and (\ref{collective_spin_isospin}). 
We obtain
\begin{subequations}
\begin{align}
& \mathcal{M}_{\mathrm{FFs}}[i \sigma^{0i}]
=2 M_{N}    Y^{i}_{1} \bm{1}_{f'\! f} \langle 1 \rangle_{B'B} \frac{\sqrt{-t}}{2M_{N}} F_{\textrm{mf}, 1}, 
\\[1ex]
& \mathcal{M}_{\mathrm{FFs}}[i \sigma^{ij}]
=2 M_{N} (\tau^{k})_{f'\! f} \langle O^{km} \rangle_{B'B} 
\cr
& \times \left[ -i \epsilon^{ijm}  3 F_{\textrm{mf}, 0} 
+i \epsilon^{ijl} Y^{lm}_{2} \frac{3t}{4M^{2}_{N}} F_{\textrm{mf}, 2} \right].
\end{align}
\end{subequations}
We evaluate the matrix elements for the spin-flavor quantum number of the
proton, $S = T = S' = T = 1/2$ and $T_{3}=T'_{3}=1/2$. In this case
only the flavor-diagonal matrices $\bm{1}$ and $\tau^3$ contribute,
and the rotational matrix elements are given in Eq.~\eqref{eq:proton_wf}. 
We obtain
\begin{subequations}
\label{eq:meanfield_FFs}
\begin{align}
& \mathcal{M}^{u+d}_{\mathrm{FFs}}[i \sigma^{0i}]=2 M_{N} \, \bm{1} \, Y^{i}_{1} \,
\frac{\sqrt{-t}}{2M_{N}} F_{\textrm{mf}, 1}, 
\\[1ex]
& \mathcal{M}^{u-d}_{\mathrm{FFs}}[i \sigma^{ij}]=2 M_{N} \bigg{[} i \epsilon^{ijm}\, \sigma^{m} \, Y_{0} \, 
F_{\textrm{mf}, 0}  
\cr
 &-i \epsilon^{ijl} \sigma^{m} \, Y^{lm}_{2} \, \frac{t}{4M^{2}_{N}} F_{\textrm{mf}, 2} \bigg{]}.
\end{align}
\end{subequations}
The $0i$ components produce a flavor-singlet structure, the $ij$ components a flavor-nonsinglet structure.

In the next step we connect the mean-field FFs to the conventional proton tensor FFs, by comparing 
the mean-field matrix elements Eq.~\eqref{eq:meanfield_FFs} with the multipole expansion of the 
proton matrix element Eq.~\eqref{eq:general_tensor_FF_largeNc}. In the large-$N_{c}$ limit the 
latter becomes
\begin{subequations}
\label{eq:general_tensor_FF_largeNc_2}
\begin{align}
&\mathcal{M}^{u\pm d}_{\mathrm{FFs}}[i\sigma^{0j}] 
\nonumber \\
&= 2M_N \, \bm{1} \, Y^{j}_{1} \, \frac{\sqrt{-t}}{2M_N} \left[ H^{u\pm d}_{T} 
+ 2 \tilde{H}^{u\pm d}_{T} + E^{u\pm d}_{T}  \right], 
\\[1ex]
&\mathcal{M}^{u\pm d}_{\mathrm{FFs}}[i\sigma^{ij}] 
\nonumber \\
&= 2M_N \left\{ i \epsilon^{ijk} \sigma^{k} \, Y_0 
\left[  H^{u\pm d}_{T} + \frac{t}{6M_{N}^2}  E^{u\pm d}_{T} \right] \right.
\nonumber \\
&\left. -i \epsilon^{ijk} \sigma^{m} \, Y^{km}_{2} \frac{t}{4M^{2}_{N}}
\left[  \frac{1}{2}  H^{u\pm d}_{T} +  E^{u\pm d}_{T} \right] \right\} .
\end{align}
\end{subequations}
Equating Eq.~\eqref{eq:general_tensor_FF_largeNc_2} and Eq.~\eqref{eq:meanfield_FFs}, we obtain
the relations between the mean-field FFs and the tensor FFs in the flavor non-singlet sector,
\begin{subequations}
\label{eq:relation_1}
\begin{align}
H^{u-d}_{T} + \frac{t}{6M^{2}_{N}}  E^{u-d}_{T}  &=F_{\textrm{mf}, 0},
\\
H^{u-d}_{T} + 2 \tilde{H}^{u-d}_{T} + E^{u-d}_{T} &=N_{1},
\\
\frac{1}{2} H^{u-d}_{T} + E^{u-d}_{T} &= F_{\textrm{mf}, 2},
\end{align}
\end{subequations}
and in the flavor-singlet sector
\begin{subequations}
\label{eq:relation_2}
\begin{align}
H^{u+d}_{T} + \frac{t}{6M^{2}_{N}}  E^{u+d}_{T} &= N_{0},
\\
H^{u+d}_{T} + 2 \tilde{H}^{u+d}_{T} + E^{u+d}_{T} &=F_{\textrm{mf}, 1},
\\
\frac{1}{2} H^{u+d}_{T} + E^{u+d}_{T} &= N_{2}.
\end{align}
\end{subequations}
In analogy to the corresponding relations for the GPDs, the functions $N_1$ in Eq.~\eqref{eq:relation_1} 
and $N_0$ and $N_2$ in Eq.~\eqref{eq:relation_2} represent isovector and isoscalar multipoles that 
are zero in leading order of the $1/N_c$ expansion.

From the relations Eqs.~(\ref{eq:relation_1}) and (\ref{eq:relation_2}) we now determine the $N_{c}$ scaling 
of the proton tensor FFs. Based on similar arguments as for the GPDs in Eq.~(\ref{ncscaling_natural}),
we posit that the natural $N_{c}$ scaling of the mean-field FFs is
\begin{align}
&\left\{ F_{\textrm{mf}, 0}, \, \frac{F_{\textrm{mf}, 1}}{M_N},
\, \frac{F_{\textrm{mf}, 2}}{M_N^2} \right\}(t) \; \sim \; N_c^1 \times \mathrm{function}(t),
\label{largenc_dimensionful_ffs}
\end{align}
and that the scaling of the dimensionless mean-field FFs is therefore
\begin{align}
&\left\{ F_{\textrm{mf}, 0}, \, F_{\textrm{mf}, 1},
\, F_{\textrm{mf}, 2} \right\}(t) 
\cr
&\sim \{N_c^1,\, N^{2}_{c},\, N^{3}_{c}  \} \times \mathrm{function}(t).
\label{largenc_dimensionless_ffs_1}
\end{align}
These FFs parametrize the leading spin-flavor components of the matrix element in the large-$N_c$ limit.
The corresponding ``other'' flavor components of the matrix element are suppressed by one power of $1/N_c$,
so that 
\begin{align}
&\left\{ N_{0}, \, N_{1}, \, N_{2} \right\}(t) 
\cr
&\sim \{N_c^0, \, N^{1}_{c}, \, N^{2}_{c}  \} \times \mathrm{function}(t).
\label{largenc_dimensionless_ffs_2}
\end{align}
The scaling behavior of the conventional FFs is established by solving Eqs.~\eqref{eq:relation_1} 
and~\eqref{eq:relation_2} with the scaling of Eqs.~\eqref{largenc_dimensionless_ffs_1} 
and \eqref{largenc_dimensionless_ffs_2} as input. We obtain
\begin{subequations}
\label{eq:Ncscaling_FFs}
\begin{align}
&\{H^{u-d}_{T}, \, \tilde{H}^{u-d}_{T}, \, E^{u-d}_{T} \}(t)
\cr
&\sim \{N^{1}_{c}, \, N^{3}_{c}, \, N^{3}_{c} \} \times \mathrm{function}(t) , 
\\[1ex]
&\{H^{u+d}_{T}, \, \tilde{H}^{u+d}_{T}, \, E^{u+d}_{T} \}(t) 
\cr
&\sim \{N^{0}_{c}, \, N^{2}_{c}, \, N^{2}_{c} \} \times \mathrm{function}(t),
\end{align}
\end{subequations}
and the non-trivial large-$N_{c}$ relation
\begin{align}
2\tilde{H}^{u-d}_{T}(t) = - E^{u-d}_{T}(t).
\label{eq:largeNc_relation_1_tensor}
\end{align}
It is natural that the tensor FFs in Eq.~(\ref{eq:Ncscaling_FFs}) scale with one power less in $N_{c}$ 
compared to the chiral-odd GPDs in Eqs.~(\ref{eq:Ncscaling}) and (\ref{eq:Ncscaling_singlet}).
The FFs are the first moments of the GPDs, and the integral over $x = \mathcal{O}(N_c^{-1})$ 
reduces the power of $N_{c}$ of the GPDs by one [see Eq.~(\ref{x_parametric})]. This 
statement can be generalized to the $m$-th moments of the GPDs,
\begin{subequations}
\begin{align}
&\int dx \, x^{m-1} \{H^{u-d}_{T}, \, \tilde{H}^{u-d}_{T}, \, E^{u-d}_{T}, \, \tilde{E}^{u-d}_{T} \}
\nonumber \\
&= \{N^{2-m}_{c}, \, N^{4-m}_{c}, \, N^{4-m}_{c}, \, N^{3-m}_{c} \} \times \mathrm{function}(t),
\nonumber \\
\\
&\int dx \, x^{m-1} \{H^{u+d}_{T}, \, \tilde{H}^{u+d}_{T}, \, E^{u+d}_{T}, \, \tilde{E}^{u+d}_{T} \}
\nonumber \\
&= \{N^{1-m}_{c}, \, N^{3-m}_{c}, \, N^{3-m}_{c}, \, N^{2-m}_{c} \} \times \mathrm{function}(t).
\nonumber \\
\end{align}
\end{subequations}
Finally, applying the $N_{c}$ scaling Eq.~(\ref{eq:Ncscaling_FFs}) of the tensor FFs to 
Eqs.~\eqref{eq:relation_1} and \eqref{eq:relation_2}, we obtain the connection between the
the tensor FFs and the mean-field FFs as
\begin{subequations}
\label{eq:relation}
\begin{align}
F_{\textrm{mf}, 0} &= H^{u-d}_{T} + \frac{t}{6M^{2}_{N}}  E^{u-d}_{T},
\\
F_{\textrm{mf}, 1} &= E^{u+d}_{T} + 2 \tilde{H}^{u+d}_{T},
\\[1ex]
F_{\textrm{mf}, 2} &= E^{u-d}_{T} .
\end{align}
\end{subequations}
\subsection{Tensor FFs in mean-field picture}
To verify the sum rules of the GPDs in the large-$N_c$ limit, we need to know how the tensor FFs are
expressed as matrix elements of quark single-particle operators in the mean field, in the same way as 
the GPDs in Sec~\ref{sec:5}. Using the same assumption as for the partonic operator in Sec~\ref{sec:5}, 
we posit that the leading isoscalar and isovector components of the matrix element of the local tensor
operator are given by
\begin{subequations}
\label{eq:general_tensor}
\begin{align}
\mathcal{M}^{u+d}_{\mathrm{FFs}}[i \sigma^{0k}] &= 2 M_{N} N_{c} \bm{1}
\cr
&\times   \sum_{n,\mathrm{occ}} \langle n| \gamma^{0} i \sigma^{0k} e^{i \bm{\Delta} \cdot \hat{\bm{x}}} 
| n \rangle, 
\\[1ex]
\mathcal{M}^{u-d}_{\mathrm{FFs}}[i \sigma^{ij}] &= -\frac{2}{3} M_{N} N_{c} \sigma^{k}
\cr
&\times \sum_{n,\mathrm{occ}} \langle n| \tau^{k} \gamma^{0} i \sigma^{ij} 
e^{i \bm{\Delta} \cdot \hat{\bm{x}}} | n \rangle. 
\end{align}
\end{subequations}
These expressions can also be obtained by integrating the matrix element of the partonic operator, 
Eq.~\eqref{eq:general}, over $x$. Here again the momentum transfer to the mean field is represented by 
the operator $e^{i \bm{\Delta} \cdot \hat{\bm{x}}}$ in the single-particle matrix elements.
The expressions of the mean-field FFs are extracted by performing a 3D multipole expansion in 
$\bm{\Delta}$ and separating the spin components of the matrix element Eq.~\eqref{eq:general_tensor}. 
We obtain
\begin{subequations}
\label{eq:Tensor_FFs}
\begin{align}
F_{\textrm{mf}, 0}(t) &= - \frac{N_{c}}{9} \sum_{n, \mathrm{occ}} 
\nonumber \\
&\times  \langle n | \gamma^{0} (\bm{\Sigma} \cdot \bm{\tau})j_{0}(|\hat{\bm{x}}| |\bm{\Delta}|) | n \rangle, 
\\[2ex]
F_{\textrm{mf}, 1}(t) &=  2M_{N} N_{c} \sum_{n, \mathrm{occ}} 
\nonumber \\
&\times  \langle n | \gamma^{0}\gamma^{5}  \Sigma^{i} Y^{i}_{1}(\Omega_{\hat{\bm{x}}})  
\frac{ i j_{1}(|\hat{\bm{x}}||\bm{\Delta}|}{ |\bm{\Delta}|} | n \rangle, 
\\[2ex]
F_{\textrm{mf}, 2}(t) &=  2M_{N}^{2}N_{c} \sum_{n, \mathrm{occ}} 
\nonumber \\
&\times  \langle n |  \gamma^{0} \Sigma^{i} \tau^{j} Y_{2}^{ij}(\Omega_{\hat{\bm{x}}})  
\frac{j_{2}(|\hat{\bm{x}}| |\bm{\Delta}|)}{|\bm{\Delta}|^{2}} | n \rangle ,
\end{align}
\end{subequations}
where $t = -|\bm{\Delta}|^2$, see Eq.~(\ref{xi_t_Delta}). In the forward limit the tensor FFs 
define the tensor charge $F_{\textrm{mf}, 0}(0)=g^{u-d}_{T}$, the anomalous tensor magnetic moment 
$F_{\textrm{mf}, 1}(0) = \kappa^{u+d}_{T}$, and the tensor quadrupole moment $F_{\textrm{mf}, 2}(0)=E_{T}(0)$.
Their first-quantized expressions are obtained by taking the limit $|\bm{\Delta}| \rightarrow 0$ in 
Eq.~(\ref{eq:Tensor_FFs}),
\begin{subequations}
\label{eq:Tensor_moment}
\begin{align}
F_{\textrm{mf}, 0}(0) & = - \frac{N_{c}}{9} \sum_{n, \mathrm{occ}}  
\langle n | \gamma^{0} (\bm{\Sigma} \cdot \bm{\tau}) | n \rangle,
\\
F_{\textrm{mf}, 1}(0) &  =  \frac{2M_{N} N_{c}}{3} \sum_{n, \mathrm{occ}} 
\langle n | \gamma^{0}\gamma^{5} i  \bm{\Sigma} \cdot \hat{\bm{x}}  | n \rangle,
\\
F_{\textrm{mf}, 2}(0) &=  \frac{2M^{2}_{N}N_{c}}{15} \sum_{n, \mathrm{occ}} 
\nonumber \\
&\hspace{-1.2cm}\times \langle n |  \gamma^{0} 
\left[(\bm{\Sigma} \cdot \hat{\bm{x}})  (\bm{\tau} \cdot \hat{\bm{x}}) 
- \frac{1}{3} (\bm{\Sigma} \cdot \bm{\tau} )\hat{\bm{x}}^{2}  \right] | n \rangle.
\end{align}
\end{subequations}
The numerical values of these quantities have been estimated in the chiral quark-soliton model in 
Refs.~\cite{Kim:1995bq,Kim:1996vk,Ledwig:2010zq}. 
\subsection{Sum rules for chiral-odd GPDs}
We are now in a position to prove the sum rules for the chiral-odd GPDs in the large-$N_c$ limit. 
To do so, we take the first moment of the mean-field GPDs in the first-quantized representation
and connect them with the tensor FFs in Eq.~(\ref{eq:Tensor_FFs}), 
making use of the symmetries of the mean field.

{\it{Monopole GPD:}} The first moment of the monopole GPD is given by
Eq.~(\ref{eq:mono_poly}) with $m = 1$,
\begin{align}
&\int dx \, G_{\textrm{mf}, 0}(x,\xi,t) = -\frac{N_{c}}{6}  \sum_{n,\mathrm{occ}} 
\cr
&\times \bigg{[}  \langle n | (\gamma^{0}\gamma^{3}) i (\bm{\gamma} 
\times \bm{\tau})^{3}  j_{0} (|\hat{\bm{x}}| \sqrt{-t})  | n \rangle 
\cr
&-\frac{15}{2}  P_{2} \left(-\frac{2\xi M_{N}}{\sqrt{-t}} \right) 
\langle n | (\gamma^{0}\gamma^{3}) i (\bm{\gamma} \times \bm{\tau})^{3}  
\cr
&\hspace{2em} \times Y^{33}_{2}(\Omega_{\hat{\bm{x}}}) j_{2} (|\hat{\bm{x}}| \sqrt{-t})  
| n \rangle  \bigg{]} ,
\label{eq:monopole_sumrule_0}
\end{align}
where the spin-flavor operator can also be expressed as [see Eq.~(\ref{spin_flavor_keven})]
\begin{align}
(\gamma^{0}\gamma^{3}) i (\bm{\gamma} \times \bm{\tau})^{3}
= \gamma^{0} (\bm{\Sigma} \cdot \bm{\tau} - \Sigma^3 \tau^3).
\end{align}
The first quantized operator in Eq.~\eqref{eq:monopole_sumrule_0} has a specific orientation 
with respect to the 3-axis. Since the mean field has spherical symmetry in combined position and isospin space, 
the averages of the oriented components can be converted to a manifestly spherically symmetric way. 
Specifically, 
\begin{subequations}
\begin{align}
\tau^{3} \Sigma^{3} &= \frac{1}{3} \bm{\tau} \cdot \bm{\Sigma} + \ldots ,  \label{eq:rotation_1} \\
Y^{33}_{2} \tau^{3} \Sigma^{3} &= \frac{2}{15} Y^{ij}_{2} \tau^{i} \Sigma^{j} + \ldots ,   \label{eq:rotation_2}
\end{align}
\end{subequations}
where the ellipsis denotes structures with $t$-channel grand spin $> 0$, which average to zero because of
the selection rule Eq.~(\ref{eq:Wigner_Ekart}). We obtain
\begin{align}
&\int dx \, G_{\textrm{mf}, 0}(x,\xi,t)
\cr
&= - \frac{N_{c}}{9} \sum_{n,\mathrm{occ}} 
\langle n | \gamma_{0} (\bm{\Sigma} \cdot \bm{\tau})  j_{0} (|\hat{\bm{x}}| \sqrt{-t})  | n \rangle
\cr
&+\left(\frac{1}{12}  + \frac{\xi^{2} M^{2}_{N} }{t} \right) N_{c} \sum_{n,\mathrm{occ}}
\cr
&\times \langle n | \gamma^{0} {\Sigma}^{i} \tau^{j} 
Y^{ij}_{2}(\Omega_{\hat{\bm{x}}})  j_{2} (|\hat{\bm{x}}| \sqrt{-t})  | n \rangle.
\label{eq:monopole_sumrule}
\end{align}
This spherically symmetric expression of the mean-field GPD moment can be compared with the first-quantized
expressions of the mean-field tensor FFs, Eq.~(\ref{eq:Tensor_FFs}),
\begin{align}
&\int dx \, G_{\textrm{mf}, 0}(x,\xi,t)
\nonumber \\
&=  F_{\textrm{mf}, 0}(t) - \left(\frac{t}{24M^{2}_{N}} + \frac{\xi^2}{2}\right) F_{\textrm{mf}, 2}(t)
\nonumber \\
&=  H^{u-d}_{T}(t) + \left(\frac{t}{8M^{2}_{N}}  -\frac{\xi^{2}}{2}\right) E^{u-d}_{T}(t) .
\label{sumrule_meanfield}
\end{align}
In the last line we have substituted the expression of the mean-field FFs in terms of the conventional
FFs of Eq.~(\ref{eq:relation}). One observes that this result agrees with the general expression of the 
first moment of $G_{\textrm{mf}, 0}$ that one obtains by expressing the mean-field GPD in terms of
the conventional GPDs. It shows the overall consistency of the approach: the first-quantized mean-field 
expressions reproduce the general sum rules obeyed by the $1/N_c$ expanded GPDs and FFs.
Interestingly, the mean-field result Eq.~(\ref{sumrule_meanfield}) indeed delivers a $\xi^2$ term in the 
first moment of the monopole GPD $G_{\textrm{mf}, 0}$, as required by the presence of the $\xi^2 E_T^{u - d}$ 
term in Eq.~(\ref{G0_conventional}). It also correctly reproduces the vanishing first moment of 
$\xi \tilde{E}^{u-d}_{T}$ vanishes.

The sum rules for the dipole and quadrupole mean-field GPDs are proved in a similar manner
in Appendix~\ref{app:sumrules}. Altogether our analysis shows that the sum rules for the
chiral-odd GPDs are correctly realized in the mean-field picture of the nucleon at large $N_c$.

\section{Conclusions and extensions \label{sec:7}}
We have performed a comprehensive study of the nonperturbative properties of the nucleon's 
chiral-odd GPDs in the large $N_c$ limit of QCD. This includes the parametric ordering of the
spin-flavor components, the polynomiality property of the moments, 
and the sum rules connecting the GPDs with the tensor FFs. 
The main findings can be summarized as follows.

\begin{itemize}
\item[(i)] {\it Multipole structure:} 
The chiral-odd GPDs contain monopole, dipole, and quadrupole structures 
in the transverse momentum transfer $\bm{\Delta}_\perp$. The presence of the quadrupole structure follows from 
longitudinal angular momentum conservation in the light-front representation and is confirmed by comparison
with the multipole expansion of the matrix elements of the local tensor operators. The quadrupole structure
is unique to the chiral-odd GPDs and absent in the chiral-even GPDs.

\item[(ii)] {\it $1/N_c$ expansion:} 
The nucleon matrix element of the chiral-odd partonic operator in the large-$N_c$ limit is characterized by 
four independent mean-field GPDs. This corrects the earlier analysis of Ref.~\cite{Schweitzer:2016jmd},
which assumed three independent GPDs because it did not include the quadrupole structure.
As a result, all four chiral-odd nucleon GPDs can be derived from large-$N_c$ mean-field GPDs
without degeneracies, and their $N_c$ scaling is determined completely.

\item[(iii)] {\it Large-$N_c$ relation:} 
The leading flavor-non-singlet GPDs $E_T^{u-d}$ and $\tilde H_T^{u-d}$ are connected by the 
non-trivial large-$N_c$ relation Eq.~(\ref{eq:largeNc_relation_1}). The relation is well satisfied 
by numerical results from recent lattice QCD calculations (see Fig.~\ref{fig:lattice}).

\item[(iv)] {\it Polynomiality:} The polynomiality of the moments of the chiral-odd GPDs is fulfilled
with the restricted realization of translational and rotational invariance in the mean-field
picture of the nucleon at large $N_c$. Our analysis exhibits precisely what elements of the 
abstract mean-field picture are responsible for ensuring polynomiality 
(discrete symmetries, triangle rule for angular momentum addition).

\item[(v)] {\it Sum rules:} The sum rules connecting the chiral-odd GPDs with the FFs
of the local tensor operator are satisfied in the large-$N_c$ limit. The presence of the quadrupole 
structure in the mean-field matrix elements is essential for ensuring the correspondence between
the matrix elements of the non-local partonic and local tensor operators.

\end{itemize}

Altogether, our theoretical study shows that the essential qualitative features of the chiral-odd GPDs 
are correctly reproduced in the general mean-field picture of the nucleon in the large-$N_c$ limit of QCD. 
This provides a basis for quantitative estimates of the chiral-odd GPDs with specific dynamical
realizations of the mean-field picture, such as the chiral quark-soliton model.
It also enables a model-independent phenomenological analysis of chiral-odd GPDs and exclusive processes 
incorporating large-$N_c$ constraints, such as the hierarchy of spin-flavor components of the GPDs. 

The studies presented here could be extended in several directions.
The spin-flavor symmetry of baryons in the large-$N_c$ limit of QCD naturally connects the 
$N \rightarrow N$ with the $N \to \Delta$ (and even $\Delta \to \Delta$) transition matrix elements
of the same QCD operator \cite{Diehl:2024bmd}. 
This connection could be used to predict the chiral-odd $N \to \Delta$
transition GPDs in terms of the chiral-odd $N \rightarrow N$ GPDs, in analogy to what was done
with the chiral-even GPDs \cite{Goeke:2001tz}. The chiral-odd $N \to \Delta$ GPDs 
are sampled in exclusive pion production processes with $N \to \Delta$ transitions
\cite{Kroll:2022roq}, assuming the chiral-odd twist-3 mechanism
that was proposed and tested for $N \rightarrow N$ exclusive pion production
\cite{Ahmad:2008hp, Goloskokov:2009ia, Goloskokov:2011rd, Goldstein:2012az}; first measurements 
have been performed at JLab \cite{CLAS:2023akb,Diehl:2024bmd}. 
Measurements of the $N \rightarrow \Delta$ transition GPDs can serve 
as additional tests of the spin-flavor structure predicted by the $1/N_c$ expansion and access
structures that are difficult to separate in $N \rightarrow N$ measurements alone. In the chiral-odd
sector such studies are particularly promising because the operators are pure non-singlets, and there is
no singlet contribution in the $N \rightarrow N$ channel that would affect the comparison between 
$N \rightarrow N$ and $N \rightarrow \Delta$.

The multipole expansion of the partonic matrix elements developed here could be applied also to 
higher-twist structures, e.g. the chiral-odd twist-3 structures connected with spin-orbit correlations
\cite{Bhoonah:2017olu}.

The predictive power of the $1/N_c$ expansion could be greatly increased by computing subleading 
corrections to the nucleon matrix elements. Subleading spin-flavor structures appear due to the 
finite angular velocity of the collective rotations of the mean field; in the abstract formulation 
of Sec.~\ref{sec:4}
they could be captured by including terms proportional to the angular velocity in the parametrization
of the mean-field matrix elements. In addition, there are $1/N_c$ corrections to the spin-flavor structures
that are non-zero in leading order; see Ref.~\cite{Christov:1995vm} for a review. 
Both types of corrections involve new dynamical input beyond the
mean-field expectation values. This input can be provided by dynamical models such as the chiral quark-soliton
model. If lattice QCD calculations were available at different values of $N_c$, the dynamical input for 
subleading $1/N_c$ corrections could be assembled model-independently, substantially expanding the reach of the 
$1/N_c$ expansion.

\appendix

\section{Polynomiality of multipole GPDs \label{app:a}}
\subsection{$\xi$-even dipole GPD}
In this appendix we demonstrate the polynomiality properties of the higher-multipole chiral-odd GPDs in
the large-$N_c$ limit. As in the analysis of the monopole GPD in Sec.~\ref{subsec:polynomiality}, 
we take the first-quantized expressions of 
the mean-field GPDs in Eq.~\eqref{eq:explicit}, compute the moments as in Eq.~(\ref{G0_moment}), 
and convert the results to a form that explicitly 
shows the polynomiality in $\xi$. For simplicity we take $t=0$ in the analysis of the 
higher multipole GPDs; the extension to finite $t$ is straightforward.

The $m$-th moment of the dipole mean-field GPD $G_{\textrm{mf}, 1}$ in Eq.~\eqref{eq:explicit} 
is obtained as
\begin{align}
&\int dx \, x^{m-1}  G_{\textrm{mf}, 1}(x,\xi,t)
\cr
&=- \frac{2 M^{2}_{N} N_{c}}{M^{m}_{N} |\bm{\Delta_{\perp}}|^{2}}  \sum_{n,\mathrm{occ}} 
\sum^{m-1}_{k=0} \left(\begin{array}{c} m-1 \\ k \end{array} \right) 
\frac{E^{m-1-k}_{n}}{2^{k}} \sum^{k}_{j=0} \left(\begin{array}{c} k \\ j \end{array} \right)
\cr
&\times \langle n | (1+\gamma^{0}\gamma^{3})  (\bm{\gamma}_{\perp} \cdot \bm{\Delta_{\perp}}) 
(\hat{p}^{3})^{j} e^{i \bm{\Delta} \cdot \bm{ \hat{x}}} (\hat{p}^{3})^{k-j} | n \rangle.
\end{align}
Under the $G_{5}$ transformation Eq.~(\ref{eq:Gparity}) the single-particle matrix element transforms as
\begin{align}
&\langle n | (1+\gamma^{0}\gamma^{3})  (\bm{\gamma}_{\perp} \cdot \bm{\Delta_{\perp}}) 
(\hat{p}^{3})^{j} e^{i \bm{\Delta} \cdot \bm{ \hat{x}}} (\hat{p}^{3})^{k-j} | n \rangle
\cr
&=  \langle n |  (\gamma^{0}\gamma^{3})^{k}  (\bm{\gamma}_{\perp} \cdot \bm{\Delta_{\perp}}) 
(\hat{p}^{3})^{j} e^{i \bm{\Delta} \cdot \bm{ \hat{x}}} (\hat{p}^{3})^{k-j} | n \rangle. 
\label{eq:dipole_int_0}
\end{align}
Here we use a notation analogous to Eq.~(\ref{gamma0_gamma_3_notation}),
\begin{align}
(\gamma^{0}\gamma^{3})^{k} =
\left\{
\begin{array}{ll}
1 & \textrm{$k$ even},
\\
\gamma^{0}\gamma^{3} & \textrm{$k$ odd};
\end{array}
\right.
\end{align}
note that the even/odd pattern is opposite to the one in the monopole GPD Eq.~\eqref{eq:mono_int_0}. 
Performing the partial-wave expansion of the function $e^{i \bm{\Delta} \cdot \hat{\bm{x}}}$
in Eq.~\eqref{eq:dipole_int_0} in the limit $t\to 0$ [see Eq.~\eqref{eq:t0limit}],
and replacing the factor $1/|\bm{\Delta}_{\perp}|^{2}$ by [see Eq.~(\ref{xi_t_Delta})]
\begin{align}
\frac{1}{|\bm{\Delta}_{\perp}|^{2}} &= \frac{1}{-t - (2\xi M_{N})^{2}} 
\stackrel{t \to 0}{=} - \frac{1}{ (2\xi M_{N})^{2}},
\label{eq:dipole_forward_kin}
\end{align}
we obtain
\begin{align} 
&\lim_{\substack{t\to 0, \ \xi \neq 0 } }\sum^{\infty}_{l=2} \langle n | 
(\gamma^{0}\gamma^{3})^{k}  (\hat{p}^{3})^{j}  i(\bm{\nabla}_{\perp} \cdot \bm{\gamma}_{\perp})
\cr
&\times  \left[  \frac{(-2i \xi M_{N} |\hat{\bm{x}}|)^{l}}{(2\xi M_{N})^{2} l!}  
P_{l}\left(\frac{\hat{x}^{3}}{|\hat{\bm{x}}|}\right)\right] (\hat{p}^{3})^{k-j} | n \rangle .
\label{eq:dipole_int_1}
\end{align}
Here the momentum transfer $\bm{\Delta}_{\perp}$ in Eq.~(\ref{eq:dipole_int_0})
has been replaced by a derivative acting on the function $e^{i \bm{\Delta} \cdot \hat{\bm{x}}}$,
\begin{align}
&\lim_{\substack{t\to 0, \ \xi \neq 0 } } \Delta_{\perp}^{i} e^{i \bm{\Delta} \cdot \hat{\bm{x}}} 
= \lim_{\substack{t\to 0, \ \xi \neq 0 } } [\hat{p}^{i}_{\perp}, e^{i \bm{\Delta} \cdot \hat{\bm{x}}}]
\nonumber \\
&=\lim_{\substack{t\to 0, \ \xi \neq 0 } } - i \nabla^{i}_{\perp} e^{i \bm{\Delta} \cdot \hat{\bm{x}}} 
\nonumber \\
&= -i \sum^{\infty}_{l=2} \nabla^{i}_{\perp} \left[\frac{(-2i\xi M_{N} |\hat{\bm{x}} |)^{l}}{l!} 
P_{l}\left(\frac{\hat{x}^{3}}{|\hat{\bm{x}}|}\right)\right].
\label{eq:derivative_1}
\end{align}
Note that the summation over $l$ in Eq.~(\ref{eq:dipole_int_1}) starts from $l=2$ because
\begin{align}
\left[\hat{p}^{i}_{\perp}, |\hat{\bm{x}}| P_{1}\left(\frac{\hat{x}^{3}}{|\hat{\bm{x}}|}\right) \right] = 0.
\label{eq:derivative_1_null}
\end{align}
Because of parity invariance Eq.\eqref{eq:parity}, the allowed partial waves in 
Eq.~\eqref{eq:dipole_int_1} are even, $l=2,4,...$, and the sum becomes
\begin{align}
\sum^{\infty}_{l=2}  [...] \to \sum^{\infty}_{l=2,4...} [...].
\end{align}
The spin part of the single-particle operator in Eq.~\eqref{eq:dipole_int_0} can be rewritten as
\begin{align}
&(\gamma^{0} \gamma^{3})^{k} \bm{\gamma}_{\perp}\cdot \bm{\nabla}_{\perp}
\cr
& = \bm{\gamma}_{\perp}\cdot \bm{\nabla}_{\perp} 
= -\gamma^{0}\gamma^{5} (\bm{\Sigma}_{\perp} \cdot \bm{\nabla}_{\perp})   
\hspace{0.4cm} (k \ \mathrm{even}), 
\cr
& =\gamma^{0} \gamma^{3} \bm{\gamma}_{\perp}\cdot \bm{\nabla}_{\perp} 
=   \gamma^{0} i (\bm{\Sigma} \times \bm{\nabla})^{3}    \quad (k \ \mathrm{odd}).
\label{eq:xi_dipole_fqo}
\end{align}
Using the grand spin selection rule and Eq.~\eqref{eq:xi_dipole_fqo}, following the same logic as in the
derivation of Eq.~\eqref{l_max_k}, the maximum value of $l$ in the sum of Eq.~\eqref{eq:dipole_int_1} 
is determined as
\begin{subequations}
\label{l_max_k_2}
\begin{alignat}{2}
l_{\mathrm{max}}(k) &= k + 2 \hspace{2em} && (\text{$k$ even}),
\\
&= k +1 && (\text{$k$ odd}).
\end{alignat}
\end{subequations}
In terms of $m$, the maximum value is given by $l_{\mathrm{max}}(k=m-1)$. It is convenient to change 
the summation variable in Eq.~\eqref{eq:dipole_int_1} from $l$ to $l - 2$, so that it represents the 
actual powers of $\xi$ in the polynomial. The maximum value of the new variable is then given by
\begin{subequations}
\label{l_max_m_2}
\begin{alignat}{2}
l_{\mathrm{max}}(m) &= m - 1 \hspace{2em} && (\text{$m$ odd}),
\\
&= m -2 && (\text{$m$ even}).
\end{alignat}
\end{subequations}
Altogether we obtain the moment of the dipole mean-field GPD $G_{\textrm{mf}, 1}$ as
\begin{align}
&\int dx \, x^{m-1}  G_{\textrm{mf}, 1}(x,\xi,0)
\cr
&=   \sum^{l_{\mathrm{max}}(m)}_{l=0,2...} \xi^{l} C_{\textrm{mf}, 1}^{ml}(t=0),
\label{eq:di_even_poly}
\end{align}
where the generalized mean-field FFs are given by
\begin{align}
&C_{\textrm{mf}, 1}^{ml}(t=0)  
\nonumber \\
&=- \frac{ N_{c}}{2M^{m}_{N}} \sum_{n,\mathrm{occ}}  \sum^{m-1}_{k=0} 
\left(\begin{array}{c} m-1 \\ k \end{array} \right) \frac{E^{m-1-k}_{n}}{2^{k}} \sum^{k}_{j=0} 
\left(\begin{array}{c} k \\ j \end{array} \right) 
\nonumber \\
&\times \langle n | (\gamma^{0}\gamma^{3})^{k}  (\hat{p}^{3})^{j} 
i(\bm{\nabla}_{\perp} \cdot \bm{\gamma}_{\perp}) 
\nonumber \\[1ex]
&\times \left[  \frac{(-2i  M_{N} |\hat{\bm{x}}|)^{l+2}}{ (l+2)!}  
P_{l+2}\left(\frac{\hat{x}^{3}}{|\hat{\bm{x}}|}\right)\right] (\hat{p}^{3})^{k-j} | n \rangle.
\end{align}
Equation~\eqref{eq:di_even_poly} shows that the moments of the dipole mean-field GPD $G_{\textrm{mf}, 1}$
are even polynomials in $\xi$, with a degree given by Eq.~(\ref{l_max_m_2}). This agrees with the polynomiality 
properties required by the identification of the mean-field GPD with the conventional GPDs, 
Eqs.~\eqref{eq:polynomiality} and (\ref{eq:large_Nc_relations}).
\subsection{$\xi$-odd dipole GPD}
The $m$-th moment of the dipole mean-field GPD $\tilde G_{\textrm{mf}, 1}$ in Eq.~\eqref{eq:explicit}
is obtained as
\begin{align}
&\int dx \, x^{m-1} \tilde{G}_{\textrm{mf}, 1}(x,\xi, t)
\cr
&=\frac{2}{3} i \frac{ M^{2}_{N} N_{c}}{M^{m}_{N} |\bm{\Delta_{\perp}}|^{2}}  
\sum_{n,\mathrm{occ}} \sum^{m-1}_{k=0} \left(\begin{array}{c} m-1 \\ k \end{array} \right) 
\frac{E^{m-1-k}_{n}}{2^{k}} \sum^{k}_{j=0} \left(\begin{array}{c} k \\ j \end{array} \right)
\cr
&\hspace{-0.2cm}\times \langle n | (1+\gamma^{0}\gamma^{3})  \tau^{3} (\hat{p}^{3})^{j}   
( \bm{\Delta} \times \bm{\gamma})^{3} e^{i \bm{\Delta} \cdot \bm{ \hat{x}}} (\hat{p}^{3})^{k-j} | n \rangle. 
\label{eq:dipole_odd_int_m1}
\end{align}
Under the $G_{5}$ transformation Eq.~(\ref{eq:Gparity}) the single-particle matrix element transforms as
\begin{align}
& \langle n | (1+\gamma^{0}\gamma^{3})  \tau^{3} (\hat{p}^{3})^{j}  (\bm{\Delta} \times \bm{\gamma})^{3} 
e^{i \bm{\Delta} \cdot \bm{ \hat{x}}} (\hat{p}^{3})^{k-j} | n \rangle
\cr
&=  \langle n | (\gamma^{0}\gamma^{3})^{k+1}  \tau^{3}  (\hat{p}^{3})^{j}   
(\bm{\Delta} \times \bm{\gamma})^{3} e^{i\bm{\Delta} \cdot \hat{\bm{x}}} (\hat{p}^{3})^{k-j} | n \rangle,
\hspace{2em}
\label{eq:dipole_odd_int_0}
\end{align}
where we use the same notation as in Eq.~(\ref{gamma0_gamma_3_notation}). 
Performing the partial-wave expansion of the function $e^{i \bm{\Delta} \cdot \hat{\bm{x}}}$ at $t=0$,
treating the factor $1/|\bm{\Delta}_\perp|^2$ as in Eq.~(\ref{eq:dipole_forward_kin}), 
we obtain
\begin{align} 
&\lim_{\substack{ t\to 0, \ \xi \neq 0 } } -\sum^{\infty}_{l=2} 
\langle n | (\gamma^{0}\gamma^{3})^{k+1}  \tau^{3}  (\hat{p}^{3})^{j} (\bm{\nabla} \times \bm{\gamma})^{3} 
\cr
&\times   \left[ \frac{(-2i \xi M_{N} |\hat{\bm{x}}|)^{l}}{(2\xi M_{N})^{2} l!}  
P_{l}\left(\frac{\hat{x}^{3}}{|\hat{\bm{x}}|}\right) \right] (\hat{p}^{3})^{k-j} | n \rangle. 
\label{eq:dipole_odd_int_1}
\end{align}
Here the momentum transfer $\Delta_{\perp}$ in the matrix element has again been replaced by a derivative
as in Eq.~(\ref{eq:derivative_1}). Parity invariance Eq.~\eqref{eq:parity} now requires that the
partial waves are odd, $l=1,3,5,...$, and the sum becomes
\begin{align}
\sum^{\infty}_{l=2}  [...] \to \sum^{\infty}_{l=3,5...} [...].
\end{align}
The sum now starts with $l=3$ because the $l = 1$ term is zero by Eq.~(\ref{eq:derivative_1_null}).
The spin part of the single-particle operator can be rewritten as
\begin{align}
&  \gamma^{0} \gamma^{3} \tau^{3} i (\bm{\nabla} \times \bm{\gamma})^{3} 
= -\tau^{3} (\bm{\Sigma}_{\perp} \cdot \bm{\nabla}_{\perp}), 
\cr
& \tau^{3} i (\bm{\nabla} \times \bm{\gamma})^{3} 
=  - \gamma^{0}  \gamma^{5} \tau^{3} i (\bm{\nabla} \times \bm{\Sigma})^{3},
\label{eq:xi_odd_dipole_fqo}
\end{align}
for even and odd $k$, respectively. 
Using the grand spin selection rule and Eq.~\eqref{eq:xi_odd_dipole_fqo}, following the same logic as in the
derivation of Eq.~\eqref{l_max_k}, the maximum value of $l$ in the sum of Eq.~\eqref{eq:dipole_odd_int_1}
is now determined as
\begin{subequations}
\label{l_max_k_3}
\begin{alignat}{2}
l_{\mathrm{max}}(k) &= k + 3 \hspace{2em} && (\text{$k$ even}),
\\
&= k +2 && (\text{$k$ odd}).
\end{alignat}
\end{subequations}
In terms of $m$, the maximum value is given by $l_{\mathrm{max}}(k=m-1)$. Changing the summation variable 
from $l$ to $l-3$, the maximum value of the new variable is then
\begin{subequations}
\label{l_max_m_3}
\begin{alignat}{2}
l_{\mathrm{max}}(m) &= m - 1 \hspace{2em} && (\text{$m$ odd}),
\\
&= m -2 && (\text{$m$ even}).
\end{alignat}
\end{subequations}
Altogether we obtain the moment of the dipole mean-field GPD $\tilde{G}_{\textrm{mf}, 1}$ as
\begin{align}
&\int dx \, x^{m-1} \tilde{G}_{\textrm{mf}, 1}(x,\xi,0)
\cr
&=  \sum^{l_{\mathrm{max}}(m)}_{l=0,2,...}  \xi^{l+1} \tilde{C}_{\textrm{mf}, 1}^{ml}(t=0),
\label{eq:di_odd_poly}
\end{align}
where the generalized mean-field FFs are now given by
\begin{align}
&\tilde{C}_{\textrm{mf}, 1}^{ml}(t=0)
\cr
&=-\frac{  N_{c}}{6M^{m}_{N}} \sum_{n,\mathrm{occ}} \sum^{m-1}_{k=0} 
\left(\begin{array}{c} m-1 \\ k \end{array} \right) \frac{E^{m-1-k}_{n}}{2^{k}} 
\sum^{k}_{j=0} \left(\begin{array}{c} k \\ j \end{array} \right) 
\nonumber \\
&\times \langle n | (\gamma^{0}\gamma^{3})^{k+1}  \tau^{3}  (\hat{p}^{3})^{j}  
(\bm{\nabla} \times \bm{\gamma})^{3} 
\nonumber \\[1ex]
&\times \left[ \frac{(-2i  M_{N} |\hat{\bm{x}}|)^{l+3}}{(l+3)!} 
P_{l+3}\left(\frac{\hat{x}^{3}}{|\hat{\bm{x}}|}\right) \right] (\hat{p}^{3})^{k-j} | n \rangle.
\label{eq:di_odd_poly_2}
\end{align}
Equation~(\ref{eq:di_odd_poly}) shows that the moments of the dipole mean-field GPD $\tilde{G}_{\textrm{mf}, 1}$ 
are odd polynomials in $\xi$, with a degree given by Eq.~(\ref{l_max_m_3}). This agrees with the polynomiality 
properties required by the identification of the mean-field GPD with the conventional GPDs, 
Eqs.~\eqref{eq:polynomiality} and (\ref{eq:large_Nc_relations}). In particular, the highest power of
$\xi$ in the mean-field GPD $\tilde{G}_{\textrm{mf}, 1}$ correctly accounts for the fact that its 
expression in terms of the conventional GPDs in Eq.~(\ref{eq:large_Nc_relations}) 
contains an ``extra'' power of $\xi$ in the $\xi E^{u-d}_{T}$ term.
\subsection{Quadrupole GPD}
The $m$-th moment of the quadrupole mean-field GPD $G_{\textrm{mf}, 2}$ in Eq.~\eqref{eq:explicit} is
obtained as
\begin{align}
&\int dx \, x^{m-1}  G_{\textrm{mf}, 2}(x,\xi, t) 
= -\frac{8}{3} \frac{ M^{3}_{N} N_{c}}{M^{m}_{N} |\bm{\Delta_{\perp}}|^{2}}  \sum_{n,\mathrm{occ}}
\cr
& \times \sum^{m-1}_{k=0} \left(\begin{array}{c} m-1 \\ k \end{array} \right) 
\frac{E^{m-1-k}_{n}}{2^{k}} \sum^{k}_{j=0} \left(\begin{array}{c} k \\ j \end{array} \right)
\nonumber \\[1ex]
&\times\langle n | (1+\gamma^{0}\gamma^{3}) i \epsilon^{3ia} \gamma^{i}  \tau^{b} 
(\hat{p}^{3})^{j}  X^{ab}_{2} e^{i \bm{\Delta} \cdot \bm{ \hat{x}}} (\hat{p}^{3})^{k-j} | n \rangle. 
\nonumber \\
\label{moment_G2}
\end{align}
Under the $G_{5}$ transformation the single-particle matrix element transforms as
\begin{align}
& \langle n | (1+\gamma^{0}\gamma^{3})  \gamma^{i}  \tau^{b} (\hat{p}^{3})^{j} 
X^{ab}_{2} e^{i \bm{\Delta} \cdot \bm{ \hat{x}}} (\hat{p}^{3})^{k-j} | n \rangle
\cr
&=  \langle n | (\gamma^{0}\gamma^{3})^{k+1}  \gamma^{i}  \tau^{b} (\hat{p}^{3})^{j}
X^{ab}_{2} e^{i \bm{\Delta} \cdot \bm{ \hat{x}}} (\hat{p}^{3})^{k-j} | n \rangle,
\label{eq:quadrupole_int_0}
\end{align}
where we use the same notation as in Eqs.~(\ref{gamma0_gamma_3_notation}) and (\ref{eq:dipole_odd_int_0}).
Performing the partial-wave expansion of the function $e^{i \bm{\Delta} \cdot \hat{\bm{x}}}$
in Eq.~\eqref{eq:quadrupole_int_0}, including the factor $1/|\bm{\Delta_{\perp}}|^{2}$, 
and taking the limit $t\to 0$, we obtain
\begin{align} 
&\lim_{\substack{ t\to 0, \ \xi \neq 0 } } -  \sum^{\infty}_{l=4} 
\left(\delta^{ac}\delta^{db}- \frac{1}{2} \delta^{cd}\delta^{ab}\right)
\cr
&\times \langle n | (\gamma^{0}\gamma^{3})^{k+1}  \gamma^{i} i \epsilon^{3ia} 
\tau^{b} (\hat{p}^{3})^{j}  \nabla^{c}_{\perp}\nabla^{d}_{\perp}
\cr
&\times \left[ \frac{(-2i\xi M_{N} |\hat{\bm{x}} |)^{l}}{(2\xi M_{N})^{4}l!} 
P_{l}\left(\frac{\hat{x}^{3}}{|\hat{\bm{x}}|}\right) \right]   (\hat{p}^{3})^{k-j} | n \rangle .
\label{eq:quadrupole_int_1}
\end{align}
Here we use the relation Eq.~(\ref{eq:dipole_forward_kin}), and $X^{ab}_{2} e^{i \bm{\Delta} \cdot \hat{\bm{x}}}$ 
in Eq.~\eqref{eq:quadrupole_int_0} has be rewritten in terms of the momentum operator:
\begin{align}
|\bm{\Delta}_{\perp}|^{2} X^{ab}_{2} e^{i \bm{\Delta} \cdot \hat{\bm{x}}}
&= \left(\delta^{ac}\delta^{bd}- \frac{1}{2} \delta^{cd}\delta^{ab}\right)
\nonumber \\
& \times \left[\hat{p}^{c}_{\perp}\hat{p}^{d}_{\perp}, e^{i \bm{\Delta} \cdot \hat{\bm{x}}}\right].
\label{eq:partial_multipole_quad}
\end{align}
This expression is the same as the derivative acting on $e^{i \bm{\Delta} \cdot \hat{\bm{x}}}$, i.e.,
\begin{align}
&\lim_{\substack{t\to 0, \ \xi \neq 0 } } |\bm{\Delta}_{\perp}|^{2} 
X^{ab}_{2} e^{i \bm{\Delta} \cdot \hat{\bm{x}}}
\nonumber \\
&= -  \sum^{\infty}_{l=4} \left(\delta^{ac}\delta^{bd}- \frac{1}{2} \delta^{cd}\delta^{ab}\right)
\nonumber \\
& \times
\nabla^{c}_{\perp}\nabla^{d}_{\perp} \left[ \frac{(-2i\xi M_{N} |\hat{\bm{x}} |)^{l}}{l!} 
P_{l}\left(\frac{\hat{x}^{3}}{|\hat{\bm{x}}|}\right)  \right]. \label{eq:derivative_2}
\end{align}
Note that the lower limit starts from $l=4$ because 
\begin{align}
\left[\left(\delta^{ac}\delta^{db} 
- \frac{1}{2} \delta^{cd}\delta^{ab}\right)\hat{p}^{c}_{\perp}\hat{p}^{d}_{\perp}, 
|\hat{\bm{x}}|^{l} P_{l}\left(\frac{\hat{x}^{3}}{|\hat{\bm{x}}|}\right)  \right] = 0,
\end{align}
for $l=1...3$. Due to the parity transformation given in Eq.\eqref{eq:parity}, the allowed partial waves 
in Eq.\eqref{eq:quadrupole_int_1} are even, i.e., $l=4,6,...$. Thus, Eq.~\eqref{eq:quadrupole_int_1} becomes
\begin{align}
\sum^{\infty}_{l=4}  [...] \to \sum^{\infty}_{l=4,6,...} [...].
\end{align}
In order to apply the grand spin selection rule we convert the 2D irreducible tensors in the single-particle
operator into 3D irreducible tensors. 
The spin part of the operator is rewritten as
\begin{align}
& \gamma^{0} \gamma^{3} \gamma^{i} i \epsilon^{3ia} \tau^{b} 
\left(\nabla^{a}_{\perp}\nabla^{b}_{\perp} -\frac{1}{2} \bm{\nabla}^{2}_{\perp}\delta^{ab}\right)
\cr
& = \left(\nabla^{a}\nabla^{b} -\frac{1}{3}  \bm{\nabla}^{2} \delta^{ab}\right) 
(\Sigma^{a}_{\perp} \tau^{b}_{\perp})
\cr
&+ \frac{1}{2} \left(\nabla^{3}\nabla^{3} -\frac{1}{3} \bm{\nabla}^{2}\right) 
(\bm{\Sigma}_{\perp}\cdot \bm{\tau}_{\perp})  
\label{eq:quad_dipole_fqo_1}
\end{align}
for even $k$, and 
\begin{align}
&\gamma^{i} i \epsilon^{3ia} \tau^{b} \left(\nabla^{a}_{\perp}\nabla^{b}_{\perp} 
-\frac{1}{2} \bm{\nabla}^{2}_{\perp}\delta^{ab}\right)
\cr
&=   i \gamma^{0}\gamma^{5} (\bm{\nabla} \times \bm{\Sigma})^{3} 
(\bm{\nabla}_{\perp} \cdot \bm{\tau}_{\perp})
\cr
&+ \frac{1}{2} i \gamma^{0}\gamma^{5} (\bm{\Sigma} \times \bm{\tau})^{3} \bm{\nabla}_{\perp}^{2}
\label{eq:quad_dipole_fqo_2}
\end{align}
for odd $k$. Using the grand spin selection rule and Eqs.~\eqref{eq:quad_dipole_fqo_1} 
and \eqref{eq:quad_dipole_fqo_2}, similar to the logic presented in the derivation of 
Eq.~\eqref{l_max_k}, the maximum value of $l$ is determined as
\begin{subequations}
\label{l_max_k_4}
\begin{alignat}{2}
l_{\mathrm{max}}(k) &= k + 4 \hspace{2em} && (\text{$k$ even}),
\\
&= k +3 && (\text{$k$ odd}).
\end{alignat}
\end{subequations}
In terms of $m$, the maximum value is given by $l_{\mathrm{max}}(k=m-1)$. Changing the 
summation variable from $l$ to $l-4$, the maximum value is then given by
\begin{subequations}
\label{l_max_m_4}
\begin{alignat}{2}
l_{\mathrm{max}}(m) &= m - 1 \hspace{2em} && (\text{$m$ odd}),
\\
&= m -2 && (\text{$m$ even}).
\end{alignat}
\end{subequations}
Altogether we obtain the moment of the mean-field GPD Eq.~(\ref{moment_G2}) as
\begin{align}
&\int dx \, x^{m-1}  G_{\textrm{mf}, 2}(x,\xi, 0)
\cr
&= \sum^{l_{\mathrm{max}}(m)}_{l=0,2...} \xi^{l} C_{\textrm{mf}, 2}^{ml}(t=0),
\label{eq:quad_poly}
\end{align}
where the generalized mean-field FFs are given by
\begin{align}
&C_{\textrm{mf}, 2}^{ml}(t=0) = \frac{ N_{c}}{6M^{m-1}_{N}} \sum_{n,\mathrm{occ}}
\cr
&\times   \sum^{m-1}_{k=0} \left(\begin{array}{c} m-1 \\ k \end{array} \right) 
\frac{E^{m-1-k}_{n}}{2^{k}} \sum^{k}_{j=0} \left(\begin{array}{c} k \\ j \end{array} \right) i \epsilon^{3ia}
\\
&\times \langle n | (\gamma^{0}\gamma^{3})^{k+1} \gamma^{i} \tau^{b} 
(\hat{p}^{3})^{j} (\nabla^{a}_{\perp}\nabla^{b}_{\perp} - \frac{1}{2} \bm{\nabla}^{2}_{\perp}\delta^{ab})
\nonumber \\[1ex]
&\times \left[ \frac{(-2i  M_{N} |\hat{\bm{x}}|)^{l+4}}{(l+4)!} 
P_{l+4}\left(\frac{\hat{x}^{3}}{|\hat{\bm{x}}|}\right)\right] (\hat{p}^{3})^{k-j} | n \rangle.
\end{align}
Equation~(\ref{eq:quad_poly}) shows that the moments of the quadrupole mean-field GPD 
$\tilde{G}_{\textrm{mf}, 2}$ are even polynomials in $\xi$, with a degree given by Eq.~(\ref{l_max_m_4}).
This agrees with the polynomiality properties required by the identification of the mean-field GPD with 
the conventional GPDs, Eqs.~\eqref{eq:polynomiality} and (\ref{eq:large_Nc_relations}).
\section{Sum rules of multipole GPDs}
\label{app:sumrules}

\subsection{$\xi$-even dipole GPD}
In this appendix we prove the sum rules for the higher-multipole chiral-odd GPDs in the large-$N_c$ limit. 
As in Sec.~\ref{sec:6}, compute the first moment of the mean-field GPDs in the first-quantized representation
of Eq.~\eqref{eq:explicit}, convert it to spherically symmetric form using the mean-field symmetry, and
compare it with the mean-field expression of the tensor FFs. For simplicity we take $t=0$ in the analysis 
of the sum rules of the higher multipole GPDs; the extension to finite $t$ is straightforward.

The first moment of the $\xi$-even dipole GPD Eq.~\eqref{eq:di_even_poly} is obtained as
\begin{align}
& \int dx \,  G_{\textrm{mf}, 1}(x,\xi,0)
\nonumber \\
&= \frac{2}{3}  M_{N} N_{c} \sum_{n,\mathrm{occ}}
\langle n | \gamma^{0} \gamma^{5} i( \bm{\Sigma} \cdot \hat{\bm{x}} )| n \rangle,
\label{eq:dipole_even_sumrule}
\end{align}
where we have used the relation Eq.~\eqref{eq:rotation_1} to convert the expression
to rotationally invariant form. Comparing Eq.~\eqref{eq:dipole_even_sumrule} with the expression 
of the anomalous tensor magnetic moment Eq.~\eqref{eq:Tensor_moment}, we verify the sum rule
\begin{align}
& \int dx \,  G_{\textrm{mf}, 1}(x,\xi,0)  = \kappa^{u+d}_{T}.
\end{align}

\subsection{$\xi$-odd dipole GPD}
The first moment of the $\xi$-odd dipole GPDs Eq.~\eqref{eq:di_odd_poly} is obtained as
\begin{align}
& \int dx \,  \tilde{G}_{\textrm{mf}, 1}(x,\xi,0) 
=  -\frac{2}{15}  M^{2}_{N} N_{c}  \xi \sum_{n, \mathrm{occ}}
\cr
&\times \langle n |  \gamma^{0} \left[(\bm{\Sigma} \cdot \hat{\bm{x}})
(\bm{\tau}\cdot \hat{\bm{x}}) - \frac{1}{3} (\bm{\Sigma} \cdot \bm{\tau} )\hat{\bm{x}}^{2}  \right] 
| n \rangle,
\label{eq:dipole_odd_sumrule}
\end{align}
where we have employed the relation
\begin{align}
Y^{3b}_{2} \tau^{3} \Sigma^{b}_{\perp} &= \frac{1}{5} Y^{ab}_{2} \tau^{a} \Sigma^{b} + ...,
\end{align}
to convert the expression to rotationally invariant form; the ellipsis denotes 
structures with $t$-channel grand spin $> 0$. Comparing Eq.~\eqref{eq:dipole_odd_sumrule}
with the expression of the tensor quadrupole moment Eq.~\eqref{eq:Tensor_moment},
we veryfy the sum rule
\begin{align}
 \int dx \, \tilde{G}_{\textrm{mf}, 1}(x,\xi, t)   = -\xi E^{u-d}_{T}(0).
\end{align}
Similar to the case of the monopole GPD in Sec.~\ref{sec:6}, we obtain the proper $\xi$ dependence
of the moment, corresponding to the nonzero moment of $\xi E^{u-d}_{T}$ and the 
vanishing moment of $\tilde{E}^{u-d}_{T}$ in Eq.~\eqref{eq:large_Nc_relations}.
\subsection{Quadrupole GPD}
The first moment of the quadrupole GPDs Eq.~\eqref{eq:quad_poly} is evaluated 
using the relation Eq.~\eqref{eq:derivative_2}, 
\begin{align}
&\int dx \,  G_{\textrm{mf}, 2}(x,\xi, 0) = \frac{ M^{2}_{N} N_{c}}{9} 
\sum_{n,\mathrm{occ}}  \langle n | \gamma^{0}\gamma^{3} \gamma^{i}    i \epsilon^{3ia}
\cr
&\times  \tau^{b} (\nabla^{a}_{\perp}\nabla^{b}_{\perp} - \frac{1}{2} \nabla^{2}_{\perp}\delta^{ab}) 
\left[ |\hat{\bm{x}}|^{4} P_{4}\left( \frac{\hat{x}^{3}}{|\hat{\bm{x}}|}\right)\right]  | n \rangle.  
\end{align}
Applying the derivatives to the expression in brackets, we obtain
\begin{align}
&\int dx \,  G_{\textrm{mf}, 2}(x,\xi, 0)
\cr
&=\frac{ M^{2}_{N} N_{c}}{3} \sum_{n,\mathrm{occ}}  
\langle n |  \bigg{[} Y_{2}^{ab}(\Omega_{\hat{\bm{x}}})\Sigma^{a}_{\perp}\tau^{b}_{\perp}
\cr
&+ \frac{1}{2} Y^{33}_{2}(\Omega_{\hat{\bm{x}}}) 
(\bm{\Sigma}_{\perp}\cdot \bm{\tau}_{\perp}) \bigg{]} \hat{\bm{x}}^{2} | n \rangle.
\end{align}
Using now the relation
\begin{align}
Y^{ab}_{2} \tau^{a}_{\perp} \Sigma^{b}_{\perp} &= \frac{7}{15} Y^{ij}_{2} \tau^{i} \Sigma^{j} + ...,
\end{align}
to convert the expression to rotationally invariant form, we obtain
\begin{align}
&\int dx \,  G_{\textrm{mf}, 2}(x,\xi, 0)  = \frac{2}{15}  M^{2}_{N} N_{c}   \sum_{n, \mathrm{occ}}
\cr
&\times \langle n |  \gamma^{0} \left[(\bm{\Sigma} \cdot \hat{\bm{x}}) 
(\bm{\tau} \cdot \hat{\bm{x}}) - \frac{1}{3} (\bm{\Sigma} \cdot \bm{\tau} )\hat{\bm{x}}^{2}  \right] 
| n \rangle.
\label{eq:quadrupole_sumrule}
\end{align}
Comparing this expression with that of the tensor quadrupole moment Eq.~\eqref{eq:Tensor_moment},
we verify the sum rule
\begin{align}
\int dx \, G_{\textrm{mf}, 2}(x,\xi, 0) = E^{u-d}_{T}(0).
\end{align}
\section*{Acknowledgments}
We thank Peter Schweitzer and Kemal Tezgin for a critical reading of an earlier version of the
article and helpful comments and suggestions.

This material is based upon work supported by the U.S.~Department of Energy, Office of Science,
Office of Nuclear Physics under contract DE-AC05-06OR23177. 

The research reported here takes place in the context of the Topical Collaboration ``3D quark-gluon
structure of hadrons: mass, spin, tomography'' (Quark-Gluon Tomography Collaboration) supported by
the U.S.~Department of Energy, Office of Science, Office of Nuclear Physics under
contract DE-SC0023646.
\bibliography{chiral_odd_gpds}

\begin{thebibliography}{72}%
\makeatletter
\providecommand \@ifxundefined [1]{%
 \@ifx{#1\undefined}
}%
\providecommand \@ifnum [1]{%
 \ifnum #1\expandafter \@firstoftwo
 \else \expandafter \@secondoftwo
 \fi
}%
\providecommand \@ifx [1]{%
 \ifx #1\expandafter \@firstoftwo
 \else \expandafter \@secondoftwo
 \fi
}%
\providecommand \natexlab [1]{#1}%
\providecommand \enquote  [1]{``#1''}%
\providecommand \bibnamefont  [1]{#1}%
\providecommand \bibfnamefont [1]{#1}%
\providecommand \citenamefont [1]{#1}%
\providecommand \href@noop [0]{\@secondoftwo}%
\providecommand \href [0]{\begingroup \@sanitize@url \@href}%
\providecommand \@href[1]{\@@startlink{#1}\@@href}%
\providecommand \@@href[1]{\endgroup#1\@@endlink}%
\providecommand \@sanitize@url [0]{\catcode `\\12\catcode `\$12\catcode
  `\&12\catcode `\#12\catcode `\^12\catcode `\_12\catcode `\%12\relax}%
\providecommand \@@startlink[1]{}%
\providecommand \@@endlink[0]{}%
\providecommand \url  [0]{\begingroup\@sanitize@url \@url }%
\providecommand \@url [1]{\endgroup\@href {#1}{\urlprefix }}%
\providecommand \urlprefix  [0]{URL }%
\providecommand \Eprint [0]{\href }%
\providecommand \doibase [0]{https://doi.org/}%
\providecommand \selectlanguage [0]{\@gobble}%
\providecommand \bibinfo  [0]{\@secondoftwo}%
\providecommand \bibfield  [0]{\@secondoftwo}%
\providecommand \translation [1]{[#1]}%
\providecommand \BibitemOpen [0]{}%
\providecommand \bibitemStop [0]{}%
\providecommand \bibitemNoStop [0]{.\EOS\space}%
\providecommand \EOS [0]{\spacefactor3000\relax}%
\providecommand \BibitemShut  [1]{\csname bibitem#1\endcsname}%
\let\auto@bib@innerbib\@empty
\bibitem [{\citenamefont {Goeke}\ \emph {et~al.}(2001)\citenamefont {Goeke},
  \citenamefont {Polyakov},\ and\ \citenamefont
  {Vanderhaeghen}}]{Goeke:2001tz}%
  \BibitemOpen
  \bibfield  {author} {\bibinfo {author} {\bibfnamefont {K.}~\bibnamefont
  {Goeke}}, \bibinfo {author} {\bibfnamefont {M.~V.}\ \bibnamefont
  {Polyakov}},\ and\ \bibinfo {author} {\bibfnamefont {M.}~\bibnamefont
  {Vanderhaeghen}},\ }\bibfield  {title} {\bibinfo {title} {{Hard exclusive
  reactions and the structure of hadrons}},\ }\href
  {https://doi.org/10.1016/S0146-6410(01)00158-2} {\bibfield  {journal}
  {\bibinfo  {journal} {Prog. Part. Nucl. Phys.}\ }\textbf {\bibinfo {volume}
  {47}},\ \bibinfo {pages} {401} (\bibinfo {year} {2001})},\ \Eprint
  {https://arxiv.org/abs/hep-ph/0106012} {arXiv:hep-ph/0106012} \BibitemShut
  {NoStop}%
\bibitem [{\citenamefont {Diehl}(2003)}]{Diehl:2003ny}%
  \BibitemOpen
  \bibfield  {author} {\bibinfo {author} {\bibfnamefont {M.}~\bibnamefont
  {Diehl}},\ }\bibfield  {title} {\bibinfo {title} {{Generalized parton
  distributions}},\ }\href {https://doi.org/10.1016/j.physrep.2003.08.002}
  {\bibfield  {journal} {\bibinfo  {journal} {Phys. Rept.}\ }\textbf {\bibinfo
  {volume} {388}},\ \bibinfo {pages} {41} (\bibinfo {year} {2003})},\ \Eprint
  {https://arxiv.org/abs/hep-ph/0307382} {arXiv:hep-ph/0307382} \BibitemShut
  {NoStop}%
\bibitem [{\citenamefont {Belitsky}\ and\ \citenamefont
  {Radyushkin}(2005)}]{Belitsky:2005qn}%
  \BibitemOpen
  \bibfield  {author} {\bibinfo {author} {\bibfnamefont {A.~V.}\ \bibnamefont
  {Belitsky}}\ and\ \bibinfo {author} {\bibfnamefont {A.~V.}\ \bibnamefont
  {Radyushkin}},\ }\bibfield  {title} {\bibinfo {title} {{Unraveling hadron
  structure with generalized parton distributions}},\ }\href
  {https://doi.org/10.1016/j.physrep.2005.06.002} {\bibfield  {journal}
  {\bibinfo  {journal} {Phys. Rept.}\ }\textbf {\bibinfo {volume} {418}},\
  \bibinfo {pages} {1} (\bibinfo {year} {2005})},\ \Eprint
  {https://arxiv.org/abs/hep-ph/0504030} {arXiv:hep-ph/0504030} \BibitemShut
  {NoStop}%
\bibitem [{\citenamefont {Boffi}\ and\ \citenamefont
  {Pasquini}(2007)}]{Boffi:2007yc}%
  \BibitemOpen
  \bibfield  {author} {\bibinfo {author} {\bibfnamefont {S.}~\bibnamefont
  {Boffi}}\ and\ \bibinfo {author} {\bibfnamefont {B.}~\bibnamefont
  {Pasquini}},\ }\bibfield  {title} {\bibinfo {title} {{Generalized parton
  distributions and the structure of the nucleon}},\ }\href
  {https://doi.org/10.1393/ncr/i2007-10025-7} {\bibfield  {journal} {\bibinfo
  {journal} {Riv. Nuovo Cim.}\ }\textbf {\bibinfo {volume} {30}},\ \bibinfo
  {pages} {387} (\bibinfo {year} {2007})},\ \Eprint
  {https://arxiv.org/abs/0711.2625} {arXiv:0711.2625 [hep-ph]} \BibitemShut
  {NoStop}%
\bibitem [{\citenamefont {Burkardt}(2000)}]{Burkardt:2000za}%
  \BibitemOpen
  \bibfield  {author} {\bibinfo {author} {\bibfnamefont {M.}~\bibnamefont
  {Burkardt}},\ }\bibfield  {title} {\bibinfo {title} {{Impact parameter
  dependent parton distributions and off forward parton distributions for
  $\zeta \rightarrow 0$}},\ }\href {https://doi.org/10.1103/PhysRevD.62.071503}
  {\bibfield  {journal} {\bibinfo  {journal} {Phys. Rev. D}\ }\textbf {\bibinfo
  {volume} {62}},\ \bibinfo {pages} {071503} (\bibinfo {year} {2000})},\
  \bibinfo {note} {[Erratum: Phys.Rev.D 66, 119903 (2002)]},\ \Eprint
  {https://arxiv.org/abs/hep-ph/0005108} {arXiv:hep-ph/0005108} \BibitemShut
  {NoStop}%
\bibitem [{\citenamefont {Burkardt}(2003)}]{Burkardt:2002hr}%
  \BibitemOpen
  \bibfield  {author} {\bibinfo {author} {\bibfnamefont {M.}~\bibnamefont
  {Burkardt}},\ }\bibfield  {title} {\bibinfo {title} {{Impact parameter space
  interpretation for generalized parton distributions}},\ }\href
  {https://doi.org/10.1142/S0217751X03012370} {\bibfield  {journal} {\bibinfo
  {journal} {Int. J. Mod. Phys. A}\ }\textbf {\bibinfo {volume} {18}},\
  \bibinfo {pages} {173} (\bibinfo {year} {2003})},\ \Eprint
  {https://arxiv.org/abs/hep-ph/0207047} {arXiv:hep-ph/0207047} \BibitemShut
  {NoStop}%
\bibitem [{\citenamefont {Diehl}(2002)}]{Diehl:2002he}%
  \BibitemOpen
  \bibfield  {author} {\bibinfo {author} {\bibfnamefont {M.}~\bibnamefont
  {Diehl}},\ }\bibfield  {title} {\bibinfo {title} {{Generalized parton
  distributions in impact parameter space}},\ }\href
  {https://doi.org/10.1007/s10052-002-1016-9} {\bibfield  {journal} {\bibinfo
  {journal} {Eur. Phys. J. C}\ }\textbf {\bibinfo {volume} {25}},\ \bibinfo
  {pages} {223} (\bibinfo {year} {2002})},\ \bibinfo {note} {[Erratum:
  Eur.Phys.J.C 31, 277--278 (2003)]},\ \Eprint
  {https://arxiv.org/abs/hep-ph/0205208} {arXiv:hep-ph/0205208} \BibitemShut
  {NoStop}%
\bibitem [{\citenamefont {Leader}\ and\ \citenamefont
  {Lorc\'e}(2014)}]{Leader:2013jra}%
  \BibitemOpen
  \bibfield  {author} {\bibinfo {author} {\bibfnamefont {E.}~\bibnamefont
  {Leader}}\ and\ \bibinfo {author} {\bibfnamefont {C.}~\bibnamefont
  {Lorc\'e}},\ }\bibfield  {title} {\bibinfo {title} {{The angular momentum
  controversy: What\textquoteright{}s it all about and does it matter?}},\
  }\href {https://doi.org/10.1016/j.physrep.2014.02.010} {\bibfield  {journal}
  {\bibinfo  {journal} {Phys. Rept.}\ }\textbf {\bibinfo {volume} {541}},\
  \bibinfo {pages} {163} (\bibinfo {year} {2014})},\ \Eprint
  {https://arxiv.org/abs/1309.4235} {arXiv:1309.4235 [hep-ph]} \BibitemShut
  {NoStop}%
\bibitem [{\citenamefont {Polyakov}\ and\ \citenamefont
  {Schweitzer}(2018)}]{Polyakov:2018zvc}%
  \BibitemOpen
  \bibfield  {author} {\bibinfo {author} {\bibfnamefont {M.~V.}\ \bibnamefont
  {Polyakov}}\ and\ \bibinfo {author} {\bibfnamefont {P.}~\bibnamefont
  {Schweitzer}},\ }\bibfield  {title} {\bibinfo {title} {{Forces inside
  hadrons: pressure, surface tension, mechanical radius, and all that}},\
  }\href {https://doi.org/10.1142/S0217751X18300259} {\bibfield  {journal}
  {\bibinfo  {journal} {Int. J. Mod. Phys. A}\ }\textbf {\bibinfo {volume}
  {33}},\ \bibinfo {pages} {1830025} (\bibinfo {year} {2018})},\ \Eprint
  {https://arxiv.org/abs/1805.06596} {arXiv:1805.06596 [hep-ph]} \BibitemShut
  {NoStop}%
\bibitem [{\citenamefont {Lorc\'e}\ \emph {et~al.}(2019)\citenamefont
  {Lorc\'e}, \citenamefont {Moutarde},\ and\ \citenamefont
  {Trawi\'nski}}]{Lorce:2018egm}%
  \BibitemOpen
  \bibfield  {author} {\bibinfo {author} {\bibfnamefont {C.}~\bibnamefont
  {Lorc\'e}}, \bibinfo {author} {\bibfnamefont {H.}~\bibnamefont {Moutarde}},\
  and\ \bibinfo {author} {\bibfnamefont {A.~P.}\ \bibnamefont {Trawi\'nski}},\
  }\bibfield  {title} {\bibinfo {title} {{Revisiting the mechanical properties
  of the nucleon}},\ }\href {https://doi.org/10.1140/epjc/s10052-019-6572-3}
  {\bibfield  {journal} {\bibinfo  {journal} {Eur. Phys. J. C}\ }\textbf
  {\bibinfo {volume} {79}},\ \bibinfo {pages} {89} (\bibinfo {year} {2019})},\
  \Eprint {https://arxiv.org/abs/1810.09837} {arXiv:1810.09837 [hep-ph]}
  \BibitemShut {NoStop}%
\bibitem [{\citenamefont {Burkert}\ \emph {et~al.}(2023)\citenamefont
  {Burkert}, \citenamefont {Elouadrhiri}, \citenamefont {Girod}, \citenamefont
  {Lorc\'e}, \citenamefont {Schweitzer},\ and\ \citenamefont
  {Shanahan}}]{Burkert:2023wzr}%
  \BibitemOpen
  \bibfield  {author} {\bibinfo {author} {\bibfnamefont {V.~D.}\ \bibnamefont
  {Burkert}}, \bibinfo {author} {\bibfnamefont {L.}~\bibnamefont
  {Elouadrhiri}}, \bibinfo {author} {\bibfnamefont {F.~X.}\ \bibnamefont
  {Girod}}, \bibinfo {author} {\bibfnamefont {C.}~\bibnamefont {Lorc\'e}},
  \bibinfo {author} {\bibfnamefont {P.}~\bibnamefont {Schweitzer}},\ and\
  \bibinfo {author} {\bibfnamefont {P.~E.}\ \bibnamefont {Shanahan}},\
  }\bibfield  {title} {\bibinfo {title} {{Colloquium: Gravitational form
  factors of the proton}},\ }\href
  {https://doi.org/10.1103/RevModPhys.95.041002} {\bibfield  {journal}
  {\bibinfo  {journal} {Rev. Mod. Phys.}\ }\textbf {\bibinfo {volume} {95}},\
  \bibinfo {pages} {041002} (\bibinfo {year} {2023})},\ \Eprint
  {https://arxiv.org/abs/2303.08347} {arXiv:2303.08347 [hep-ph]} \BibitemShut
  {NoStop}%
\bibitem [{\citenamefont {Barone}\ \emph {et~al.}(2002)\citenamefont {Barone},
  \citenamefont {Drago},\ and\ \citenamefont {Ratcliffe}}]{Barone:2001sp}%
  \BibitemOpen
  \bibfield  {author} {\bibinfo {author} {\bibfnamefont {V.}~\bibnamefont
  {Barone}}, \bibinfo {author} {\bibfnamefont {A.}~\bibnamefont {Drago}},\ and\
  \bibinfo {author} {\bibfnamefont {P.~G.}\ \bibnamefont {Ratcliffe}},\
  }\bibfield  {title} {\bibinfo {title} {{Transverse polarisation of quarks in
  hadrons}},\ }\href {https://doi.org/10.1016/S0370-1573(01)00051-5} {\bibfield
   {journal} {\bibinfo  {journal} {Phys. Rept.}\ }\textbf {\bibinfo {volume}
  {359}},\ \bibinfo {pages} {1} (\bibinfo {year} {2002})},\ \Eprint
  {https://arxiv.org/abs/hep-ph/0104283} {arXiv:hep-ph/0104283} \BibitemShut
  {NoStop}%
\bibitem [{\citenamefont {Burkardt}(2005)}]{Burkardt:2005hp}%
  \BibitemOpen
  \bibfield  {author} {\bibinfo {author} {\bibfnamefont {M.}~\bibnamefont
  {Burkardt}},\ }\bibfield  {title} {\bibinfo {title} {{Transverse deformation
  of parton distributions and transversity decomposition of angular
  momentum}},\ }\href {https://doi.org/10.1103/PhysRevD.72.094020} {\bibfield
  {journal} {\bibinfo  {journal} {Phys. Rev. D}\ }\textbf {\bibinfo {volume}
  {72}},\ \bibinfo {pages} {094020} (\bibinfo {year} {2005})},\ \Eprint
  {https://arxiv.org/abs/hep-ph/0505189} {arXiv:hep-ph/0505189} \BibitemShut
  {NoStop}%
\bibitem [{\citenamefont {Diehl}\ and\ \citenamefont
  {Hagler}(2005)}]{Diehl:2005jf}%
  \BibitemOpen
  \bibfield  {author} {\bibinfo {author} {\bibfnamefont {M.}~\bibnamefont
  {Diehl}}\ and\ \bibinfo {author} {\bibfnamefont {P.}~\bibnamefont {Hagler}},\
  }\bibfield  {title} {\bibinfo {title} {{Spin densities in the transverse
  plane and generalized transversity distributions}},\ }\href
  {https://doi.org/10.1140/epjc/s2005-02342-6} {\bibfield  {journal} {\bibinfo
  {journal} {Eur. Phys. J. C}\ }\textbf {\bibinfo {volume} {44}},\ \bibinfo
  {pages} {87} (\bibinfo {year} {2005})},\ \Eprint
  {https://arxiv.org/abs/hep-ph/0504175} {arXiv:hep-ph/0504175} \BibitemShut
  {NoStop}%
\bibitem [{\citenamefont {Ahmad}\ \emph {et~al.}(2009)\citenamefont {Ahmad},
  \citenamefont {Goldstein},\ and\ \citenamefont {Liuti}}]{Ahmad:2008hp}%
  \BibitemOpen
  \bibfield  {author} {\bibinfo {author} {\bibfnamefont {S.}~\bibnamefont
  {Ahmad}}, \bibinfo {author} {\bibfnamefont {G.~R.}\ \bibnamefont
  {Goldstein}},\ and\ \bibinfo {author} {\bibfnamefont {S.}~\bibnamefont
  {Liuti}},\ }\bibfield  {title} {\bibinfo {title} {{Nucleon Tensor Charge from
  Exclusive $\pi^0$ Electroproduction}},\ }\href
  {https://doi.org/10.1103/PhysRevD.79.054014} {\bibfield  {journal} {\bibinfo
  {journal} {Phys. Rev. D}\ }\textbf {\bibinfo {volume} {79}},\ \bibinfo
  {pages} {054014} (\bibinfo {year} {2009})},\ \Eprint
  {https://arxiv.org/abs/0805.3568} {arXiv:0805.3568 [hep-ph]} \BibitemShut
  {NoStop}%
\bibitem [{\citenamefont {Goloskokov}\ and\ \citenamefont
  {Kroll}(2010)}]{Goloskokov:2009ia}%
  \BibitemOpen
  \bibfield  {author} {\bibinfo {author} {\bibfnamefont {S.~V.}\ \bibnamefont
  {Goloskokov}}\ and\ \bibinfo {author} {\bibfnamefont {P.}~\bibnamefont
  {Kroll}},\ }\bibfield  {title} {\bibinfo {title} {{An attempt to understand
  exclusive $\pi^+$ electroproduction}},\ }\href
  {https://doi.org/10.1140/epjc/s10052-009-1178-9} {\bibfield  {journal}
  {\bibinfo  {journal} {Eur. Phys. J. C}\ }\textbf {\bibinfo {volume} {65}},\
  \bibinfo {pages} {137} (\bibinfo {year} {2010})},\ \Eprint
  {https://arxiv.org/abs/0906.0460} {arXiv:0906.0460 [hep-ph]} \BibitemShut
  {NoStop}%
\bibitem [{\citenamefont {Goloskokov}\ and\ \citenamefont
  {Kroll}(2011)}]{Goloskokov:2011rd}%
  \BibitemOpen
  \bibfield  {author} {\bibinfo {author} {\bibfnamefont {S.~V.}\ \bibnamefont
  {Goloskokov}}\ and\ \bibinfo {author} {\bibfnamefont {P.}~\bibnamefont
  {Kroll}},\ }\bibfield  {title} {\bibinfo {title} {{Transversity in hard
  exclusive electroproduction of pseudoscalar mesons}},\ }\href
  {https://doi.org/10.1140/epja/i2011-11112-6} {\bibfield  {journal} {\bibinfo
  {journal} {Eur. Phys. J. A}\ }\textbf {\bibinfo {volume} {47}},\ \bibinfo
  {pages} {112} (\bibinfo {year} {2011})},\ \Eprint
  {https://arxiv.org/abs/1106.4897} {arXiv:1106.4897 [hep-ph]} \BibitemShut
  {NoStop}%
\bibitem [{\citenamefont {Goldstein}\ \emph {et~al.}(2012)\citenamefont
  {Goldstein}, \citenamefont {Gonzalez~Hernandez},\ and\ \citenamefont
  {Liuti}}]{Goldstein:2012az}%
  \BibitemOpen
  \bibfield  {author} {\bibinfo {author} {\bibfnamefont {G.~R.}\ \bibnamefont
  {Goldstein}}, \bibinfo {author} {\bibfnamefont {J.~O.}\ \bibnamefont
  {Gonzalez~Hernandez}},\ and\ \bibinfo {author} {\bibfnamefont
  {S.}~\bibnamefont {Liuti}},\ }\bibfield  {title} {\bibinfo {title} {{Easy as
  $\pi^0$: On the Interpretation of Recent Electroproduction Results}},\ }\href
  {https://doi.org/10.1088/0954-3899/39/11/115001} {\bibfield  {journal}
  {\bibinfo  {journal} {J. Phys. G}\ }\textbf {\bibinfo {volume} {39}},\
  \bibinfo {pages} {115001} (\bibinfo {year} {2012})},\ \Eprint
  {https://arxiv.org/abs/1201.6088} {arXiv:1201.6088 [hep-ph]} \BibitemShut
  {NoStop}%
\bibitem [{\citenamefont {Bedlinskiy}\ \emph {et~al.}(2012)\citenamefont
  {Bedlinskiy} \emph {et~al.}}]{CLAS:2012cna}%
  \BibitemOpen
  \bibfield  {author} {\bibinfo {author} {\bibfnamefont {I.}~\bibnamefont
  {Bedlinskiy}} \emph {et~al.} (\bibinfo {collaboration} {CLAS}),\ }\bibfield
  {title} {\bibinfo {title} {{Measurement of Exclusive $\pi^0$
  Electroproduction Structure Functions and their Relationship to Transversity
  GPDs}},\ }\href {https://doi.org/10.1103/PhysRevLett.109.112001} {\bibfield
  {journal} {\bibinfo  {journal} {Phys. Rev. Lett.}\ }\textbf {\bibinfo
  {volume} {109}},\ \bibinfo {pages} {112001} (\bibinfo {year} {2012})},\
  \Eprint {https://arxiv.org/abs/1206.6355} {arXiv:1206.6355 [hep-ex]}
  \BibitemShut {NoStop}%
\bibitem [{\citenamefont {Bedlinskiy}\ \emph {et~al.}(2014)\citenamefont
  {Bedlinskiy} \emph {et~al.}}]{CLAS:2014jpc}%
  \BibitemOpen
  \bibfield  {author} {\bibinfo {author} {\bibfnamefont {I.}~\bibnamefont
  {Bedlinskiy}} \emph {et~al.} (\bibinfo {collaboration} {CLAS}),\ }\bibfield
  {title} {\bibinfo {title} {{Exclusive $\pi^0$ electroproduction at $W > 2$
  GeV with CLAS}},\ }\href {https://doi.org/10.1103/PhysRevC.90.039901}
  {\bibfield  {journal} {\bibinfo  {journal} {Phys. Rev. C}\ }\textbf {\bibinfo
  {volume} {90}},\ \bibinfo {pages} {025205} (\bibinfo {year} {2014})},\
  \bibinfo {note} {[Addendum: Phys.Rev.C 90, 039901 (2014)]},\ \Eprint
  {https://arxiv.org/abs/1405.0988} {arXiv:1405.0988 [nucl-ex]} \BibitemShut
  {NoStop}%
\bibitem [{\citenamefont {Bedlinskiy}\ \emph {et~al.}(2017)\citenamefont
  {Bedlinskiy} \emph {et~al.}}]{CLAS:2017jjr}%
  \BibitemOpen
  \bibfield  {author} {\bibinfo {author} {\bibfnamefont {I.}~\bibnamefont
  {Bedlinskiy}} \emph {et~al.} (\bibinfo {collaboration} {CLAS}),\ }\bibfield
  {title} {\bibinfo {title} {{Exclusive $\eta$ electroproduction at $W > 2$ GeV
  with CLAS and transversity generalized parton distributions}},\ }\href
  {https://doi.org/10.1103/PhysRevC.95.035202} {\bibfield  {journal} {\bibinfo
  {journal} {Phys. Rev. C}\ }\textbf {\bibinfo {volume} {95}},\ \bibinfo
  {pages} {035202} (\bibinfo {year} {2017})},\ \Eprint
  {https://arxiv.org/abs/1703.06982} {arXiv:1703.06982 [nucl-ex]} \BibitemShut
  {NoStop}%
\bibitem [{\citenamefont {Kim}\ \emph {et~al.}(2024)\citenamefont {Kim} \emph
  {et~al.}}]{CLAS:2023wda}%
  \BibitemOpen
  \bibfield  {author} {\bibinfo {author} {\bibfnamefont {A.}~\bibnamefont
  {Kim}} \emph {et~al.} (\bibinfo {collaboration} {CLAS}),\ }\bibfield  {title}
  {\bibinfo {title} {{Beam spin asymmetry measurements of deeply virtual
  $\pi^0$ production with CLAS12}},\ }\href
  {https://doi.org/10.1016/j.physletb.2024.138459} {\bibfield  {journal}
  {\bibinfo  {journal} {Phys. Lett. B}\ }\textbf {\bibinfo {volume} {849}},\
  \bibinfo {pages} {138459} (\bibinfo {year} {2024})},\ \Eprint
  {https://arxiv.org/abs/2307.07874} {arXiv:2307.07874 [nucl-ex]} \BibitemShut
  {NoStop}%
\bibitem [{\citenamefont {Alexeev}\ \emph {et~al.}(2020)\citenamefont {Alexeev}
  \emph {et~al.}}]{COMPASS:2019fea}%
  \BibitemOpen
  \bibfield  {author} {\bibinfo {author} {\bibfnamefont {M.~G.}\ \bibnamefont
  {Alexeev}} \emph {et~al.} (\bibinfo {collaboration} {COMPASS}),\ }\bibfield
  {title} {\bibinfo {title} {{Measurement of the cross section for hard
  exclusive $\pi^0$ muoproduction on the proton}},\ }\href
  {https://doi.org/10.1016/j.physletb.2020.135454} {\bibfield  {journal}
  {\bibinfo  {journal} {Phys. Lett. B}\ }\textbf {\bibinfo {volume} {805}},\
  \bibinfo {pages} {135454} (\bibinfo {year} {2020})},\ \Eprint
  {https://arxiv.org/abs/1903.12030} {arXiv:1903.12030 [hep-ex]} \BibitemShut
  {NoStop}%
\bibitem [{\citenamefont {Abdul~Khalek}\ \emph {et~al.}(2022)\citenamefont
  {Abdul~Khalek} \emph {et~al.}}]{AbdulKhalek:2021gbh}%
  \BibitemOpen
  \bibfield  {author} {\bibinfo {author} {\bibfnamefont {R.}~\bibnamefont
  {Abdul~Khalek}} \emph {et~al.},\ }\bibfield  {title} {\bibinfo {title}
  {{Science Requirements and Detector Concepts for the Electron-Ion Collider}:
  {EIC Yellow Report}},\ }\href
  {https://doi.org/10.1016/j.nuclphysa.2022.122447} {\bibfield  {journal}
  {\bibinfo  {journal} {Nucl. Phys. A}\ }\textbf {\bibinfo {volume} {1026}},\
  \bibinfo {pages} {122447} (\bibinfo {year} {2022})},\ \Eprint
  {https://arxiv.org/abs/2103.05419} {arXiv:2103.05419 [physics.ins-det]}
  \BibitemShut {NoStop}%
\bibitem [{\citenamefont {Aoki}\ \emph {et~al.}(2021)\citenamefont {Aoki} \emph
  {et~al.}}]{Aoki:2021cqa}%
  \BibitemOpen
  \bibfield  {author} {\bibinfo {author} {\bibfnamefont {K.}~\bibnamefont
  {Aoki}} \emph {et~al.},\ }\href@noop {} {\bibinfo {title} {{Extension of the
  J-PARC Hadron Experimental Facility: Third White Paper}}} (\bibinfo {year}
  {2021}),\ \Eprint {https://arxiv.org/abs/2110.04462} {arXiv:2110.04462
  [nucl-ex]} \BibitemShut {NoStop}%
\bibitem [{\citenamefont {Enberg}\ \emph {et~al.}(2006)\citenamefont {Enberg},
  \citenamefont {Pire},\ and\ \citenamefont {Szymanowski}}]{Enberg:2006he}%
  \BibitemOpen
  \bibfield  {author} {\bibinfo {author} {\bibfnamefont {R.}~\bibnamefont
  {Enberg}}, \bibinfo {author} {\bibfnamefont {B.}~\bibnamefont {Pire}},\ and\
  \bibinfo {author} {\bibfnamefont {L.}~\bibnamefont {Szymanowski}},\
  }\bibfield  {title} {\bibinfo {title} {{Transversity GPD in photo- and
  electroproduction of two vector mesons}},\ }\href
  {https://doi.org/10.1140/epjc/s2006-02545-3} {\bibfield  {journal} {\bibinfo
  {journal} {Eur. Phys. J. C}\ }\textbf {\bibinfo {volume} {47}},\ \bibinfo
  {pages} {87} (\bibinfo {year} {2006})},\ \Eprint
  {https://arxiv.org/abs/hep-ph/0601138} {arXiv:hep-ph/0601138} \BibitemShut
  {NoStop}%
\bibitem [{\citenamefont {El~Beiyad}\ \emph {et~al.}(2010)\citenamefont
  {El~Beiyad}, \citenamefont {Pire}, \citenamefont {Segond}, \citenamefont
  {Szymanowski},\ and\ \citenamefont {Wallon}}]{ElBeiyad:2010pji}%
  \BibitemOpen
  \bibfield  {author} {\bibinfo {author} {\bibfnamefont {M.}~\bibnamefont
  {El~Beiyad}}, \bibinfo {author} {\bibfnamefont {B.}~\bibnamefont {Pire}},
  \bibinfo {author} {\bibfnamefont {M.}~\bibnamefont {Segond}}, \bibinfo
  {author} {\bibfnamefont {L.}~\bibnamefont {Szymanowski}},\ and\ \bibinfo
  {author} {\bibfnamefont {S.}~\bibnamefont {Wallon}},\ }\bibfield  {title}
  {\bibinfo {title} {{Photoproduction of a $\pi \rho_T$ pair with a large
  invariant mass and transversity generalized parton distribution}},\ }\href
  {https://doi.org/10.1016/j.physletb.2010.02.086} {\bibfield  {journal}
  {\bibinfo  {journal} {Phys. Lett. B}\ }\textbf {\bibinfo {volume} {688}},\
  \bibinfo {pages} {154} (\bibinfo {year} {2010})},\ \Eprint
  {https://arxiv.org/abs/1001.4491} {arXiv:1001.4491 [hep-ph]} \BibitemShut
  {NoStop}%
\bibitem [{\citenamefont {'t~Hooft}(1974)}]{tHooft:1973alw}%
  \BibitemOpen
  \bibfield  {author} {\bibinfo {author} {\bibfnamefont {G.}~\bibnamefont
  {'t~Hooft}},\ }\bibfield  {title} {\bibinfo {title} {{A Planar Diagram Theory
  for Strong Interactions}},\ }\href
  {https://doi.org/10.1016/0550-3213(74)90154-0} {\bibfield  {journal}
  {\bibinfo  {journal} {Nucl. Phys. B}\ }\textbf {\bibinfo {volume} {72}},\
  \bibinfo {pages} {461} (\bibinfo {year} {1974})}\BibitemShut {NoStop}%
\bibitem [{\citenamefont {Witten}(1979)}]{Witten:1979kh}%
  \BibitemOpen
  \bibfield  {author} {\bibinfo {author} {\bibfnamefont {E.}~\bibnamefont
  {Witten}},\ }\bibfield  {title} {\bibinfo {title} {{Baryons in the $1/N$
  Expansion}},\ }\href {https://doi.org/10.1016/0550-3213(79)90232-3}
  {\bibfield  {journal} {\bibinfo  {journal} {Nucl. Phys. B}\ }\textbf
  {\bibinfo {volume} {160}},\ \bibinfo {pages} {57} (\bibinfo {year}
  {1979})}\BibitemShut {NoStop}%
\bibitem [{\citenamefont {Coleman}\ and\ \citenamefont
  {Witten}(1980)}]{Coleman:1980mx}%
  \BibitemOpen
  \bibfield  {author} {\bibinfo {author} {\bibfnamefont {S.~R.}\ \bibnamefont
  {Coleman}}\ and\ \bibinfo {author} {\bibfnamefont {E.}~\bibnamefont
  {Witten}},\ }\bibfield  {title} {\bibinfo {title} {{Chiral-Symmetry Breakdown
  in Large-$N$ Chromodynamics}},\ }\href
  {https://doi.org/10.1103/PhysRevLett.45.100} {\bibfield  {journal} {\bibinfo
  {journal} {Phys. Rev. Lett.}\ }\textbf {\bibinfo {volume} {45}},\ \bibinfo
  {pages} {100} (\bibinfo {year} {1980})}\BibitemShut {NoStop}%
\bibitem [{\citenamefont {Gervais}\ and\ \citenamefont
  {Sakita}(1984)}]{Gervais:1983wq}%
  \BibitemOpen
  \bibfield  {author} {\bibinfo {author} {\bibfnamefont {J.-L.}\ \bibnamefont
  {Gervais}}\ and\ \bibinfo {author} {\bibfnamefont {B.}~\bibnamefont
  {Sakita}},\ }\bibfield  {title} {\bibinfo {title} {{Large-$N$ QCD Baryon
  Dynamics: Exact Results from Its Relation to the Static Strong Coupling
  Theory}},\ }\href {https://doi.org/10.1103/PhysRevLett.52.87} {\bibfield
  {journal} {\bibinfo  {journal} {Phys. Rev. Lett.}\ }\textbf {\bibinfo
  {volume} {52}},\ \bibinfo {pages} {87} (\bibinfo {year} {1984})}\BibitemShut
  {NoStop}%
\bibitem [{\citenamefont {Dashen}\ \emph {et~al.}(1994)\citenamefont {Dashen},
  \citenamefont {Jenkins},\ and\ \citenamefont {Manohar}}]{Dashen:1993jt}%
  \BibitemOpen
  \bibfield  {author} {\bibinfo {author} {\bibfnamefont {R.~F.}\ \bibnamefont
  {Dashen}}, \bibinfo {author} {\bibfnamefont {E.~E.}\ \bibnamefont
  {Jenkins}},\ and\ \bibinfo {author} {\bibfnamefont {A.~V.}\ \bibnamefont
  {Manohar}},\ }\bibfield  {title} {\bibinfo {title} {{$1/N_c$ expansion for
  baryons}},\ }\href {https://doi.org/10.1103/PhysRevD.51.2489} {\bibfield
  {journal} {\bibinfo  {journal} {Phys. Rev. D}\ }\textbf {\bibinfo {volume}
  {49}},\ \bibinfo {pages} {4713} (\bibinfo {year} {1994})},\ \bibinfo {note}
  {[Erratum: Phys.Rev.D 51, 2489 (1995)]},\ \Eprint
  {https://arxiv.org/abs/hep-ph/9310379} {arXiv:hep-ph/9310379} \BibitemShut
  {NoStop}%
\bibitem [{\citenamefont {Diakonov}\ \emph {et~al.}(1988)\citenamefont
  {Diakonov}, \citenamefont {Petrov},\ and\ \citenamefont
  {Pobylitsa}}]{Diakonov:1987ty}%
  \BibitemOpen
  \bibfield  {author} {\bibinfo {author} {\bibfnamefont {D.}~\bibnamefont
  {Diakonov}}, \bibinfo {author} {\bibfnamefont {V.~Y.}\ \bibnamefont
  {Petrov}},\ and\ \bibinfo {author} {\bibfnamefont {P.~V.}\ \bibnamefont
  {Pobylitsa}},\ }\bibfield  {title} {\bibinfo {title} {{A Chiral Theory of
  Nucleons}},\ }\href {https://doi.org/10.1016/0550-3213(88)90443-9} {\bibfield
   {journal} {\bibinfo  {journal} {Nucl. Phys. B}\ }\textbf {\bibinfo {volume}
  {306}},\ \bibinfo {pages} {809} (\bibinfo {year} {1988})}\BibitemShut
  {NoStop}%
\bibitem [{\citenamefont {Wakamatsu}\ and\ \citenamefont
  {Yoshiki}(1991)}]{Wakamatsu:1990ud}%
  \BibitemOpen
  \bibfield  {author} {\bibinfo {author} {\bibfnamefont {M.}~\bibnamefont
  {Wakamatsu}}\ and\ \bibinfo {author} {\bibfnamefont {H.}~\bibnamefont
  {Yoshiki}},\ }\bibfield  {title} {\bibinfo {title} {{A chiral quark model of
  the nucleon}},\ }\href {https://doi.org/10.1016/0375-9474(91)90263-6}
  {\bibfield  {journal} {\bibinfo  {journal} {Nucl. Phys. A}\ }\textbf
  {\bibinfo {volume} {524}},\ \bibinfo {pages} {561} (\bibinfo {year}
  {1991})}\BibitemShut {NoStop}%
\bibitem [{\citenamefont {Christov}\ \emph {et~al.}(1996)\citenamefont
  {Christov}, \citenamefont {Blotz}, \citenamefont {Kim}, \citenamefont
  {Pobylitsa}, \citenamefont {Watabe}, \citenamefont {Meissner}, \citenamefont
  {Ruiz~Arriola},\ and\ \citenamefont {Goeke}}]{Christov:1995vm}%
  \BibitemOpen
  \bibfield  {author} {\bibinfo {author} {\bibfnamefont {C.~V.}\ \bibnamefont
  {Christov}}, \bibinfo {author} {\bibfnamefont {A.}~\bibnamefont {Blotz}},
  \bibinfo {author} {\bibfnamefont {H.-C.}\ \bibnamefont {Kim}}, \bibinfo
  {author} {\bibfnamefont {P.}~\bibnamefont {Pobylitsa}}, \bibinfo {author}
  {\bibfnamefont {T.}~\bibnamefont {Watabe}}, \bibinfo {author} {\bibfnamefont
  {T.}~\bibnamefont {Meissner}}, \bibinfo {author} {\bibfnamefont
  {E.}~\bibnamefont {Ruiz~Arriola}},\ and\ \bibinfo {author} {\bibfnamefont
  {K.}~\bibnamefont {Goeke}},\ }\bibfield  {title} {\bibinfo {title} {{Baryons
  as nontopological chiral solitons}},\ }\href
  {https://doi.org/10.1016/0146-6410(96)00057-9} {\bibfield  {journal}
  {\bibinfo  {journal} {Prog. Part. Nucl. Phys.}\ }\textbf {\bibinfo {volume}
  {37}},\ \bibinfo {pages} {91} (\bibinfo {year} {1996})},\ \Eprint
  {https://arxiv.org/abs/hep-ph/9604441} {arXiv:hep-ph/9604441} \BibitemShut
  {NoStop}%
\bibitem [{\citenamefont {Schweitzer}\ and\ \citenamefont
  {Weiss}(2016)}]{Schweitzer:2016jmd}%
  \BibitemOpen
  \bibfield  {author} {\bibinfo {author} {\bibfnamefont {P.}~\bibnamefont
  {Schweitzer}}\ and\ \bibinfo {author} {\bibfnamefont {C.}~\bibnamefont
  {Weiss}},\ }\bibfield  {title} {\bibinfo {title} {{Spin-flavor structure of
  chiral-odd generalized parton distributions in the large- N$_c$ limit}},\
  }\href {https://doi.org/10.1103/PhysRevC.94.045202} {\bibfield  {journal}
  {\bibinfo  {journal} {Phys. Rev. C}\ }\textbf {\bibinfo {volume} {94}},\
  \bibinfo {pages} {045202} (\bibinfo {year} {2016})},\ \Eprint
  {https://arxiv.org/abs/1606.08388} {arXiv:1606.08388 [hep-ph]} \BibitemShut
  {NoStop}%
\bibitem [{\citenamefont {Mattis}(1989)}]{Mattis:1988hf}%
  \BibitemOpen
  \bibfield  {author} {\bibinfo {author} {\bibfnamefont {M.~P.}\ \bibnamefont
  {Mattis}},\ }\bibfield  {title} {\bibinfo {title} {{The $I(t) = J(t)$ Rule in
  Action}},\ }\href {https://doi.org/10.1103/PhysRevD.39.994} {\bibfield
  {journal} {\bibinfo  {journal} {Phys. Rev. D}\ }\textbf {\bibinfo {volume}
  {39}},\ \bibinfo {pages} {994} (\bibinfo {year} {1989})}\BibitemShut
  {NoStop}%
\bibitem [{\citenamefont {Mattis}\ and\ \citenamefont
  {Mukherjee}(1988)}]{Mattis:1988hg}%
  \BibitemOpen
  \bibfield  {author} {\bibinfo {author} {\bibfnamefont {M.~P.}\ \bibnamefont
  {Mattis}}\ and\ \bibinfo {author} {\bibfnamefont {M.}~\bibnamefont
  {Mukherjee}},\ }\bibfield  {title} {\bibinfo {title} {{The $I(t) = J(t)$
  Rule: A New Large $N_c$ Selection Rule for Meson - Baryon Scattering}},\
  }\href {https://doi.org/10.1103/PhysRevLett.61.1344} {\bibfield  {journal}
  {\bibinfo  {journal} {Phys. Rev. Lett.}\ }\textbf {\bibinfo {volume} {61}},\
  \bibinfo {pages} {1344} (\bibinfo {year} {1988})}\BibitemShut {NoStop}%
\bibitem [{\citenamefont {Lebed}(2006)}]{Lebed:2006us}%
  \BibitemOpen
  \bibfield  {author} {\bibinfo {author} {\bibfnamefont {R.~F.}\ \bibnamefont
  {Lebed}},\ }\bibfield  {title} {\bibinfo {title} {{The Large $N_c$
  baryon-meson $I(t) = J(t)$ rule holds for three flavors}},\ }\href
  {https://doi.org/10.1016/j.physletb.2006.06.014} {\bibfield  {journal}
  {\bibinfo  {journal} {Phys. Lett. B}\ }\textbf {\bibinfo {volume} {639}},\
  \bibinfo {pages} {68} (\bibinfo {year} {2006})},\ \Eprint
  {https://arxiv.org/abs/hep-ph/0603150} {arXiv:hep-ph/0603150} \BibitemShut
  {NoStop}%
\bibitem [{\citenamefont {Sheikh}\ \emph {et~al.}(2021)\citenamefont {Sheikh},
  \citenamefont {Dobaczewski}, \citenamefont {Ring}, \citenamefont {Robledo},\
  and\ \citenamefont {Yannouleas}}]{Sheikh:2019qdz}%
  \BibitemOpen
  \bibfield  {author} {\bibinfo {author} {\bibfnamefont {J.~A.}\ \bibnamefont
  {Sheikh}}, \bibinfo {author} {\bibfnamefont {J.}~\bibnamefont {Dobaczewski}},
  \bibinfo {author} {\bibfnamefont {P.}~\bibnamefont {Ring}}, \bibinfo {author}
  {\bibfnamefont {L.~M.}\ \bibnamefont {Robledo}},\ and\ \bibinfo {author}
  {\bibfnamefont {C.}~\bibnamefont {Yannouleas}},\ }\bibfield  {title}
  {\bibinfo {title} {{Symmetry restoration in mean-field approaches}},\ }\href
  {https://doi.org/10.1088/1361-6471/ac288a} {\bibfield  {journal} {\bibinfo
  {journal} {J. Phys. G}\ }\textbf {\bibinfo {volume} {48}},\ \bibinfo {pages}
  {123001} (\bibinfo {year} {2021})},\ \Eprint
  {https://arxiv.org/abs/1901.06992} {arXiv:1901.06992 [nucl-th]} \BibitemShut
  {NoStop}%
\bibitem [{\citenamefont {Goeke}\ \emph {et~al.}(2007)\citenamefont {Goeke},
  \citenamefont {Grabis}, \citenamefont {Ossmann}, \citenamefont {Polyakov},
  \citenamefont {Schweitzer}, \citenamefont {Silva},\ and\ \citenamefont
  {Urbano}}]{Goeke:2007fp}%
  \BibitemOpen
  \bibfield  {author} {\bibinfo {author} {\bibfnamefont {K.}~\bibnamefont
  {Goeke}}, \bibinfo {author} {\bibfnamefont {J.}~\bibnamefont {Grabis}},
  \bibinfo {author} {\bibfnamefont {J.}~\bibnamefont {Ossmann}}, \bibinfo
  {author} {\bibfnamefont {M.~V.}\ \bibnamefont {Polyakov}}, \bibinfo {author}
  {\bibfnamefont {P.}~\bibnamefont {Schweitzer}}, \bibinfo {author}
  {\bibfnamefont {A.}~\bibnamefont {Silva}},\ and\ \bibinfo {author}
  {\bibfnamefont {D.}~\bibnamefont {Urbano}},\ }\bibfield  {title} {\bibinfo
  {title} {{Nucleon form-factors of the energy momentum tensor in the chiral
  quark-soliton model}},\ }\href {https://doi.org/10.1103/PhysRevD.75.094021}
  {\bibfield  {journal} {\bibinfo  {journal} {Phys. Rev. D}\ }\textbf {\bibinfo
  {volume} {75}},\ \bibinfo {pages} {094021} (\bibinfo {year} {2007})},\
  \Eprint {https://arxiv.org/abs/hep-ph/0702030} {arXiv:hep-ph/0702030}
  \BibitemShut {NoStop}%
\bibitem [{\citenamefont {Gockeler}\ \emph {et~al.}(2005)\citenamefont
  {Gockeler}, \citenamefont {Hagler}, \citenamefont {Horsley}, \citenamefont
  {Pleiter}, \citenamefont {Rakow}, \citenamefont {Schafer}, \citenamefont
  {Schierholz},\ and\ \citenamefont {Zanotti}}]{Gockeler:2005cj}%
  \BibitemOpen
  \bibfield  {author} {\bibinfo {author} {\bibfnamefont {M.}~\bibnamefont
  {Gockeler}}, \bibinfo {author} {\bibfnamefont {P.}~\bibnamefont {Hagler}},
  \bibinfo {author} {\bibfnamefont {R.}~\bibnamefont {Horsley}}, \bibinfo
  {author} {\bibfnamefont {D.}~\bibnamefont {Pleiter}}, \bibinfo {author}
  {\bibfnamefont {P.~E.~L.}\ \bibnamefont {Rakow}}, \bibinfo {author}
  {\bibfnamefont {A.}~\bibnamefont {Schafer}}, \bibinfo {author} {\bibfnamefont
  {G.}~\bibnamefont {Schierholz}},\ and\ \bibinfo {author} {\bibfnamefont
  {J.~M.}\ \bibnamefont {Zanotti}} (\bibinfo {collaboration} {QCDSF, UKQCD}),\
  }\bibfield  {title} {\bibinfo {title} {{Quark helicity flip generalized
  parton distributions from two-flavor lattice QCD}},\ }\href
  {https://doi.org/10.1016/j.physletb.2005.09.002} {\bibfield  {journal}
  {\bibinfo  {journal} {Phys. Lett. B}\ }\textbf {\bibinfo {volume} {627}},\
  \bibinfo {pages} {113} (\bibinfo {year} {2005})},\ \Eprint
  {https://arxiv.org/abs/hep-lat/0507001} {arXiv:hep-lat/0507001} \BibitemShut
  {NoStop}%
\bibitem [{\citenamefont {G\"ockeler}\ \emph {et~al.}(2007)\citenamefont
  {G\"ockeler}, \citenamefont {H\"agler}, \citenamefont {Horsley},
  \citenamefont {Nakamura}, \citenamefont {Pleiter}, \citenamefont {Rakow},
  \citenamefont {Sch\"afer}, \citenamefont {Schierholz}, \citenamefont
  {St\"uben},\ and\ \citenamefont {Zanotti}}]{QCDSF:2006tkx}%
  \BibitemOpen
  \bibfield  {author} {\bibinfo {author} {\bibfnamefont {M.}~\bibnamefont
  {G\"ockeler}}, \bibinfo {author} {\bibfnamefont {P.}~\bibnamefont
  {H\"agler}}, \bibinfo {author} {\bibfnamefont {R.}~\bibnamefont {Horsley}},
  \bibinfo {author} {\bibfnamefont {Y.}~\bibnamefont {Nakamura}}, \bibinfo
  {author} {\bibfnamefont {D.}~\bibnamefont {Pleiter}}, \bibinfo {author}
  {\bibfnamefont {P.~E.~L.}\ \bibnamefont {Rakow}}, \bibinfo {author}
  {\bibfnamefont {A.}~\bibnamefont {Sch\"afer}}, \bibinfo {author}
  {\bibfnamefont {G.}~\bibnamefont {Schierholz}}, \bibinfo {author}
  {\bibfnamefont {H.}~\bibnamefont {St\"uben}},\ and\ \bibinfo {author}
  {\bibfnamefont {J.~M.}\ \bibnamefont {Zanotti}} (\bibinfo {collaboration}
  {QCDSF, UKQCD}),\ }\bibfield  {title} {\bibinfo {title} {{Transverse spin
  structure of the nucleon from lattice QCD simulations}},\ }\href
  {https://doi.org/10.1103/PhysRevLett.98.222001} {\bibfield  {journal}
  {\bibinfo  {journal} {Phys. Rev. Lett.}\ }\textbf {\bibinfo {volume} {98}},\
  \bibinfo {pages} {222001} (\bibinfo {year} {2007})},\ \Eprint
  {https://arxiv.org/abs/hep-lat/0612032} {arXiv:hep-lat/0612032} \BibitemShut
  {NoStop}%
\bibitem [{\citenamefont {Park}\ \emph {et~al.}(2022)\citenamefont {Park},
  \citenamefont {Gupta}, \citenamefont {Yoon}, \citenamefont {Mondal},
  \citenamefont {Bhattacharya}, \citenamefont {Jang}, \citenamefont {Jo\'o},\
  and\ \citenamefont {Winter}}]{Park:2021ypf}%
  \BibitemOpen
  \bibfield  {author} {\bibinfo {author} {\bibfnamefont {S.}~\bibnamefont
  {Park}}, \bibinfo {author} {\bibfnamefont {R.}~\bibnamefont {Gupta}},
  \bibinfo {author} {\bibfnamefont {B.}~\bibnamefont {Yoon}}, \bibinfo {author}
  {\bibfnamefont {S.}~\bibnamefont {Mondal}}, \bibinfo {author} {\bibfnamefont
  {T.}~\bibnamefont {Bhattacharya}}, \bibinfo {author} {\bibfnamefont {Y.-C.}\
  \bibnamefont {Jang}}, \bibinfo {author} {\bibfnamefont {B.}~\bibnamefont
  {Jo\'o}},\ and\ \bibinfo {author} {\bibfnamefont {F.}~\bibnamefont {Winter}}
  (\bibinfo {collaboration} {Nucleon Matrix Elements (NME)}),\ }\bibfield
  {title} {\bibinfo {title} {{Precision nucleon charges and form factors using
  (2+1)-flavor lattice QCD}},\ }\href
  {https://doi.org/10.1103/PhysRevD.105.054505} {\bibfield  {journal} {\bibinfo
   {journal} {Phys. Rev. D}\ }\textbf {\bibinfo {volume} {105}},\ \bibinfo
  {pages} {054505} (\bibinfo {year} {2022})},\ \Eprint
  {https://arxiv.org/abs/2103.05599} {arXiv:2103.05599 [hep-lat]} \BibitemShut
  {NoStop}%
\bibitem [{\citenamefont {Alexandrou}\ \emph {et~al.}(2022)\citenamefont
  {Alexandrou}, \citenamefont {Cichy}, \citenamefont {Constantinou},
  \citenamefont {Hadjiyiannakou}, \citenamefont {Jansen}, \citenamefont
  {Scapellato},\ and\ \citenamefont {Steffens}}]{Alexandrou:2021bbo}%
  \BibitemOpen
  \bibfield  {author} {\bibinfo {author} {\bibfnamefont {C.}~\bibnamefont
  {Alexandrou}}, \bibinfo {author} {\bibfnamefont {K.}~\bibnamefont {Cichy}},
  \bibinfo {author} {\bibfnamefont {M.}~\bibnamefont {Constantinou}}, \bibinfo
  {author} {\bibfnamefont {K.}~\bibnamefont {Hadjiyiannakou}}, \bibinfo
  {author} {\bibfnamefont {K.}~\bibnamefont {Jansen}}, \bibinfo {author}
  {\bibfnamefont {A.}~\bibnamefont {Scapellato}},\ and\ \bibinfo {author}
  {\bibfnamefont {F.}~\bibnamefont {Steffens}},\ }\bibfield  {title} {\bibinfo
  {title} {{Transversity GPDs of the proton from lattice QCD}},\ }\href
  {https://doi.org/10.1103/PhysRevD.105.034501} {\bibfield  {journal} {\bibinfo
   {journal} {Phys. Rev. D}\ }\textbf {\bibinfo {volume} {105}},\ \bibinfo
  {pages} {034501} (\bibinfo {year} {2022})},\ \Eprint
  {https://arxiv.org/abs/2108.10789} {arXiv:2108.10789 [hep-lat]} \BibitemShut
  {NoStop}%
\bibitem [{\citenamefont {Alexandrou}\ \emph {et~al.}(2023)\citenamefont
  {Alexandrou} \emph {et~al.}}]{Alexandrou:2022dtc}%
  \BibitemOpen
  \bibfield  {author} {\bibinfo {author} {\bibfnamefont {C.}~\bibnamefont
  {Alexandrou}} \emph {et~al.},\ }\bibfield  {title} {\bibinfo {title}
  {{Moments of the nucleon transverse quark spin densities using lattice
  QCD}},\ }\href {https://doi.org/10.1103/PhysRevD.107.054504} {\bibfield
  {journal} {\bibinfo  {journal} {Phys. Rev. D}\ }\textbf {\bibinfo {volume}
  {107}},\ \bibinfo {pages} {054504} (\bibinfo {year} {2023})},\ \Eprint
  {https://arxiv.org/abs/2202.09871} {arXiv:2202.09871 [hep-lat]} \BibitemShut
  {NoStop}%
\bibitem [{\citenamefont {Hagler}(2004)}]{Hagler:2004yt}%
  \BibitemOpen
  \bibfield  {author} {\bibinfo {author} {\bibfnamefont {P.}~\bibnamefont
  {Hagler}},\ }\bibfield  {title} {\bibinfo {title} {{Form-factor decomposition
  of generalized parton distributions at leading twist}},\ }\href
  {https://doi.org/10.1016/j.physletb.2004.05.014} {\bibfield  {journal}
  {\bibinfo  {journal} {Phys. Lett. B}\ }\textbf {\bibinfo {volume} {594}},\
  \bibinfo {pages} {164} (\bibinfo {year} {2004})},\ \Eprint
  {https://arxiv.org/abs/hep-ph/0404138} {arXiv:hep-ph/0404138} \BibitemShut
  {NoStop}%
\bibitem [{\citenamefont {Brodsky}\ \emph {et~al.}(1998)\citenamefont
  {Brodsky}, \citenamefont {Pauli},\ and\ \citenamefont
  {Pinsky}}]{Brodsky:1997de}%
  \BibitemOpen
  \bibfield  {author} {\bibinfo {author} {\bibfnamefont {S.~J.}\ \bibnamefont
  {Brodsky}}, \bibinfo {author} {\bibfnamefont {H.-C.}\ \bibnamefont {Pauli}},\
  and\ \bibinfo {author} {\bibfnamefont {S.~S.}\ \bibnamefont {Pinsky}},\
  }\bibfield  {title} {\bibinfo {title} {{Quantum chromodynamics and other
  field theories on the light cone}},\ }\href
  {https://doi.org/10.1016/S0370-1573(97)00089-6} {\bibfield  {journal}
  {\bibinfo  {journal} {Phys. Rept.}\ }\textbf {\bibinfo {volume} {301}},\
  \bibinfo {pages} {299} (\bibinfo {year} {1998})},\ \Eprint
  {https://arxiv.org/abs/hep-ph/9705477} {arXiv:hep-ph/9705477} \BibitemShut
  {NoStop}%
\bibitem [{\citenamefont {Cosyn}\ and\ \citenamefont
  {Weiss}(2020)}]{Cosyn:2020kwu}%
  \BibitemOpen
  \bibfield  {author} {\bibinfo {author} {\bibfnamefont {W.}~\bibnamefont
  {Cosyn}}\ and\ \bibinfo {author} {\bibfnamefont {C.}~\bibnamefont {Weiss}},\
  }\bibfield  {title} {\bibinfo {title} {{Polarized electron-deuteron
  deep-inelastic scattering with spectator nucleon tagging}},\ }\href
  {https://doi.org/10.1103/PhysRevC.102.065204} {\bibfield  {journal} {\bibinfo
   {journal} {Phys. Rev. C}\ }\textbf {\bibinfo {volume} {102}},\ \bibinfo
  {pages} {065204} (\bibinfo {year} {2020})},\ \Eprint
  {https://arxiv.org/abs/2006.03033} {arXiv:2006.03033 [hep-ph]} \BibitemShut
  {NoStop}%
\bibitem [{\citenamefont {Kim}\ \emph {et~al.}(2023{\natexlab{a}})\citenamefont
  {Kim}, \citenamefont {Sun}, \citenamefont {Fu},\ and\ \citenamefont
  {Kim}}]{Kim:2022wkc}%
  \BibitemOpen
  \bibfield  {author} {\bibinfo {author} {\bibfnamefont {J.-Y.}\ \bibnamefont
  {Kim}}, \bibinfo {author} {\bibfnamefont {B.-D.}\ \bibnamefont {Sun}},
  \bibinfo {author} {\bibfnamefont {D.}~\bibnamefont {Fu}},\ and\ \bibinfo
  {author} {\bibfnamefont {H.-C.}\ \bibnamefont {Kim}},\ }\bibfield  {title}
  {\bibinfo {title} {{Mechanical structure of a spin-1 particle}},\ }\href
  {https://doi.org/10.1103/PhysRevD.107.054007} {\bibfield  {journal} {\bibinfo
   {journal} {Phys. Rev. D}\ }\textbf {\bibinfo {volume} {107}},\ \bibinfo
  {pages} {054007} (\bibinfo {year} {2023}{\natexlab{a}})},\ \Eprint
  {https://arxiv.org/abs/2208.01240} {arXiv:2208.01240 [hep-ph]} \BibitemShut
  {NoStop}%
\bibitem [{\citenamefont {Kim}\ \emph {et~al.}(2023{\natexlab{b}})\citenamefont
  {Kim}, \citenamefont {Won}, \citenamefont {Goity},\ and\ \citenamefont
  {Weiss}}]{Kim:2023xvw}%
  \BibitemOpen
  \bibfield  {author} {\bibinfo {author} {\bibfnamefont {J.-Y.}\ \bibnamefont
  {Kim}}, \bibinfo {author} {\bibfnamefont {H.-Y.}\ \bibnamefont {Won}},
  \bibinfo {author} {\bibfnamefont {J.~L.}\ \bibnamefont {Goity}},\ and\
  \bibinfo {author} {\bibfnamefont {C.}~\bibnamefont {Weiss}},\ }\bibfield
  {title} {\bibinfo {title} {{QCD angular momentum in $N \rightarrow \Delta$
  transitions}},\ }\href {https://doi.org/10.1016/j.physletb.2023.138083}
  {\bibfield  {journal} {\bibinfo  {journal} {Phys. Lett. B}\ }\textbf
  {\bibinfo {volume} {844}},\ \bibinfo {pages} {138083} (\bibinfo {year}
  {2023}{\natexlab{b}})},\ \Eprint {https://arxiv.org/abs/2304.08575}
  {arXiv:2304.08575 [hep-ph]} \BibitemShut {NoStop}%
\bibitem [{\citenamefont {Kim}(2022)}]{Kim:2022bia}%
  \BibitemOpen
  \bibfield  {author} {\bibinfo {author} {\bibfnamefont {J.-Y.}\ \bibnamefont
  {Kim}},\ }\bibfield  {title} {\bibinfo {title} {{Electromagnetic multipole
  structure of a spin-one particle: Abel tomography case}},\ }\href
  {https://doi.org/10.1103/PhysRevD.106.014022} {\bibfield  {journal} {\bibinfo
   {journal} {Phys. Rev. D}\ }\textbf {\bibinfo {volume} {106}},\ \bibinfo
  {pages} {014022} (\bibinfo {year} {2022})},\ \Eprint
  {https://arxiv.org/abs/2204.08248} {arXiv:2204.08248 [hep-ph]} \BibitemShut
  {NoStop}%
\bibitem [{\citenamefont {Lorc\'e}(2020)}]{Lorce:2020onh}%
  \BibitemOpen
  \bibfield  {author} {\bibinfo {author} {\bibfnamefont {C.}~\bibnamefont
  {Lorc\'e}},\ }\bibfield  {title} {\bibinfo {title} {{Charge Distributions of
  Moving Nucleons}},\ }\href {https://doi.org/10.1103/PhysRevLett.125.232002}
  {\bibfield  {journal} {\bibinfo  {journal} {Phys. Rev. Lett.}\ }\textbf
  {\bibinfo {volume} {125}},\ \bibinfo {pages} {232002} (\bibinfo {year}
  {2020})},\ \Eprint {https://arxiv.org/abs/2007.05318} {arXiv:2007.05318
  [hep-ph]} \BibitemShut {NoStop}%
\bibitem [{\citenamefont {Witten}(1983)}]{Witten:1983tx}%
  \BibitemOpen
  \bibfield  {author} {\bibinfo {author} {\bibfnamefont {E.}~\bibnamefont
  {Witten}},\ }\bibfield  {title} {\bibinfo {title} {{Current Algebra, Baryons,
  and Quark Confinement}},\ }\href
  {https://doi.org/10.1016/0550-3213(83)90064-0} {\bibfield  {journal}
  {\bibinfo  {journal} {Nucl. Phys. B}\ }\textbf {\bibinfo {volume} {223}},\
  \bibinfo {pages} {433} (\bibinfo {year} {1983})}\BibitemShut {NoStop}%
\bibitem [{\citenamefont {Berestetskii}\ \emph {et~al.}(1982)\citenamefont
  {Berestetskii}, \citenamefont {Lifshitz},\ and\ \citenamefont
  {Pitaevskii}}]{Berestetskii:1982qgu}%
  \BibitemOpen
  \bibfield  {author} {\bibinfo {author} {\bibfnamefont {V.~B.}\ \bibnamefont
  {Berestetskii}}, \bibinfo {author} {\bibfnamefont {E.~M.}\ \bibnamefont
  {Lifshitz}},\ and\ \bibinfo {author} {\bibfnamefont {L.~P.}\ \bibnamefont
  {Pitaevskii}},\ }\href@noop {} {\emph {\bibinfo {title} {{Quantum
  Electrodynamics}}}},\ \bibinfo {series} {Course of Theoretical Physics},
  Vol.~\bibinfo {volume} {4}\ (\bibinfo  {publisher} {Pergamon Press},\
  \bibinfo {address} {Oxford},\ \bibinfo {year} {1982})\BibitemShut {NoStop}%
\bibitem [{\citenamefont {Pobylitsa}\ and\ \citenamefont
  {Polyakov}(2000)}]{Pobylitsa:2000tt}%
  \BibitemOpen
  \bibfield  {author} {\bibinfo {author} {\bibfnamefont {P.~V.}\ \bibnamefont
  {Pobylitsa}}\ and\ \bibinfo {author} {\bibfnamefont {M.~V.}\ \bibnamefont
  {Polyakov}},\ }\bibfield  {title} {\bibinfo {title} {{New positivity bounds
  on parton distributions in multicolored QCD}},\ }\href
  {https://doi.org/10.1103/PhysRevD.62.097502} {\bibfield  {journal} {\bibinfo
  {journal} {Phys. Rev. D}\ }\textbf {\bibinfo {volume} {62}},\ \bibinfo
  {pages} {097502} (\bibinfo {year} {2000})},\ \Eprint
  {https://arxiv.org/abs/hep-ph/0004094} {arXiv:hep-ph/0004094} \BibitemShut
  {NoStop}%
\bibitem [{\citenamefont {Diakonov}\ \emph {et~al.}(1996)\citenamefont
  {Diakonov}, \citenamefont {Petrov}, \citenamefont {Pobylitsa}, \citenamefont
  {Polyakov},\ and\ \citenamefont {Weiss}}]{Diakonov:1996sr}%
  \BibitemOpen
  \bibfield  {author} {\bibinfo {author} {\bibfnamefont {D.}~\bibnamefont
  {Diakonov}}, \bibinfo {author} {\bibfnamefont {V.}~\bibnamefont {Petrov}},
  \bibinfo {author} {\bibfnamefont {P.}~\bibnamefont {Pobylitsa}}, \bibinfo
  {author} {\bibfnamefont {M.~V.}\ \bibnamefont {Polyakov}},\ and\ \bibinfo
  {author} {\bibfnamefont {C.}~\bibnamefont {Weiss}},\ }\bibfield  {title}
  {\bibinfo {title} {{Nucleon parton distributions at low normalization point
  in the large $N_c$ limit}},\ }\href
  {https://doi.org/10.1016/S0550-3213(96)00486-5} {\bibfield  {journal}
  {\bibinfo  {journal} {Nucl. Phys. B}\ }\textbf {\bibinfo {volume} {480}},\
  \bibinfo {pages} {341} (\bibinfo {year} {1996})},\ \Eprint
  {https://arxiv.org/abs/hep-ph/9606314} {arXiv:hep-ph/9606314} \BibitemShut
  {NoStop}%
\bibitem [{\citenamefont {Varshalovich}\ \emph {et~al.}(1988)\citenamefont
  {Varshalovich}, \citenamefont {Moskalev},\ and\ \citenamefont
  {Khersonskii}}]{Varshalovich:1988ifq}%
  \BibitemOpen
  \bibfield  {author} {\bibinfo {author} {\bibfnamefont {D.~A.}\ \bibnamefont
  {Varshalovich}}, \bibinfo {author} {\bibfnamefont {A.~N.}\ \bibnamefont
  {Moskalev}},\ and\ \bibinfo {author} {\bibfnamefont {V.~K.}\ \bibnamefont
  {Khersonskii}},\ }\href {https://doi.org/10.1142/0270} {\emph {\bibinfo
  {title} {{Quantum Theory of Angular Momentum}: {Irreducible Tensors,
  Spherical Harmonics, Vector Coupling Coefficients, 3nj Symbols}}}}\ (\bibinfo
   {publisher} {World Scientific Publishing Company},\ \bibinfo {year}
  {1988})\BibitemShut {NoStop}%
\bibitem [{\citenamefont {Tezgin}\ \emph {et~al.}(2024)\citenamefont {Tezgin},
  \citenamefont {Maynard},\ and\ \citenamefont {Schweitzer}}]{Tezgin:2024tfh}%
  \BibitemOpen
  \bibfield  {author} {\bibinfo {author} {\bibfnamefont {K.}~\bibnamefont
  {Tezgin}}, \bibinfo {author} {\bibfnamefont {B.}~\bibnamefont {Maynard}},\
  and\ \bibinfo {author} {\bibfnamefont {P.}~\bibnamefont {Schweitzer}},\
  }\bibfield  {title} {\bibinfo {title} {{Chiral-odd GPDs in the bag model}},\
  }\href {https://doi.org/10.1103/PhysRevD.110.054028} {\bibfield  {journal}
  {\bibinfo  {journal} {Phys. Rev. D}\ }\textbf {\bibinfo {volume} {110}},\
  \bibinfo {pages} {054028} (\bibinfo {year} {2024})},\ \Eprint
  {https://arxiv.org/abs/2404.11563} {arXiv:2404.11563 [hep-ph]} \BibitemShut
  {NoStop}%
\bibitem [{\citenamefont {Pasquini}\ \emph {et~al.}(2005)\citenamefont
  {Pasquini}, \citenamefont {Pincetti},\ and\ \citenamefont
  {Boffi}}]{Pasquini:2005dk}%
  \BibitemOpen
  \bibfield  {author} {\bibinfo {author} {\bibfnamefont {B.}~\bibnamefont
  {Pasquini}}, \bibinfo {author} {\bibfnamefont {M.}~\bibnamefont {Pincetti}},\
  and\ \bibinfo {author} {\bibfnamefont {S.}~\bibnamefont {Boffi}},\ }\bibfield
   {title} {\bibinfo {title} {{Chiral-odd generalized parton distributions in
  constituent quark models}},\ }\href
  {https://doi.org/10.1103/PhysRevD.72.094029} {\bibfield  {journal} {\bibinfo
  {journal} {Phys. Rev. D}\ }\textbf {\bibinfo {volume} {72}},\ \bibinfo
  {pages} {094029} (\bibinfo {year} {2005})},\ \Eprint
  {https://arxiv.org/abs/hep-ph/0510376} {arXiv:hep-ph/0510376} \BibitemShut
  {NoStop}%
\bibitem [{\citenamefont {Kim}\ and\ \citenamefont {Weiss}()}]{inprep}%
  \BibitemOpen
  \bibfield  {author} {\bibinfo {author} {\bibfnamefont {J.-Y.}\ \bibnamefont
  {Kim}}\ and\ \bibinfo {author} {\bibfnamefont {C.}~\bibnamefont {Weiss}},\
  }\href@noop {} {\bibinfo {title} {{in prepration}}}\BibitemShut {NoStop}%
\bibitem [{\citenamefont {Schweitzer}\ \emph {et~al.}(2002)\citenamefont
  {Schweitzer}, \citenamefont {Boffi},\ and\ \citenamefont
  {Radici}}]{Schweitzer:2002nm}%
  \BibitemOpen
  \bibfield  {author} {\bibinfo {author} {\bibfnamefont {P.}~\bibnamefont
  {Schweitzer}}, \bibinfo {author} {\bibfnamefont {S.}~\bibnamefont {Boffi}},\
  and\ \bibinfo {author} {\bibfnamefont {M.}~\bibnamefont {Radici}},\
  }\bibfield  {title} {\bibinfo {title} {{Polynomiality of unpolarized off
  forward distribution functions and the $D$ term in the chiral quark soliton
  model}},\ }\href {https://doi.org/10.1103/PhysRevD.66.114004} {\bibfield
  {journal} {\bibinfo  {journal} {Phys. Rev. D}\ }\textbf {\bibinfo {volume}
  {66}},\ \bibinfo {pages} {114004} (\bibinfo {year} {2002})},\ \Eprint
  {https://arxiv.org/abs/hep-ph/0207230} {arXiv:hep-ph/0207230} \BibitemShut
  {NoStop}%
\bibitem [{\citenamefont {Schweitzer}\ \emph {et~al.}(2003)\citenamefont
  {Schweitzer}, \citenamefont {Colli},\ and\ \citenamefont
  {Boffi}}]{Schweitzer:2003ms}%
  \BibitemOpen
  \bibfield  {author} {\bibinfo {author} {\bibfnamefont {P.}~\bibnamefont
  {Schweitzer}}, \bibinfo {author} {\bibfnamefont {M.}~\bibnamefont {Colli}},\
  and\ \bibinfo {author} {\bibfnamefont {S.}~\bibnamefont {Boffi}},\ }\bibfield
   {title} {\bibinfo {title} {{Polynomiality of helicity off forward
  distribution functions in the chiral quark soliton model}},\ }\href
  {https://doi.org/10.1103/PhysRevD.67.114022} {\bibfield  {journal} {\bibinfo
  {journal} {Phys. Rev. D}\ }\textbf {\bibinfo {volume} {67}},\ \bibinfo
  {pages} {114022} (\bibinfo {year} {2003})},\ \Eprint
  {https://arxiv.org/abs/hep-ph/0303166} {arXiv:hep-ph/0303166} \BibitemShut
  {NoStop}%
\bibitem [{\citenamefont {Diakonov}\ \emph {et~al.}(1997)\citenamefont
  {Diakonov}, \citenamefont {Petrov}, \citenamefont {Pobylitsa}, \citenamefont
  {Polyakov},\ and\ \citenamefont {Weiss}}]{Diakonov:1997vc}%
  \BibitemOpen
  \bibfield  {author} {\bibinfo {author} {\bibfnamefont {D.}~\bibnamefont
  {Diakonov}}, \bibinfo {author} {\bibfnamefont {V.~Y.}\ \bibnamefont
  {Petrov}}, \bibinfo {author} {\bibfnamefont {P.~V.}\ \bibnamefont
  {Pobylitsa}}, \bibinfo {author} {\bibfnamefont {M.~V.}\ \bibnamefont
  {Polyakov}},\ and\ \bibinfo {author} {\bibfnamefont {C.}~\bibnamefont
  {Weiss}},\ }\bibfield  {title} {\bibinfo {title} {{Unpolarized and polarized
  quark distributions in the large $N_c$ limit}},\ }\href
  {https://doi.org/10.1103/PhysRevD.56.4069} {\bibfield  {journal} {\bibinfo
  {journal} {Phys. Rev. D}\ }\textbf {\bibinfo {volume} {56}},\ \bibinfo
  {pages} {4069} (\bibinfo {year} {1997})},\ \Eprint
  {https://arxiv.org/abs/hep-ph/9703420} {arXiv:hep-ph/9703420} \BibitemShut
  {NoStop}%
\bibitem [{\citenamefont {Landau}\ and\ \citenamefont
  {Lifshits}(1991)}]{Landau:1991wop}%
  \BibitemOpen
  \bibfield  {author} {\bibinfo {author} {\bibfnamefont {L.~D.}\ \bibnamefont
  {Landau}}\ and\ \bibinfo {author} {\bibfnamefont {E.~M.}\ \bibnamefont
  {Lifshits}},\ }\href@noop {} {\emph {\bibinfo {title} {{Quantum Mechanics}:
  {Non-Relativistic Theory}}}},\ \bibinfo {series} {Course of Theoretical
  Physics}, Vol.\ \bibinfo {volume} {v.3}\ (\bibinfo  {publisher}
  {Butterworth-Heinemann},\ \bibinfo {address} {Oxford},\ \bibinfo {year}
  {1991})\BibitemShut {NoStop}%
\bibitem [{\citenamefont {Kim}\ \emph {et~al.}(1996{\natexlab{a}})\citenamefont
  {Kim}, \citenamefont {Polyakov},\ and\ \citenamefont {Goeke}}]{Kim:1995bq}%
  \BibitemOpen
  \bibfield  {author} {\bibinfo {author} {\bibfnamefont {H.-C.}\ \bibnamefont
  {Kim}}, \bibinfo {author} {\bibfnamefont {M.~V.}\ \bibnamefont {Polyakov}},\
  and\ \bibinfo {author} {\bibfnamefont {K.}~\bibnamefont {Goeke}},\ }\bibfield
   {title} {\bibinfo {title} {{Nucleon tensor charges in the SU(2) chiral
  quark-soliton model}},\ }\href {https://doi.org/10.1103/PhysRevD.53.R4715}
  {\bibfield  {journal} {\bibinfo  {journal} {Phys. Rev. D}\ }\textbf {\bibinfo
  {volume} {53}},\ \bibinfo {pages} {4715} (\bibinfo {year}
  {1996}{\natexlab{a}})},\ \Eprint {https://arxiv.org/abs/hep-ph/9509283}
  {arXiv:hep-ph/9509283} \BibitemShut {NoStop}%
\bibitem [{\citenamefont {Kim}\ \emph {et~al.}(1996{\natexlab{b}})\citenamefont
  {Kim}, \citenamefont {Polyakov},\ and\ \citenamefont {Goeke}}]{Kim:1996vk}%
  \BibitemOpen
  \bibfield  {author} {\bibinfo {author} {\bibfnamefont {H.-C.}\ \bibnamefont
  {Kim}}, \bibinfo {author} {\bibfnamefont {M.~V.}\ \bibnamefont {Polyakov}},\
  and\ \bibinfo {author} {\bibfnamefont {K.}~\bibnamefont {Goeke}},\ }\bibfield
   {title} {\bibinfo {title} {{Tensor charges of the nucleon in the SU(3)
  chiral quark soliton model}},\ }\href
  {https://doi.org/10.1016/0370-2693(96)01066-0} {\bibfield  {journal}
  {\bibinfo  {journal} {Phys. Lett. B}\ }\textbf {\bibinfo {volume} {387}},\
  \bibinfo {pages} {577} (\bibinfo {year} {1996}{\natexlab{b}})},\ \Eprint
  {https://arxiv.org/abs/hep-ph/9604442} {arXiv:hep-ph/9604442} \BibitemShut
  {NoStop}%
\bibitem [{\citenamefont {Ledwig}\ \emph {et~al.}(2010)\citenamefont {Ledwig},
  \citenamefont {Silva},\ and\ \citenamefont {Kim}}]{Ledwig:2010zq}%
  \BibitemOpen
  \bibfield  {author} {\bibinfo {author} {\bibfnamefont {T.}~\bibnamefont
  {Ledwig}}, \bibinfo {author} {\bibfnamefont {A.}~\bibnamefont {Silva}},\ and\
  \bibinfo {author} {\bibfnamefont {H.-C.}\ \bibnamefont {Kim}},\ }\bibfield
  {title} {\bibinfo {title} {{Anomalous tensor magnetic moments and form
  factors of the proton in the self-consistent chiral quark-soliton model}},\
  }\href {https://doi.org/10.1103/PhysRevD.82.054014} {\bibfield  {journal}
  {\bibinfo  {journal} {Phys. Rev. D}\ }\textbf {\bibinfo {volume} {82}},\
  \bibinfo {pages} {054014} (\bibinfo {year} {2010})},\ \Eprint
  {https://arxiv.org/abs/1007.1355} {arXiv:1007.1355 [hep-ph]} \BibitemShut
  {NoStop}%
\bibitem [{\citenamefont {Diehl}\ \emph {et~al.}(2024)\citenamefont {Diehl}
  \emph {et~al.}}]{Diehl:2024bmd}%
  \BibitemOpen
  \bibfield  {author} {\bibinfo {author} {\bibfnamefont {S.}~\bibnamefont
  {Diehl}} \emph {et~al.},\ }\bibfield  {title} {\bibinfo {title} {{Exploring
  Baryon Resonances with Transition Generalized Parton Distributions: Status
  and Perspectives}},\ }\href@noop {} {\  (\bibinfo {year} {2024})},\ \Eprint
  {https://arxiv.org/abs/2405.15386} {arXiv:2405.15386 [hep-ph]} \BibitemShut
  {NoStop}%
\bibitem [{\citenamefont {Kroll}\ and\ \citenamefont
  {Passek-Kumeri\v{c}ki}(2023)}]{Kroll:2022roq}%
  \BibitemOpen
  \bibfield  {author} {\bibinfo {author} {\bibfnamefont {P.}~\bibnamefont
  {Kroll}}\ and\ \bibinfo {author} {\bibfnamefont {K.}~\bibnamefont
  {Passek-Kumeri\v{c}ki}},\ }\bibfield  {title} {\bibinfo {title} {{Transition
  GPDs and exclusive electroproduction of $\pi$-$\Delta$(1232) final states}},\
  }\href {https://doi.org/10.1103/PhysRevD.107.054009} {\bibfield  {journal}
  {\bibinfo  {journal} {Phys. Rev. D}\ }\textbf {\bibinfo {volume} {107}},\
  \bibinfo {pages} {054009} (\bibinfo {year} {2023})},\ \Eprint
  {https://arxiv.org/abs/2211.09474} {arXiv:2211.09474 [hep-ph]} \BibitemShut
  {NoStop}%
\bibitem [{\citenamefont {Diehl}\ \emph {et~al.}(2023)\citenamefont {Diehl}
  \emph {et~al.}}]{CLAS:2023akb}%
  \BibitemOpen
  \bibfield  {author} {\bibinfo {author} {\bibfnamefont {S.}~\bibnamefont
  {Diehl}} \emph {et~al.} (\bibinfo {collaboration} {CLAS}),\ }\bibfield
  {title} {\bibinfo {title} {{First Measurement of Hard Exclusive $\pi^-
  \Delta^{++}$ Electroproduction Beam-Spin Asymmetries off the Proton}},\
  }\href {https://doi.org/10.1103/PhysRevLett.131.021901} {\bibfield  {journal}
  {\bibinfo  {journal} {Phys. Rev. Lett.}\ }\textbf {\bibinfo {volume} {131}},\
  \bibinfo {pages} {021901} (\bibinfo {year} {2023})},\ \Eprint
  {https://arxiv.org/abs/2303.11762} {arXiv:2303.11762 [hep-ex]} \BibitemShut
  {NoStop}%
\bibitem [{\citenamefont {Bhoonah}\ and\ \citenamefont
  {Lorc\'e}(2017)}]{Bhoonah:2017olu}%
  \BibitemOpen
  \bibfield  {author} {\bibinfo {author} {\bibfnamefont {A.}~\bibnamefont
  {Bhoonah}}\ and\ \bibinfo {author} {\bibfnamefont {C.}~\bibnamefont
  {Lorc\'e}},\ }\bibfield  {title} {\bibinfo {title} {{Quark transverse
  spin\textendash{}orbit correlations}},\ }\href
  {https://doi.org/10.1016/j.physletb.2017.10.003} {\bibfield  {journal}
  {\bibinfo  {journal} {Phys. Lett. B}\ }\textbf {\bibinfo {volume} {774}},\
  \bibinfo {pages} {435} (\bibinfo {year} {2017})},\ \Eprint
  {https://arxiv.org/abs/1703.08322} {arXiv:1703.08322 [hep-ph]} \BibitemShut
  {NoStop}%
\end{thebibliography}%
\end{document}